\documentclass[longauth]{aa-nolinenum}

\usepackage{amsmath}
\usepackage{atbegshi}
\usepackage{booktabs}
\usepackage{color}
\usepackage{graphicx}
\usepackage[breaklinks=true, hidelinks]{hyperref}
\usepackage{natbib}
\usepackage{pdflscape}
\usepackage{silence}
\usepackage{siunitx}
\usepackage{soul} 
\usepackage[varg]{txfonts}

\AtBeginDocument{\AtBeginShipoutNext{\AtBeginShipoutDiscard}}

\makeatletter
\renewcommand*\aa@pageof{, page \thepage{} of \pageref*{LastPage}}
\makeatother

\newcommand{\change}[1]{#1}

\newcommand{\HI}{\ion{H}{i}\xspace}
\newcommand{\HII}{\ion{H}{ii}\xspace}
\newcommand{\hmol}{H$_2$\xspace}

\newcommand{\jwst}{JWST}
\newcommand{\mum}{\textmu{}m\xspace}
\newcommand{\ionwav}[3]{[\ion{#1}{#2}]~#3~\mum}
\newcommand{\lineunit}{erg s$^{-1}$ cm$^{-2}$ sr$^{-1}$}

\begin{document}

\maketitle


\title{PDRs4All VIII: Mid-IR emission line inventory of the Orion Bar}

\author{%
  Dries Van De Putte\inst{\ref{STScI}} \and
  Raphael Meshaka \inst{\ref{SaclayIAS}, \ref{SorbonneLERMA}} \and
  Boris Trahin \inst{\ref{SaclayIAS}, \ref{STScI}} \and
  Emilie Habart \inst{\ref{SaclayIAS}} \and
  Els Peeters \inst{\ref{WOntarioPA}, \ref{WOntarioIESE}, \ref{CarlSagan}} \and
  Olivier Berné \inst{\ref{ToulouseIRAP}} \and
  Felipe Alarcón \inst{\ref{Michigan}} \and
  Amélie Canin \inst{\ref{ToulouseIRAP}} \and
  Ryan Chown \inst{\ref{WOntarioPA}, \ref{WOntarioIESE}, \ref{OhioState}} \and
  Ilane Schroetter \inst{\ref{ToulouseIRAP}} \and
  Ameek Sidhu \inst{\ref{WOntarioPA}, \ref{WOntarioIESE}} \and
  Christiaan Boersma \inst{\ref{AMES}} \and
  Emeric Bron \inst{\ref{SorbonneLERMA}} \and
  Emmanuel Dartois \inst{\ref{SaclayISM}} \and
  Javier R. Goicoechea \inst{\ref{MadridFisFundamental}} \and
  Karl D. Gordon \inst{\ref{STScI}, \ref{Ghent}} \and
  Takashi Onaka \inst{\ref{TokyoAstro}} \and
  Alexander G.~G.~M. Tielens \inst{\ref{Leiden}, \ref{Maryland}} \and
  Laurent Verstraete \inst{\ref{SaclayIAS}} \and
  Mark G. Wolfire \inst{\ref{Maryland}} \and 
  Alain Abergel \inst{\ref{SaclayIAS}}  \and
  Edwin A. Bergin \inst{\ref{Michigan}} \and
  Jeronimo Bernard-Salas \inst{\ref{GrasseACRIST}, \ref{GrasseINCLASS}} \and
  Jan Cami \inst{\ref{WOntarioPA}, \ref{WOntarioIESE}, \ref{CarlSagan}} \and
  Sara Cuadrado \inst{\ref{MadridFisFundamental}} \and
  Daniel Dicken \inst{\ref{Edinburgh}} \and
  Meriem Elyajouri \inst{\ref{SaclayIAS}} \and
  Asunción Fuente \inst{\ref{CSIC}} \and
  Christine Joblin \inst{\ref{ToulouseIRAP}} \and
  Baria Khan \inst{\ref{WOntarioPA}} \and
  Ozan Lacinbala \inst{\ref{LeuvenQSP}} \and
  David Languignon \inst{\ref{SorbonneLERMA}} \and
  Romane Le Gal \inst{\ref{GrenobleIPAG}, \ref{IRAM}} \and
  Alexandros Maragkoudakis \inst{\ref{AMES}} \and
  Yoko Okada \inst{\ref{Koln}} \and
  Sofia Pasquini \inst{\ref{WOntarioPA}} \and
  Marc W. Pound \inst{\ref{Maryland}} \and
  Massimo Robberto \inst{\ref{STScI}, \ref{JHU}} \and
  Markus Röllig \inst{\ref{FrankfurtPhys}, \ref{Goethe}} \and
  Bethany Schefter \inst{\ref{WOntarioPA}, \ref{WOntarioIESE}} \and
  Thiébaut Schirmer \inst{\ref{SaclayIAS}, \ref{ChalmersSpace}} \and
  Benoit Tabone \inst{\ref{SaclayIAS}} \and
  Sílvia Vicente \inst{\ref{Lisboa}} \and
  Marion Zannese \inst{\ref{SaclayIAS}} \and
  Sean W.J. Colgan \inst{\ref{AMES}} \and
  Jinhua He \inst{\ref{YunnanObs}, \ref{ChineseSouthAmCenter}, \ref{Chile}} \and
  Gaël Rouillé \inst{\ref{Jena}} \and
  Aditya Togi \inst{\ref{TexasState}} \and
  Isabel Aleman \inst{\ref{ItajubaFisQui}, \ref{SaoPauloMath}} \and
  Rebecca Auchettl \inst{\ref{VictoriaANSTO}} \and
  Giuseppe Antonio Baratta \inst{\ref{INAFcatania}} \and
  Salma Bejaoui \inst{\ref{AMES}} \and
  Partha P. Bera \inst{\ref{AMES}, \ref{BayArea}} \and
  John H. Black \inst{\ref{ChalmersSpace}} \and
  Francois Boulanger \inst{\ref{SorbonneLab}} \and
  Jordy Bouwman \inst{\ref{BoulderLab}, \ref{BoulderChem}, \ref{BoulderIMPACT}} \and
  Bernhard Brandl \inst{\ref{Leiden}, \ref{DelftAero}} \and
  Philippe Brechignac \inst{\ref{SaclayISM}} \and
  Sandra Brünken \inst{\ref{Radboud}} \and
  Mridusmita Buragohain \inst{\ref{HyderabadDST}} \and
  Andrew Burkhardt \inst{\ref{Wellesley}} \and
  Alessandra Candian \inst{\ref{Pannekoek}} \and
  Stéphanie Cazaux \inst{\ref{DelftTech}} \and
  Jose Cernicharo \inst{\ref{MadridFisFundamental}} \and
  Marin Chabot \inst{\ref{SaclayLabphys}} \and
  Shubhadip Chakraborty \inst{\ref{BangaloreChem}, \ref{Rennes}} \and
  Jason Champion \inst{\ref{ToulouseIRAP}} \and
  Ilsa R. Cooke \inst{\ref{BritishColumbiaChem}} \and
  Audrey Coutens \inst{\ref{ToulouseIRAP}} \and
  Nick L.J. Cox \inst{\ref{GrasseACRIST}, \ref{GrasseINCLASS}} \and
  Karine Demyk \inst{\ref{ToulouseIRAP}} \and
  Jennifer Donovan Meyer \inst{\ref{NRAO}} \and
  Sacha Foschino \inst{\ref{ToulouseIRAP}} \and
  Pedro García-Lario \inst{\ref{MadridESAC}} \and
  Maryvonne Gerin \inst{\ref{SorbonneObs}} \and
  Carl A. Gottlieb \inst{\ref{CambridgeHarvarSmithsonianAstro}} \and
  Pierre Guillard \inst{\ref{Sorbonne}, \ref{France}} \and
  Antoine Gusdorf \inst{\ref{SorbonneLab},\ref{SorbonneObs}} \and
  Patrick Hartigan \inst{\ref{Rice}} \and
  Eric Herbst \inst{\ref{Virginia}} \and
  Liv Hornekaer \inst{\ref{Aarhus}} \and
  Lina Issa \inst{\ref{ToulouseIRAP}} \and 
  Cornelia Jäger\inst{\ref{Jena}} \and
  Eduardo Janot-Pacheco\inst{\ref{SaoPauloAstro}} \and
  Olga Kannavou \inst{\ref{SaclayIAS}} \and
  Michael Kaufman\inst{\ref{SanJose}} \and
  Francisca Kemper\inst{\ref{ICECSIC}, \ref{ICREA}, \ref{IEEC}} \and
  Sarah Kendrew\inst{\ref{STScIESA}} \and
  Maria S. Kirsanova\inst{\ref{Russia}} \and
  Pamela Klaassen\inst{\ref{Edinburgh}} \and
  Sun Kwok\inst{\ref{BritishColumbiaEarth}} \and
  Álvaro Labiano \inst{\ref{MadridUKESA}} \and
  Thomas S.-Y. Lai \inst{\ref{IPAC}} \and
  Bertrand Le Floch \inst{\ref{BordeauxLab}} \and
  Franck Le Petit \inst{\ref{SorbonneLERMA}} \and
  Aigen Li \inst{\ref{Missouri}} \and
  Hendrik Linz \inst{\ref{Heidelberg}} \and
  Cameron J. Mackie \inst{\ref{BerkelyChem}, \ref{BerkelyPitzer}} \and
  Suzanne C. Madden \inst{\ref{SaclayAIMCEACNRS}} \and
  Jo\"elle Mascetti \inst{\ref{BordeauxMol}} \and
  Brett A. McGuire \inst{\ref{NRAO}, \ref{MITChem}} \and
  Pablo Merino \inst{\ref{MadridMateriales}} \and
  Elisabetta R. Micelotta \inst{\ref{Helsinki}} \and
  Jon A. Morse \inst{\ref{AstronetX}} \and
  Giacomo Mulas \inst{\ref{INAFcagliari}, \ref{ToulouseIRAP}} \and
  Naslim Neelamkodan \inst{\ref{Emirates}} \and
  Ryou Ohsawa \inst{\ref{TokyoNationalObs}} \and
  Alain Omont \inst{\ref{SorbonneObs}} \and
  Roberta Paladini \inst{\ref{IPAC}} \and
  Maria Elisabetta Palumbo \inst{\ref{INAFcatania}} \and
  Amit Pathak \inst{\ref{Banaras}} \and
  Yvonne J. Pendleton \inst{\ref{Florida}} \and
  Annemieke Petrignani \inst{\ref{AmsterdamVanthoff}} \and
  Thomas Pino \inst{\ref{SaclayISM}} \and
  Elena Puga \inst{\ref{STScIESA}} \and
  Naseem Rangwala \inst{\ref{AMES}} \and
  Mathias Rapacioli \inst{\ref{ToulouseChem}} \and
  Jeonghee Rho \inst{\ref{CarlSagan},\ref{SNU} }  \and  
  Alessandra Ricca \inst{\ref{AMES}, \ref{CarlSagan}} \and
  Julia Roman-Duval \inst{\ref{STScI}} \and
  Joseph~Roser \inst{\ref{CarlSagan},\ref{AMES}} \and
  Evelyne Roueff \inst{\ref{SorbonneLERMA}} \and
  Farid Salama \inst{\ref{AMES}} \and
  Dinalva A. Sales \inst{\ref{RioGrande}} \and
  Karin Sandstrom \inst{\ref{CaliforniaAstro}} \and
  Peter Sarre \inst{\ref{NottinghamChem}} \and
  Ella Sciamma-O'Brien \inst{\ref{AMES}} \and
  Kris Sellgren \inst{\ref{OhioState}} \and
  Sachindev S. Shenoy \inst{\ref{BoulderSpace}} \and
  David Teyssier \inst{\ref{MadridESAC}} \and
  Richard D. Thomas \inst{\ref{Stockholm}} \and
  Adolf N. Witt \inst{\ref{Toledo}} \and
  Alwyn Wootten \inst{\ref{NRAO}} \and
  Nathalie Ysard \inst{\ref{SaclayIAS}} \and
  Henning Zettergren \inst{\ref{Stockholm}} \and
  Yong Zhang \inst{\ref{Sunyatsen}} \and
  Ziwei E. Zhang \inst{\ref{RIKEN}} \and     
  Junfeng Zhen \inst{\ref{DeepSpace}}
}

\institute{%
  Space Telescope Science Institute, 3700 San Martin Drive, Baltimore, MD 21218, USA
  \label{STScI} \and
  Institut d'Astrophysique Spatiale, Université Paris-Saclay, CNRS,  Bâtiment 121, 91405 Orsay Cedex, France
  \label{SaclayIAS}        \and 
  LERMA, Observatoire de Paris, PSL Research University, CNRS, Sorbonne Universités, F-92190 Meudon, France
  \label{SorbonneLERMA} \and
  Department of Physics \& Astronomy, The University of Western Ontario, London ON N6A 3K7, Canada
  \label{WOntarioPA}     \and 
  Institute for Earth and Space Exploration, The University of Western Ontario, London ON N6A 3K7, Canada
  \label{WOntarioIESE}          \and 
  Carl Sagan Center, SETI Institute, 339 Bernardo Avenue, Suite 200, Mountain View, CA 94043, USA
  \label{CarlSagan}         \and
  Institut de Recherche en Astrophysique et Planétologie, Université Toulouse III - Paul Sabatier, CNRS, CNES, 9 Av. du colonel Roche, 31028 Toulouse Cedex 04, France
  \label{ToulouseIRAP}         \and 
  Department of Astronomy, University of Michigan, 1085 South University Avenue, Ann Arbor, MI 48109, USA
  \label{Michigan} \and
  Astronomy Department, Ohio State University, Columbus, OH 43210 USA
  \label{OhioState} \and
  NASA Ames Research Center, MS 245-6, Moffett Field, CA 94035-1000, USA
  \label{AMES}         \and
  Institut des Sciences Moléculaires d'Orsay, Université Paris-Saclay, CNRS, Bâtiment 520, 91405 Orsay Cedex, France
  \label{SaclayISM} \and
  Instituto de Física Fundamental (CSIC), Calle Serrano 121-123, 28006, Madrid, Spain
  \label{MadridFisFundamental} \and
  Sterrenkundig Observatorium, Universiteit Gent, Gent, Belgium
  \label{Ghent} \and 
  Department of Astronomy, Graduate School of Science, The University of Tokyo, 7-3-1 Bunkyo-ku, Tokyo 113-0033, Japan
  \label{TokyoAstro} \and
  Leiden Observatory, Leiden University, P.O. Box 9513, 2300 RA Leiden, The Netherlands
  \label{Leiden} \and
  Astronomy Department, University of Maryland, College Park, MD 20742, USA
  \label{Maryland} \and
  ACRI-ST, Centre d’Etudes et de Recherche de Grasse (CERGA), 10 Av. Nicolas Copernic, F-06130 Grasse, France
  \label{GrasseACRIST} \and
  INCLASS Common Laboratory., 10 Av. Nicolas Copernic, 06130 Grasse, France
  \label{GrasseINCLASS} \and
  UK Astronomy Technology Centre, Royal Observatory Edinburgh, Blackford Hill EH9 3HJ, UK
  \label{Edinburgh} \and
  Centro de Astrobiolog\'{\i}a (CSIC-INTA),Ctra de Torrej\'on a Ajaalvir, km 4, 28850, Torrej\'on de Ardoz, Spain
  \label{CSIC} \and
  Quantum Solid State Physics (QSP), Celestijnenlaan 200d - box 2414, 3001 Leuven, Belgium
  \label{LeuvenQSP} \and
  Institut de Planétologie et d'Astrophysique de Grenoble (IPAG), Université Grenoble Alpes, CNRS, F-38000 Grenoble, France
  \label{GrenobleIPAG} \and
  Institut de Radioastronomie Millimétrique (IRAM), 300 Rue de la Piscine, F-38406 Saint-Martin d'Hères, France
  \label{IRAM} \and
  I. Physikalisches Institut der Universität zu Köln, Zülpicher Stra{\ss}e 77, 50937 Köln, Germany
  \label{Koln} \and
  Johns Hopkins University, 3400 N. Charles Street, Baltimore, MD, 21218, USA
  \label{JHU} \and 
  Physikalischer Verein - Gesellschaft für Bildung und Wissenschaft, Robert-Mayer-Str. 2, 60325 Frankfurt, Germany
  \label{FrankfurtPhys} \and 
  Goethe-Universität, Physikalisches Institut, Frankfurt am Main, Germany
  \label{Goethe} \and
  Department of Space, Earth and Environment, Chalmers University of Technology, Onsala Space Observatory, SE-439 92 Onsala, Sweden
  \label{ChalmersSpace} \and
  Instituto de Astrofísica e Ciências do Espaço, Tapada da Ajuda, Edifício Leste, 2° Piso, P-1349-018 Lisboa, Portugal
  \label{Lisboa} \and 
  Yunnan Observatories, Chinese Academy of Sciences, 396 Yangfangwang, Guandu District, Kunming, 650216, China
  \label{YunnanObs} \and
  Chinese Academy of Sciences South America Center for Astronomy, National Astronomical Observatories, CAS, Beijing 100101, China
  \label{ChineseSouthAmCenter} \and
  Departamento de Astronomía, Universidad de Chile, Casilla 36-D, Santiago, Chile
  \label{Chile} \and
  Laboratory Astrophysics Group of the Max Planck Institute for Astronomy at the Friedrich Schiller University Jena, Institute of Solid State Physics, Helmholtzweg 3, 07743 Jena, Germany
  \label{Jena} \and
  Department of Physics, Texas State University, San Marcos, TX 78666 USA
  \label{TexasState} \and
  Instituto de Física e Química, Universidade Federal de Itajubá, Av. BPS 1303, Pinheirinho, 37500-903, Itajubá, MG, Brazil
  \label{ItajubaFisQui} \and
  Institute of Mathematics and Statistics, University of São Paulo, Rua do Matão, 1010, Cidade Universitária, Butantã, 05508-090, São Paulo, SP, Brazil
  \label{SaoPauloMath} \and
  Australian Synchrotron, Australian Nuclear Science and Technology Organisation (ANSTO), Victoria, Australia
  \label{VictoriaANSTO} \and
  INAF - Osservatorio Astrofisico di Catania, Via Santa Sofia 78, 95123 Catania, Italy
  \label{INAFcatania} \and
  Bay Area Environmental Research Institute, Moffett Field, CA 94035, USA
  \label{BayArea} \and
  Laboratoire de Physique de l'École Normale Supérieure, ENS, Université PSL, CNRS, Sorbonne Université, Université de Paris, 75005, Paris, France
  \label{SorbonneLab} \and
  Laboratory for Atmospheric and Space Physics, University of Colorado, Boulder, CO 80303, USA
  \label{BoulderLab} \and
  Department of Chemistry, University of Colorado, Boulder, CO 80309, USA
  \label{BoulderChem} \and
  Institute for Modeling Plasma, Atmospheres, and Cosmic Dust (IMPACT), University of Colorado, Boulder, CO 80303, USA
  \label{BoulderIMPACT} \and
  Faculty of Aerospace Engineering, Delft University of Technology, Kluyverweg 1, 2629 HS Delft, The Netherlands
  \label{DelftAero} \and
  Radboud University, Institute for Molecules and Materials, FELIX Laboratory, Toernooiveld 7, 6525 ED Nijmegen, the Netherlands
  \label{Radboud} \and
  School of Physics, University of Hyderabad, Hyderabad, Telangana 500046, India
  \label{HyderabadDST} \and 
  Department of Physics, Wellesley College, 106 Central Street, Wellesley, MA 02481, USA
  \label{Wellesley} \and
  Anton Pannekoek Institute for Astronomy, University of Amsterdam, The Netherlands
  \label{Pannekoek} \and
  Delft University of Technology, Delft, The Netherlands
  \label{DelftTech} \and
  Laboratoire de Physique des deux infinis Irène Joliot-Curie, Université Paris-Saclay, CNRS/IN2P3, Bâtiment 104, 91405 Orsay Cedex, France
  \label{SaclayLabphys} \and
  Department of Chemistry, GITAM school of Science, GITAM Deemed to be University, Bangalore, India
  \label{BangaloreChem} \and
  Institut de Physique de Rennes, UMR CNRS 6251, Université de Rennes 1, Campus de Beaulieu, 35042 Rennes Cedex, France
  \label{Rennes} \and
  Department of Chemistry, The University of British Columbia, Vancouver, British Columbia, Canada
  \label{BritishColumbiaChem} \and
  National Radio Astronomy Observatory (NRAO), 520 Edgemont Road, Charlottesville, VA 22903, USA
  \label{NRAO} \and
  European Space Astronomy Centre (ESAC/ESA), Villanueva de la Cañada, E-28692 Madrid, Spain
  \label{MadridESAC} \and
  Observatoire de Paris, PSL University, Sorbonne Université, LERMA, 75014, Paris, France
  \label{SorbonneObs} \and
  Harvard-Smithsonian Center for Astrophysics, 60 Garden Street, Cambridge MA 02138, USA
  \label{CambridgeHarvarSmithsonianAstro} \and
  Sorbonne Université, CNRS, UMR 7095, Institut d’Astrophysique de Paris, 98bis bd Arago, 75014 Paris, France
  \label{Sorbonne} \and
  Institut Universitaire de France, Ministère de l'Enseignement Supérieur et de la Recherche, 1 rue Descartes, 75231 Paris Cedex 05, France
  \label{France} \and
  Department of Physics and Astronomy, Rice University, Houston TX, 77005-1892, USA
  \label{Rice} \and
  Departments of Chemistry and Astronomy, University of Virginia, Charlottesville, Virginia 22904, USA
  \label{Virginia} \and
  InterCat and Dept. Physics and Astron., Aarhus University, Ny Munkegade 120, 8000 Aarhus C, Denmark
  \label{Aarhus} \and
  Instituto de Astronomia, Geofísica e Ciências Atmosféricas, Universidade de São Paulo, 05509-090 São Paulo, SP, Brazil
  \label{SaoPauloAstro} \and
  Department of Physics and Astronomy, San José State University, San Jose, CA 95192, USA
  \label{SanJose} \and
  Institut de Ciencies de l’Espai (ICE, CSIC), Can Magrans, s/n, E-08193 Bellaterra, Barcelona, Spain  
  \label{ICECSIC} \and
  ICREA, Pg. Lluís Companys 23, E-08010 Barcelona, Spain  
  \label{ICREA} \and
  Institut d’Estudis Espacials de Catalunya (IEEC), E-08034 Barcelona, Spain  
  \label{IEEC} \and
  European Space Agency, Space Telescope Science Institute, 3700 San Martin Drive, Baltimore MD 21218, USA
  \label{STScIESA} \and
  Institute of Astronomy, Russian Academy of Sciences, 119017, Pyatnitskaya str., 48 , Moscow, Russia
  \label{Russia} \and
  Department of Earth, Ocean, \& Atmospheric Sciences, University of British Columbia, Canada V6T 1Z4
  \label{BritishColumbiaEarth} \and
  Telespazio UK for ESA, ESAC, E-28692 Villanueva de la Cañada, Madrid, Spain
  \label{MadridUKESA} \and
  IPAC, California Institute of Technology, Pasadena, CA, USA
  \label{IPAC} \and
  Laboratoire d'Astrophysique de Bordeaux, Univ. Bordeaux, CNRS, B18N, allée Geoffroy Saint-Hilaire, 33615 Pessac, France 
  \label{BordeauxLab} \and
  Department of Physics and Astronomy, University of Missouri, Columbia, MO 65211, USA
  \label{Missouri} \and
  Max Planck Institute for Astronomy, Königstuhl 17, 69117 Heidelberg, Germany
  \label{Heidelberg} \and
  Chemical Sciences Division, Lawrence Berkeley National Laboratory, Berkeley, California, USA
  \label{BerkelyChem} \and
  Kenneth S.~Pitzer Center for Theoretical Chemistry, Department of Chemistry, University of California -- Berkeley, Berkeley, California, USA
  \label{BerkelyPitzer} \and
  AIM, CEA, CNRS, Université Paris-Saclay, Université Paris Diderot, Sorbonne Paris Cité, 91191 Gif-sur-Yvette, France
  \label{SaclayAIMCEACNRS} \and
  Institut des Sciences Moléculaires, CNRS, Université de Bordeaux, 33405 Talence, France
  \label{BordeauxMol} \and
  Department of Chemistry, Massachusetts Institute of Technology, Cambridge, MA 02139, USA
  \label{MITChem} \and
  Instituto de Ciencia de Materiales de Madrid (CSIC), Sor Juana Ines de la Cruz 3, E28049, Madrid, Spain
  \label{MadridMateriales} \and
  Department of Physics, PO Box 64, 00014 University of Helsinki, Finland
  \label{Helsinki} \and
  AstronetX PBC, 55 Post Rd W FL 2, Westport, CT 06880  USA
  \label{AstronetX} 
  \and
  INAF - Osservatorio Astronomico di Cagliari, Via della Scienza 5, 09047 Selargius (CA), Italy
  \label{INAFcagliari} \and
  Department of Physics, College of Science, United Arab Emirates University (UAEU), Al-Ain, 15551, UAE
  \label{Emirates} \and
  National Astronomical Observatory of Japan, National Institutes of Natural Science, 2-21-1 Osawa, Mitaka, Tokyo 181-8588, Japan
  \label{TokyoNationalObs} \and
  Department of Physics, Institute of Science, Banaras Hindu University, Varanasi 221005, India
  \label{Banaras} \and
  University of Central Florida, Orlando, FL 32765
  \label{Florida} \and
  Van’t Hoff Institute for Molecular Sciences, University of Amsterdam, Science Park 904, 1098 XH, Amsterdam, The Netherlands
  \label{AmsterdamVanthoff} \and
  Laboratoire de Chimie et Physique Quantiques LCPQ/IRSAMC, UMR5626, Université de Toulouse (UPS) and CNRS, Toulouse, France
  \label{ToulouseChem} \and
  Department of Physics and Astronomy, Seoul National University, Gwanak-ro 1, Gwanak-gu, Seoul, 08826, South Korea
  \label{SNU} \and
  Instituto de Matemática, Estatística e Física, Universidade Federal do Rio Grande, 96201-900, Rio Grande, RS, Brazil
  \label{RioGrande} \and
  Center for Astrophysics and Space Sciences, Department of Physics, University of California, San Diego, 9500 Gilman Drive, La Jolla, CA 92093, USA
  \label{CaliforniaAstro} \and
  School of Chemistry, The University of Nottingham, University Park, Nottingham, NG7 2RD, United Kingdom
  \label{NottinghamChem} \and
  Space Science Institute, 4765 Walnut St., R203, Boulder, CO 80301
  \label{BoulderSpace} \and
  Department of Physics, Stockholm University, SE-10691 Stockholm, Sweden
  \label{Stockholm} \and
  Ritter Astrophysical Research Center, University of Toledo, Toledo, OH 43606, USA
  \label{Toledo}          \and
  School of Physics and Astronomy, Sun Yat-sen University, 2 Da Xue Road, Tangjia, Zhuhai 519000,  Guangdong Province, China
  \label{Sunyatsen} \and
  Star and Planet Formation Laboratory, 0-0 S, RIKEN Cluster for Pioneering Research, Hirosawa 2-1, Wako, Saitama 351-0198, Japan
  \label{RIKEN} \and
  Institute of Deep Space Sciences, Deep Space Exploration Laboratory, Hefei 230026, China
  \label{DeepSpace}
}

\date{Submitted 2024/01/22; Under review.}

\abstract
{
  Mid-infrared emission features are important probes for the properties of ionized gas, and hot or warm molecular gas which is difficult to probe at other wavelengths.
  The Orion Bar photodissociation region (PDR) is a bright, nearby, and frequently studied target containing large amounts of gas under these conditions.
  Under the ``PDRs4All'' Early Release Science Program for \jwst, a part of the Orion Bar was observed with MIRI IFU spectroscopy, and these high-sensitivity IR spectroscopic images of very high angular resolution (0.2\arcsec) provide a rich observational inventory of the mid-IR emission lines, while resolving the \HII region, the ionization front, and multiple dissociation fronts.
}
{
  We list, identify, and measure the most prominent gas emission lines in the Orion Bar, as observed by the new MIRI IFU data.
  An initial analysis summarizes the physical conditions of the gas and demonstrates the potential of these new data and future IFU observations with \jwst.
}
{
  The MIRI IFU mosaic spatially resolves the substructure of the PDR, its footprint cutting perpendicularly across the ionization front and three dissociation fronts.
  We perform an up-to-date data reduction, and extract five spectra that represent the ionized, atomic, and molecular gas layers.
  We identify the observed lines through a comparison with theoretical line lists derived from atomic data and simulated PDR models.
  The identified species and transitions are summarized in the main table of this work, with measurements of the line intensities and central wavelengths.
}
{
  We identified \change{around} 100 lines and report an additional \change{18} lines that remain unidentified.
  A majority consists of \HI recombination lines arising from the ionized gas layer bordering the PDR.
  The \HI line ratios are well matched by emissivity coefficients from H recombination theory, but deviate up to 10\% due contamination by \ion{He}{i} lines.
  We report the observed emission lines of various ionization stages of Ne, P, S, Cl, Ar, Fe, and Ni.
  We show how the \ion{Ne}{iii}/\ion{Ne}{ii}, \ion{S}{iv}/\ion{S}{iii}, and \ion{Ar}{iii}/\ion{Ar}{ii} ratios trace the conditions the ionized layer bordering the PDR, while \ion{Fe}{iii}/\ion{Fe}{ii} and \ion{Ni}{iii}/\ion{Ni}{ii} exhibit a different behavior, as there are significant contributions to \ion{Fe}{ii} and \ion{Ni}{ii} from the neutral PDR gas.
 We observe the pure-rotational H$_2$ lines in the vibrational ground state from 0-0~$S(1)$ to 0-0~$S(8)$, and in the first vibrationally excited state from 1-1~$S(5)$ to 1-1~$S(9)$.
 We derive H$_2$ excitation diagrams, and for the three observed dissociation fronts, the rotational excitation can be approximated with one thermal ($\sim$700~K) component representative of an average gas temperature, and one non-thermal component ($\sim$2700~K) probing the effect of UV pumping.
  We compare these results to an existing model for the Orion Bar PDR, and find that the \change{predicted excitation} matches the data qualitatively, while adjustments to the parameters of the PDR model are required to reproduce the intensity of the 0-0 $S(6)$ to $S(8)$ lines. 
}
{}


\maketitle
\clearpage
\setcounter{page}{1}

\section{Introduction}

Photodissociation regions (PDRs) are ideal targets for probing strong variations in the conditions of the interstellar medium (ISM).
They appear at the borders of clouds that are illuminated by far-ultraviolet (FUV) photons originating from massive young stars that formed in the cloud, or from the interstellar radiation field (ISRF). 
The transition from the \HII region to the PDR regime starts where the extreme ultraviolet (EUV, $> 13.6$ eV) radiation is no longer sufficient to maintain the ionization, and FUV radiation ($< 13.6$ eV) drives the local physics instead.
The PDR physics is driven by the balance between the FUV flux and the attenuation by the medium, resulting in sharp spatial transitions between ionized, atomic, and molecular hydrogen layers.
When PDRs are located in nearby clouds, they offer the observational opportunity to spatially resolve the transitions between these layers in a single object \citep{1997ARA&A..35..179H, 2022ARA&A..60..247W}.

The Orion Bar is a frequently studied PDR \citep[e.g.,][]{1993Sci...262...86T, 1995A&A...294..792H, 1996A&A...313..633V, 2006ApJ...637..823K, 2009ApJ...693..285P, 2016Natur.537..207G}
with a density of several \num{e4} to \SI{e6}{\per\cubic\cm}, typical for high density PDRs.
It is subject to a high intensity FUV radiation field with $G_0$ in the range $(2.2$-$7.1)\times10^4$ \citep{pdrs4all-pasp, Habart:im, Peeters:nirspec}, where $G_0$ represents a dimensionless parameter defining the FUV intensity with respect to the ISRF intensity defined by \citet{1968BAN....19..421H}.
The proximity \citep[414 pc,][]{2007A&A...474..515M} and very high surface brightness of the Bar allow for infrared observations at excellent signal-to-noise ratio (S/N) levels.
Recent observations at high spatial resolution of the HCO$^+$ $J=$ 4-3 line \citep{2016Natur.537..207G} and vibrationally excited \hmol lines \citep{2023A&A...673A.149H}, revealed the many filamentary substructures at the surface of the Bar.
Underlying these substructures, the Bar has a simple and approximately linear geometry, a perpendicular orientation with respect to the radiation field, and a nearly edge-on orientation with respect to the observer, so that it has served as an exemplary benchmark target for theoretical PDR models \citep[e.g.,][]{1993Sci...262...86T, 1995A&A...303..541J, 2000ApJ...540..886Y, 2009ApJ...701..677S, 2018Joblin}.

The PDRs4all Early Release Science program for \jwst\ (ERS 1288; \citealt{pdrs4all-pasp}), observed the Orion Bar with NIRCam and MIRI imaging, as well as IFU spectroscopy with NIRSpec and MIRI Medium Resolution Spectroscopy (MRS).
The imaging data have led to new insights in the detailed geometry of this PDR, and suggest a terrace-like structure with three consecutive dissociation fronts as described in \cite{Habart:im}.
The rich near-IR to mid-IR (MIR) spectroscopic data produced by this program are discussed in a series of papers focusing on the NIRSpec IFU observations \citep{Peeters:nirspec}, the Aromatic Infrared Bands (AIB) in both the NIRSpec and MIRI data \citep{Chown:AIB}, the spatial variations of dust properties \citep{2024arXiv240101221E}, and the spectra from a dense disk in the same field of view as the PDR \citep{Berne:disk2024}.
In this work we focus on the emission lines observed in the MIRI spectra, which cover the 4.9-28~\mum range.

Emission lines in the MIR are a key probe for the local conditions of the gas in PDRs and the surrounding environment.
Constraints on the physical conditions of the gas can be obtained by direct observations of the MIR lines, and by comparing the intensity of the lines to theoretical models.
By studying these lines towards PDRs, where the ionized, neutral atomic, and molecular layers are spatially resolved, such diagnostics can be calibrated better.
They can then be used for studies of distant star-forming galaxies, where PDRs contribute significantly to the emission \citep{1990ApJ...358..116W, 2007ApJ...667..149F, 2012A&A...548A..20C}, but where the ISM regimes can not be spatially resolved. 
For example, the MIR fine-structure lines of different ionization stages of metals ([\ion{Fe}{ii}], [\ion{Ar}{iii}], [\ion{S}{iv}], ...) and the hydrogen recombination lines can be used to constrain the gas-phase abundances,
excitation conditions (e.g., [\ion{Fe}{ii}] lines),
and radiation field hardness (e.g., \ionwav{Ne}{iii}{15.5} / \ionwav{Ne}{ii}{12.8})
within the ionized gas \citep[e.g.,][]{2003A&A...403..829V, 2009ApJS..184..230B, 2011MNRAS.410.1320R, 2016JKAS...49..109K}.
In the warm atomic gas, excitation of [\ion{Fe}{ii}] can provide diagnostics, while in the warm molecular gas close to the dissociation front in a PDR, the MIR contains rotational \hmol lines as a key tracer.
The ratios between pure-rotational \hmol lines constrain the kinetic gas temperature and density, as well as the local strength of the UV radiation field \citep{2005SSRv..119...71H, 2006ApJ...644..283K, 2011ApJ...741...45S}.
An updated and more versatile ``PDR toolbox'' was developed in the context of the PDRs4All collaboration \citep{2023AJ....165...25P}, which enables the user to derive constraints from a variety of diagnostics in a straightforward manner.
It supports diagnostics in the MIRI wavelength range such as [\ion{Fe}{ii}] and [\ion{Ar}{iii}] lines, and the rotational \hmol lines.
By providing the line list in this work and the PDR toolbox to the community, the science products of the PDRs4All project will make future PDR studies based on MIR data more straightforward and consistent.

The MIRI IFU spectra we focus on in this article were extracted for five exemplary regions, consistent with the apertures originally defined for the analysis by \citet{Peeters:nirspec}. 
These five apertures each capture a specific regime in the PDR front: the ionized region, the atomic region, and three individual dissociation fronts.
The scope of this work is limited to these five regions of the Orion Bar, with the main goal being to clarify which emission lines are expected in each regime.
The data reduction and extraction of the spectra are described in Sect.~\ref{sec:data}.
In Sect.~\ref{sec:lines}, the identification, intensity measurements, and central wavelength measurements of the observed emission lines are described, and the results are presented in the main table of this work.
Future work is expected to make full use of the spatial resolution of \jwst, and the high S/N of these data so that analyses at single-spaxel resolution (0.2\arcsec) are possible.
The provided line list will enable such efforts for both the PDRs4All Orion Bar data and future data of other objects dominated by PDR emission.
In Sect.~\ref{sec:analysis}, we develop an analysis concentrating in particular on the \HI recombination lines emitted in the ionized region, the pure-rotational \hmol lines of the three dissociation fronts, and a number of metal fine-structure lines.
This initial analysis summarizes the nature of the gas conditions that produce the rich spectral contents of the MIRI MRS observations.

\section{Data}
\label{sec:data}

\begin{figure*}[tb]
    \centering
    \includegraphics{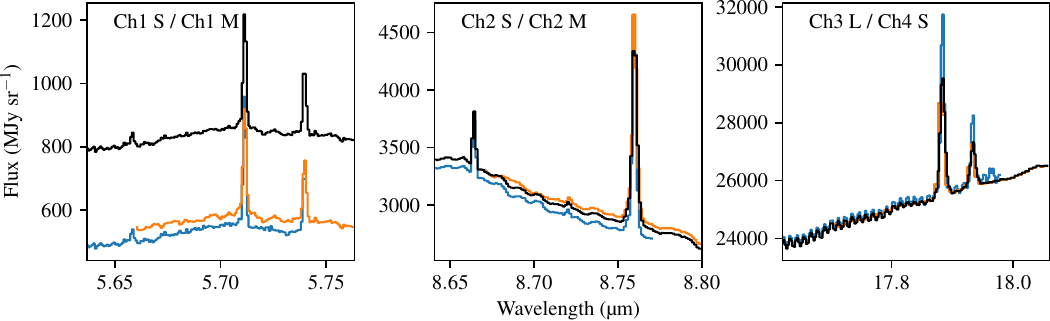}
    \caption{Examples of flux differences in overlap regions between MIRI bands.
    Blue and orange curves: individual segments, with the bands indicated at the top of each panel.
    These were extracted from the band cubes produced by the pipeline, in this example for the \HII template aperture. 
    Black line: stitched spectrum. It is offset with respect to the blue/orange line due to the additive offsets applied to the 12 segments, which are cumulative with Ch~2~LONG used as a reference.}
    \label{fig:stitch}
\end{figure*}

\subsection{Reduction}
The MIRI MRS observations were processed from uncalibrated data through the official pipeline, using 1.12.5 of the \texttt{jwst}  Python package. 
The pipeline and reference files from the Calibration Reference Data System (CRDS) have been continually updated since commissioning, and at the time we performed our latest data reduction we used the CRDS context \texttt{jwst\_1147.pmap}.
The wavelength calibration at the time of writing 
is accurate up to a few \si{\km\per\s} at short wavelengths, and about 30 \si{\km\per\s} at the longest wavelength \citep{2023A&A...675A.111A}. 
The spectral resolution ranges from  $R \sim 3500$ to 1500 depending on the wavelength, or about 85 - 200 \si{\km\per\s} in terms of radial velocity \citep{2021A&A...656A..57L}.
Concerning the astrometric calibration, the pointing accuracy is about 0.45\arcsec\ without target acquisition, and the typical accuracy of the assigned coordinate system is about 0.3\arcsec, where the main source of uncertainty is the guide star catalog \citep{2023arXiv230701025P}.
For reference, the spaxel size varies between 0.2 and 0.3\arcsec.
The spectrophotometric calibration was initially based on a single standard star \citep{gordon2022fluxcal}, but data from additional stars were recently introduced.
The inclusion of these additional calibration data (since \texttt{jwst\_1094.pmap}), led to a much improved matching of the continuum flux between the four channels of MIRI MRS, with only minor flux offsets in the overlap regions between the channels (see also Sect.~\ref{sec:extraction} and Fig.~\ref{fig:stitch}).
Together with support for time-dependent calibrations (since \texttt{jwst} version 1.11.0), the changes much improved the Channel 4 calibration in the case of the Orion Bar data, removing most local flux oscillations or broad artifacts so that the hot dust and telescope continuum resemble a smooth blackbody spectrum.
As reported in the \jwst\ user documentation\footnote{https://jwst-docs.stsci.edu/jwst-calibration-pipeline-caveats/jwst-miri-mrs-pipeline-caveats}, there may remain a 10\% systematic uncertainty in the calibration, based on the deviations that are observed  when using different calibration stars.
There is also a spectral leak artifact, where second-order light at 6~\mum arrives at a different part of the detector, causing a spurious signal at 12.2~\mum in the spectra.
We did not correct for this leak, as it is broad and does not interfere with our measurements of the emission line intensity.

We made additional corrections to the default pipeline steps to improve the quality of the final products.
In the following section and in Fig.~\ref{fig:stitch}, we refer to the four channels of MIRI MRS as Ch~1, 2, 3, or 4, and to the band setting for each channel as SHORT (S), MEDIUM (M), or LONG (L).
In the first stage of the pipeline (\texttt{Detector1}), which applies detector-level corrections to the exposures, we modified the threshold for the \texttt{jump} step to 3.0 standard deviations, instead of the default value of 4.0, to improve the cosmic ray detection.
One of the corrections in the second stage of the spectroscopic pipeline (\texttt{Spec2}), is removing the fringes resulting from coherent reflections within the detectors \citep{2020A&A...641A.150A}.
The default fringe correction (\texttt{fringe}) divides the images by a static extended source fringe flat and removes most of the fringe amplitude, while residual fringing remains primarily in Ch~2 LONG, which is known to have a poor fringe mitigation at this date.
Therefore, we also applied the optional \texttt{residual\_fringe} step of \texttt{Spec2}, which further reduces fringing by fitting sinusoidal functions at the detector level, and works with constrained spatial frequencies in order to avoid removing physical features.

The \texttt{Spec3} stage then combines the multiple exposures (dither pattern and mosaic positions of both science and background), and builds a 3D cube as the final product.
The dedicated background exposures of the program were reduced up to \texttt{Spec2} using the same workflows described above, to obtain the 1D collapsed spectra produced through the \texttt{extract\_1d} step at the end of this pipeline stage. 
The 1D background spectra were then used in the \texttt{master\_background} step of \texttt{Spec3}, as recommended for bright extended sources (as opposed to a pixel-based background subtraction available in \texttt{Spec2}).
We applied the \texttt{outlier\_detection} step as it did not introduce new artifacts and resulted in an improved identification and removal of the remaining cosmic rays and bad pixels.
Moreover, the \texttt{tweakreg} step was skipped and we took advantage of the MIRI images taken in parallel with the MRS observations to improve the WCS alignment.
Because the parallel imaging was performed simultaneously, the telescope had an identical orientation during both the MIRI MRS observations and the MIRI imaging.
The larger fields of view of the images allow for the use of the Gaia DR3 astrometric catalog \citep{2023A&A...674A...1G} for WCS correction purposes, by performing a source detection in the F770W images.
Finally, we configured the \texttt{cube\_build} step of \texttt{Spec3} to produce the 12 individual data cubes corresponding to each of the four channels channels and three bands, to better correct for any offsets between the bands as described in the next section.

\subsection{Extraction and additional corrections}
\label{sec:extraction}

\begin{figure*}[tb]
    \centering
    \includegraphics[width=\linewidth]{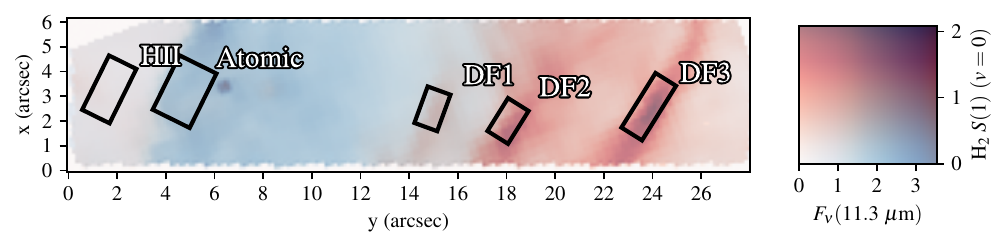}
    \caption{Apertures from which the five template spectra were extracted, overlayed on two slices of the MIRI MRS data cube.
    The horizontal axis is aligned with the y-coordinate of the IFU and the 9$\times$1 mosaic strip.
    The blue component is the total flux at 11.3~\mum ($10^4$ MJy sr$^{-1}$), which contains some continuum, but mainly traces the 11.3~\mum AIB and peaks in the atomic region.
    The red component highlights the locations of the dissociation fronts, using the intensity of the H$_2$ $S(1)$ line at 17.0~\mum ($10^{-3}$ erg s$^{-1}$ cm$^{-2}$ sr$^{-1}$).
    The two-dimensional color legend shows where one component (red and blue) or both components (dark purple) are bright.
    }
    \label{fig:templates}
\end{figure*}

We focus on 5 spectra extracted from the IFU data cubes, obtained by averaging the flux over apertures covering different spatial locations. 
The chosen apertures are defined by \citet{Peeters:nirspec}, and capture the five main zones representative of the physics observed in this section of the Orion Bar (Fig.~\ref{fig:templates}).
The resulting spectra are called the ``template spectra'' in this work as well as \citet{Peeters:nirspec}. 
We will refer to the zones as ``\HII'' for the \HII region, ``Atomic'' for the atomic zone near the ionization front, and ``DF1'', ``DF2'', ``DF3'' for the ``dissociation fronts'', the three \HI/\hmol transitions that appear as bright filaments in maps of the \hmol emission.
It should be noted that the S/N and resolution of the data are high enough to apply line diagnostics in a detailed spaxel-per-spaxel manner, that spatially resolves the structures in the PDR.
The scope of this work is focused on making a line list based on the template spectra, and showing a first analysis that demonstrates what future highly resolved analyses could include.
For a detailed description of the contents of the five apertures and a discussion of the PDR structure based on the imaging and NIRSpec IFU data, we refer the reader to the respective papers of \citet{Habart:im} and \citet{Peeters:nirspec}.

Some remaining oscillations in the spectrum are an artifact of undersampling issues and the cube building method  \citep{2023AJ....166...45L}.
For the purposes of detecting individual AIB emission features, \citet{Chown:AIB} applied an additional empirical correction based on data of the calibration star 10 Lac from the CALSPEC archive \citep{2014PASP..126..711B}.
This further reduced the fringes, but also introduced artifacts that resemble lines or cause lines to exhibit double peaks.
Because these artifacts interfere with the identification of real lines, we do not use the spectra of \citet{Chown:AIB} for this work.

Even with the absolute flux calibration, the many corrections in the pipeline, and the stacking of the spectra over the template apertures, there are still minor flux mismatches in the overlap regions of the 12 spectral segments produced by MIRI MRS.
We measured each of the 11 jumps in the continuum, by computing the median for each pair of neighboring segments over the region where they overlap, and taking the difference.
The typical values for the jumps are no more than 5\%, or a few 10 MJy sr$^{-1}$.
These jumps in the flux density do not affect the extraction of the line intensities, as we apply a local continuum measurement  for each line individually (see Sect.~\ref{sec:intensity}).
However, to improve the continuum for lines that are located in the overlap regions and to deliver higher quality template spectra, we perform a spectral stitching correction.

To match the continua over all 12 segments, we compute 12 absolute offsets by taking the cumulative sum of the 11 relative offsets calculated above.
In this calculation, Ch~2~LONG is used as a reference segment, and we set its offset to zero by construction.
For the other segments, the offsets are applied as an additive constant to the flux.
While the individual jumps are rather small, the cumulative offsets do result in some corrections $\geq 100$ MJy sr$^{-1}$, especially in Ch~1 and Ch~4.
Three examples of the overlap regions and the stitching method are shown in Fig.~\ref{fig:stitch}.
To merge the continuum-matched segments, a sliding weighted average is used for each overlap region for a gradual transition. 
The segment weights change linearly from 1 to 0 for the blueward segment, and from 0 to 1 for the redward segment, as the wavelengths are processed from the blue end to the red end of the overlap region.
For the overlap regions between channels, the wavelength resolution of the longer channel is always lower.
In this case, the grid of the short wavelength segment is used, and the long-wavelength segment is interpolated onto this grid before the sliding average is applied.

To determine if any multiplicative corrections might be needed because of potential calibration issues, we measured the intensities of lines that are present in the overlap regions, and inspected the ratio of the intensities obtained individually from each pair of segments.
We selected the \HII template spectrum for this task, since the emission lines in the overlap regions are brightest there. 
For the line at 5.711~\mum (Fig.~\ref{fig:stitch}), the intensity ratio of Ch~1~M to Ch~1~S is 0.95.
Similarly, we find a ratio of 1.02 for the line at 6.565~\mum (Ch~1~L/Ch~1~M), and 0.93 at 8.760~\mum (Ch~2~M/Ch~2~S).
The Ch~3~L to Ch~4~S transition has a continuum that matches well without any correction (right panel of Fig.~\ref{fig:stitch}), but there are two lines present at 17.883 and 17.933~\mum for which the ratios are 0.95 and 0.99.
The last two intensity ratios are different, even though the lines are present in the same overlap region.
These differences indicate that a mix of multiplicative and additive offsets could be needed, to match both the line intensity and the continuum level.
Since not every overlap region has lines, and since a cumulative multiplicative correction across the segments would potentially inflate the fractional systematic uncertainties, we do not apply a multiplicative correction.
Instead, we suggest that for a group of lines in one and the same segment, one should consider a fractional systematic uncertainty of a few percent which affects all those lines multiplicatively with the same factor.
Taking the standard deviation of the segment-to-segment line intensity ratios measured above, we estimate that the uncertainty on this systematic factor is
around 3\%.

\section{Lines}
\label{sec:lines}

\subsection{Identification}
\label{sec:identification}

\begin{figure*}[tp]
\centering
    \includegraphics[trim={0 .5cm 0 0},clip]{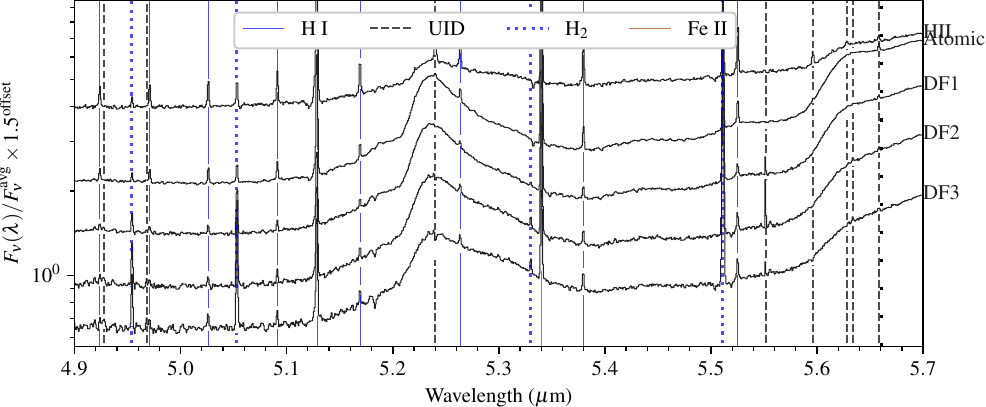}
    \includegraphics[trim={0 .5cm 0 0},clip]{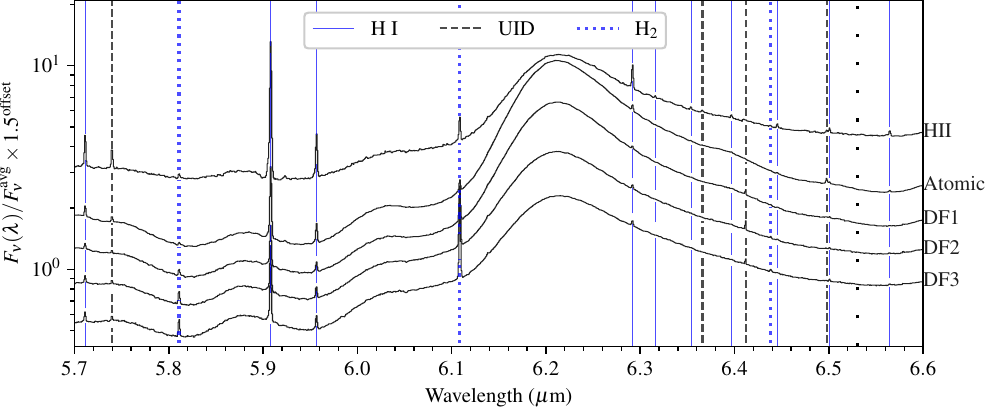}
    \includegraphics{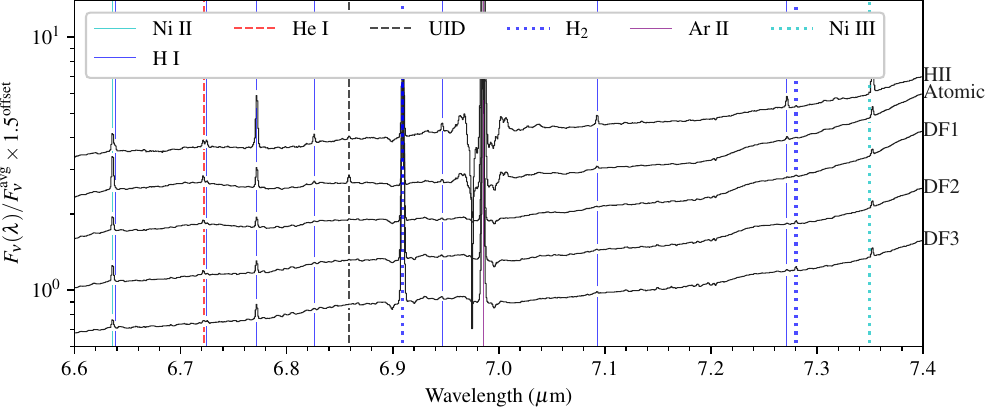}
        \caption{Overview of template spectra and lines, continued in Figs.~\ref{fig:lines2} and \ref{fig:lines3}.
        Before plotting, the segments were normalized individually, each divided by a constant factor which is the average flux of that spectrum as integrated over the wavelength range of the plot.
        Then, multiplicative offsets using a factor of 1.5 were applied for clarity.
        Vertical lines and legend: identified lines and species.
        Dashed black lines: unidentified lines (UID).
        Sparse dotted black lines: end of a MIRI MRS band, where changes in the spectral resolution occur.
        We note a peculiar detail: broad absorption or emission or artifact features around 7.0~\mum near the \ion{Ar}{ii} line, which remain unidentified.}
    \label{fig:lines}
\end{figure*}
    
\begin{figure*}[t]
\centering
    \includegraphics[trim={0 .5cm 0 0},clip]{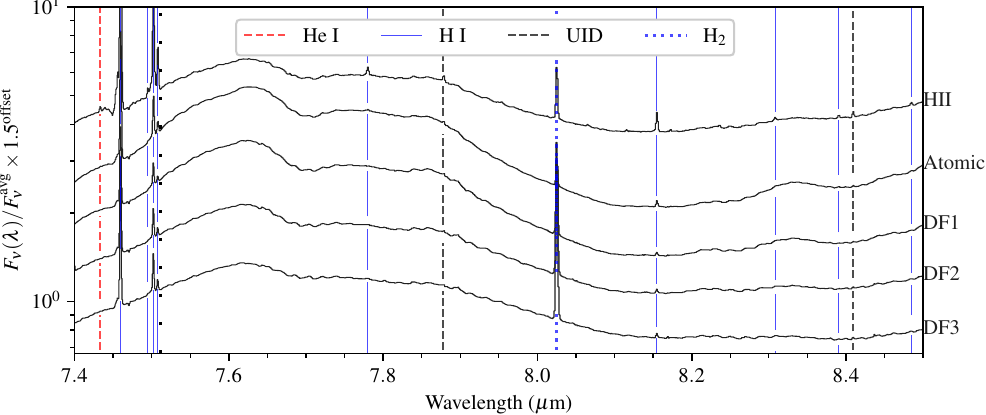}
    \includegraphics[trim={0 .5cm 0 0},clip]{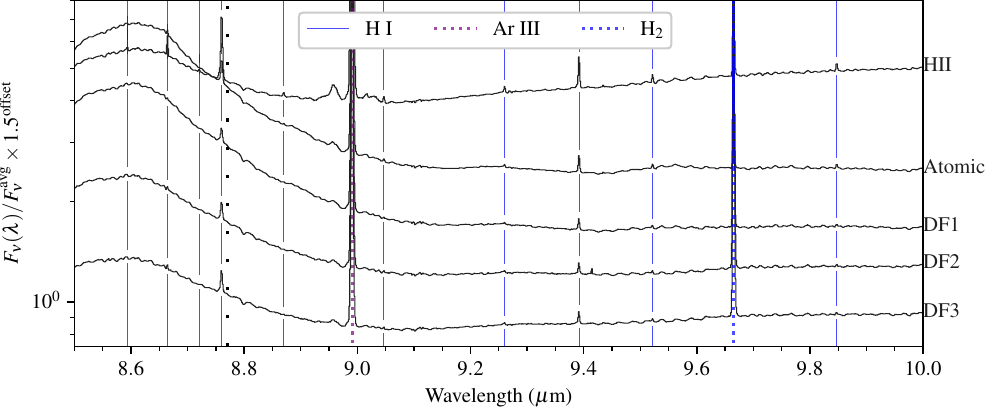}
    \includegraphics{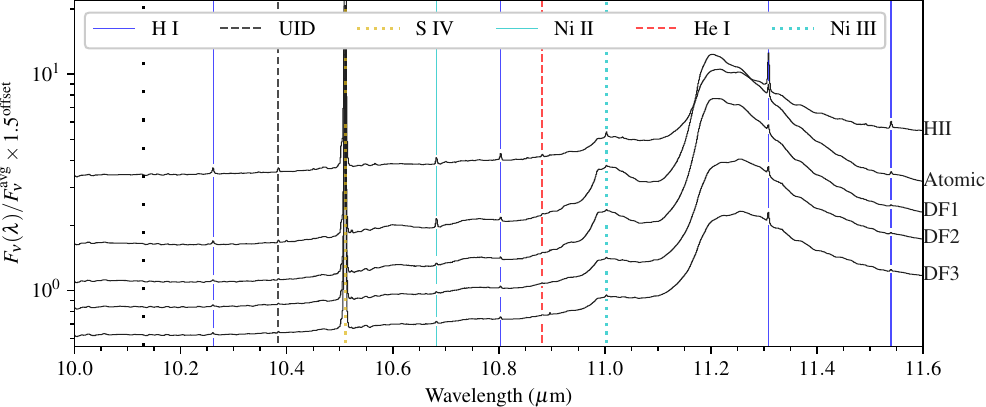}
    \caption{Figure~\ref{fig:lines} continued.
    The oscillations in the top panel are a good example of remaining fringes even after the fringe flat and residual fringe corrections in the pipeline.}
    \label{fig:lines2}
\end{figure*}

\begin{figure*}[t]
\centering
    \includegraphics[trim={0 .5cm 0 0},clip]{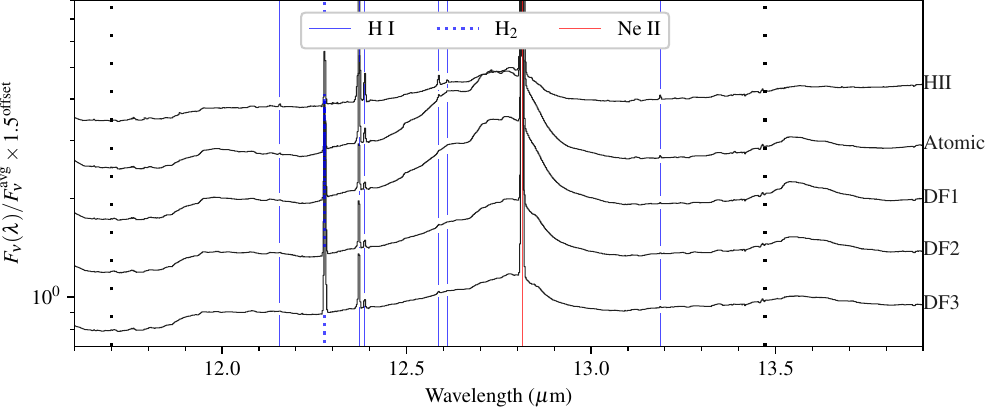}
    \includegraphics[trim={0 .5cm 0 0},clip]{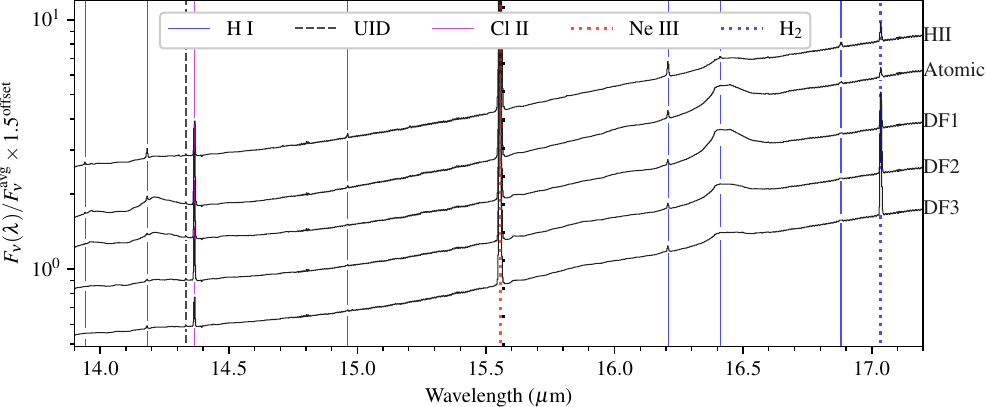}
    \includegraphics{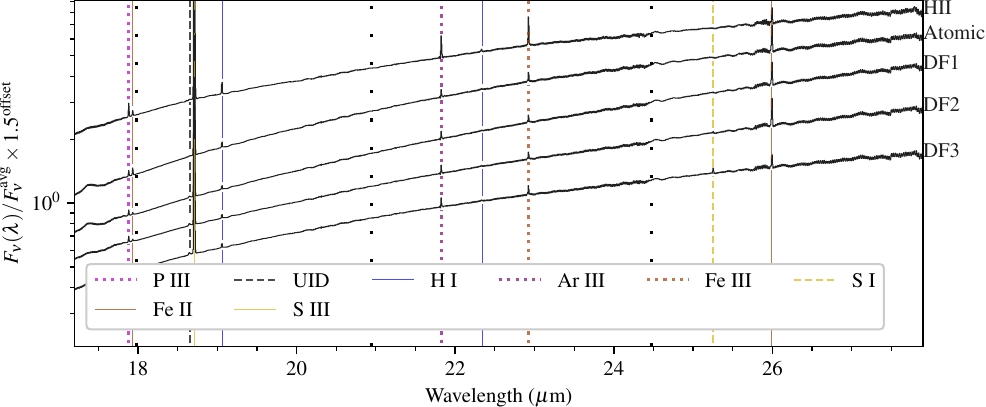}
    \caption{Figure~\ref{fig:lines} continued.}
    \label{fig:lines3}
\end{figure*}

We have composed a list of the lines present in the five extracted spectra, omitting any features which we deem to be artifacts, or where local fluctuations in the spectral baseline make it hard to identify or measure a potential line.
To distinguish features from artifacts, we compared the spectrum to observations of the 10 Lac calibration star, and any oscillations or peaks present in this observed spectrum (residual fringes and other artifact patterns) are considered suspicious and not a real feature of the Orion Bar.
We also exclude the candidates of which the amplitude is similar to that of the local fluctuations and noise patterns, by measuring how this local noise affects the line measurement (see Sect.~\ref{sec:intensity}).
We used a S/N $> 3$ as a rule of thumb to include a line in the list, but we made a few exceptions to complete a series of lines for a species, such as the \HI lines listed in Sect.~\ref{sec:hi}.

To identify the lines, our starting point was the line list originally created for the PDRs4all program \citep{pdrs4all-pasp, Peeters:nirspec}, available as a Science Enabling Product on the PDRs4All website\footnote{https://pdrs4all.org}.
This line list was composed by combining results from the Cloudy ionization model \citep{2017RMxAA..53..385F}, and the Meudon PDR model \citep{2006ApJS..164..506L}.
These models also predict the strength of the lines, and we used this information to filter out lines deemed too weak for detection, which is those with a predicted brightness lower than $10^{-7}$~\lineunit.
This line list was revised based on the spectra obtained as part of the observing program, once they were available.
We match the simulated lines in the list with the central wavelengths we observe in the five template spectra.
As an extra check for the accuracy of the identifications, we compared the measured intensities and their ratios (Sect.~\ref{sec:intensity}) to the intensities predicted by a Cloudy model (Sect.~\ref{sec:hiratios}) or a PDR model (Sect.~\ref{sec:meudon}).
The primary reason for performing this check was to verify the identity of lines for which the observations reveal a small wavelength shift compared to the observations (see also Sect.~\ref{sec:deltawav}).
For example, when the wavelength of a suspected H recombination line is slightly different from the theoretical value, the identification can be deemed more confident when the intensity ratios of the H lines match the theory.
For the \hmol lines, the order of magnitude was checked against the PDR model  included in the line list.
A ratio mismatch was found in a few cases, for example due to the presence of overlapping lines.
Overview plots of the spectrum and the line identifications are shown in Figs.~\ref{fig:lines},~\ref{fig:lines2}, and~\ref{fig:lines3}.
The wavelengths and transitions corresponding to the indicated lines were compiled in the main table of this work, which also lists the energy levels, Einstein coefficients, intensities with uncertainty estimates, and observed central wavelengths (Table~\ref{tab:lines}).

\subsubsection{Hydrogen recombination lines}
\label{sec:hi}

Roughly half of the identified lines are \HI recombination lines that originate from the \HII region between the star and the PDR (\HII template), and the ionized layer that borders the PDR lying between the observer and the neutral gas (Atomic and DF templates).
In the MIRI wavelength range, atomic hydrogen transitions with a lower principal quantum number $n_l$ of 5 or higher can be observed, with the series of higher $n_l$ having more but fainter lines.
The brightest \HI line is the 5-6 transition (shorthand for $n = 5 \leftarrow n = 6$ or Pfund $\alpha$) at 7.46~\mum.
For the Humphreys series ($n_l = 6$), we observe all four bright lines, with the upper state $n_u$ ranging from 10 (5.13~\mum) to 7 (12.37~\mum).
All 15 available lines with $n_l = 7$ are observed from 7-23 (4.92~\mum) to 7-8 (19.062~\mum).

The remaining \HI lines we observe have $n_l$ ranging from 8 to 10, and are significantly fainter. 
Because of variations in the noise and artifacts throughout the spectra, some of the fainter lines are not detected or not reliably measured.
We detect the $n_l = 8$ series from 8-29 (6.316~\mum) to 8-10 (16.21~\mum).
The 8-12 (10.503~\mum) transition was not measured because it is hidden in the wing of the extremely bright \ionwav{S}{IV}{10.510} transition.
The first transition of this series, 8-9 at 27.80~\mum, is not detected because the current MIRI calibration is not accurate enough to distinguish it from the very high continuum flux at that wavelength.
We measured the $n_l = 9$ series from 9-27 (8.308~\mum) to 9-11 (22.34~\mum), where the 9-10 line is not available as it is outside the MIRI MRS wavelength range.
Finally, were report the $n_l = 10$ series from 10-15 (16.41~\mum), up to 10-20 (12.16~\mum).
The 10-14 and 10-13 transitions are in the wavelength range but we do not list them, since they are too weak to be reliably measured, and because 10-13 (22.33~\mum) overlaps with 9-11 at 22.34~\mum while the spectral resolution is not sufficient to separate their contributions.
Several transitions with higher $n_u$ are still detectable, but their low intensity makes many of them too faint to measure at a S/N $> 3$, and we opt to omit the series beyond this point.

Because of the lower transition rates and more sparsely populated upper levels, some \HI\ lines are faint enough to be affected by the data quality, typically when the intensity $\lesssim \num{e-5}$ \lineunit.
The intensity and uncertainty measurements are described in Sect.~\ref{sec:intensity}.
Care has to be taken with 6-8, 7-11, and 8-16, which form a closely spaced trio near 7.5~\mum. Similarly, 7-9, 8-23, 8-22, and 8-17 are in close proximity to other lines, and for such cases we employed a decomposition as explained in Sect.~\ref{sec:intensity}, with an example shown in Fig.~\ref{fig:overlap}.

\subsubsection{Molecular hydrogen lines}
\label{sec:h2lines}

The brightest \hmol lines are the pure-rotational lines ($J \rightarrow J - 2$) in the vibrational ground state ($v = 0$).
We observe all the pure-rotational transitions available in the MIRI wavelength range, which range from lower rotational quantum number $J = 1$ to $J = 8$, in other words all 8 transitions denoted as 0-0~$S(1)$ to 0-0~$S(8)$.
Only the $S(0)$ line at 28.22 \mum is missing.
The calibrated wavelength range of MIRI MRS extends just far enough to include the wavelength of this line, but the continuum level is very high and the S/N and is too low to enable a detection.
The intensity of these lines is several \num{e-3} \lineunit, and these values are used in the \hmol excitation analysis presented in Sect.~\ref{sec:analysis}.
We also detect pure rotational lines in the $v = 1$ vibrationally excited state, from 1-1~$S(5)$ to 1-1~$S(9)$.
While transitions down to 1-1~$S(1)$ lie within the available wavelength range, they are too faint to detect.
The NIRSpec observations of this program show the continuation of the $v = 0$ and $v = 1$ ladders towards higher $J$, up to 0-0~$S(19)$ and 1-1~$S(17)$ \citep{Peeters:nirspec}.

\change{There are also a few $v = \text{1-0}$ rovibrational transitions in the wavelength range, that have the same upper level as the $v = \text{1-1}$ pure rotational transitions mentioned above.
For example, 1-1 $S(6)$ ($v = 1, J = 8 \rightarrow v = 1, J = 6$) has the same upper level as 1-0 $O(10)$ ($v = 1, J = 8 \rightarrow v = 0, J = 10$).
We inspected the shape of the spectrum in detail, at the wavelengths of the 1-0 $O(10)$ to 1-0 $O(13)$ transitions (5.0590, 5.6294, 6.3086, and 7.1267 \mum, respectively).
There is a tentative detection for each of these transitions upon inspection by eye, but the lines are not prominent enough for inclusion in Table~\ref{tab:lines} due to the smaller Einstein coefficient.
Another tentative detection, is the rovibrational line 4-3 $O(8)$ at 5.0981 \mum, but there are no clear detections of other vibrationally excited series such as 3-2, 2-1, 3-3, or 2-2.}

In work by \citet{2000A&A...356..705R}, unresolved observations of the Orion Molecular Cloud outflow (OMC-1) with the Short Wavelength Spectrometer (SWS) of the Infrared Space Observatory (ISO), detected the bright pure-rotational emission driven by shocks, from 0-0 $S(1)$ to $S(25)$
and from 1-1 $S(5)$ to $S(17)$.
Observations of the Orion Bar with ISO-SWS were obtained by F.~Bertoldi (priv.\ comm.)  and used by \citet{2004Habart} and \citet{2018Joblin}, and these detected the pure-rotational lines from 0-0 $S(0)$ to $S(5)$. 
Previous deep spectroscopic observations in the NIR revealed a large set of rovibrational lines with $v = 0$ to $v = 10$ in the Orion Bar, using the Immersion Grating Infrared Spectrometer of McDonald Observatory \citet{2017ApJ...838..152K}.
Spatially resolved observations of the MIR pure-rotational lines were previously obtained through ground-based observations, where the 0-0 $S(1)$, $S(2)$, and $S(4)$ lines were mapped at a 2\arcsec\ resolution \citep{2005ApJ...630..368A}.
The \jwst\ data of our program provide the first maps of the Orion Bar for a complete set of \hmol lines, extending well past 0-0 $S(5)$, at a resolution that resolves the individual dissociation fronts.
An in-depth spatially dependent analysis of the \hmol excitation from the MIR to NIR will be presented by Sidhu et al.\ (in preparation).

\subsubsection{Fine-structure lines}
\label{sec:metalionlines}

All metal lines we detect are forbidden fine-structure lines.
Based on ISO-SWS observations, \citet{2000A&A...356..705R} separated which lines originate from the foreground ionized and PDR region, and those from an outflow of OMC-1 excited by powerful shocks.
Their Table~2 lists the lines for the Orion Bar PDR, and we detect the same set of lines with a few exceptions.
The MIRI wavelength range does not extend to \ionwav{Si}{ii}{34.8}, but we do detect \ionwav{S}{i}{25.2} in DF3, and the detailed spatial distribution of sulfur will be discussed in a separate paper (Fuente et al.\  in preparation).
The newly detected and identified lines compared to these ISO data are \ionwav{Ni}{ii}{10.68}, \ionwav{Ni}{iii}{at both 7.35~\mum and 11.00}, and \ionwav{Cl}{ii}{14.37}.
The \ion{Cl}{ii} line was also detected by \citet{2011MNRAS.410.1320R} using the Spitzer Space Telescope.
Most of these lines originate only from the ionized layer, while others come from both the ionized and atomic layers between the observer and the molecular part of the Orion Bar.
This is discussed further in Sect.~\ref{sec:ionratios}.

The brightest lines have intensities of several \num{e-2} \lineunit, are the forbidden lines of \ion{Ar}{ii}, \ion{Ar}{iii}, \ion{S}{iii}, \ion{S}{iv}, \ion{Ne}{ii}, and \ion{Ne}{iii}.
They are measured without saturation issues and are present in all five apertures.
Because of their brightness, they are frequently used as tracers for \ion{H}{ii} regions, and as diagnostics for the ionic abundance ratios or the level of excitation for example, including in extragalactic objects \citep[e.g.,][]{2007MNRAS.377.1407R}.
The lines of \ion{Fe}{ii}, \ion{Fe}{iii}, \ion{Ni}{ii}, \ion{Ni}{iii}, \ion{Cl}{ii}, and \ion{P}{iii} have intensities down to \num{e-3} \lineunit.
Of these, the species with the weakest fine-structure lines are \ion{Ni}{ii} and \ion{Ni}{iii}, but they are still detected all five templates.
The S/N for most of the weak lines is still $\gtrsim 10$, meaning that the metal lines have a very high S/N in these data, and the uncertainty will be dominated by the systematics in the calibration and our spectral order matching factors (see Sect.~\ref{sec:data}).
The exceptions are \ionwav{Ni}{iii}{11.0023} and \ionwav{S}{i}{25.25} for which the S/N is lower than 3 for some of the measurements.
In Sect. \ref{sec:ionratios}, we use the brightest Ar, Ne, S, Fe, and Ni lines to trace the variation in radiation field hardness between the five template regions.

A caveat is that for [\ion{Ni}{iii}], the wavelength as given by the Atomic Spectra Database (ASD, \citealt{NIST_ASD}) of the National Institute of Standards and Technology (NIST) is 7.3492~\mum, while the measured wavelength is 7.3524~\mum, which differs by 0.0032~\mum.
A smaller difference exists for \ionwav{Ni}{iii}{11.0023}, which is instead observed at 11.0035~\mum.
It is likely that this offset occurs because the atomic data for these transitions are not sufficiently accurate, since only a Ritz wavelength and no observed wavelength is listed on the ASD.
The Ni ionization state is known to be similar to the Fe ionization state \citep{1996A&A...315L.269L} because of their very similar ionization potentials (see also Sect.~\ref{sec:ionratios} and Table~\ref{tab:potential}).
Considering that the \ion{Fe}{ii} and \ion{Fe}{iii} ions are both present, the \ion{Ni}{ii} and \ion{Ni}{iii} ions should also be present if Ni is abundant enough.
When we assign [\ion{Ni}{iii}] to the lines at 7.35 and 11.00 \mum, the observed intensity for these lines is consistent with the expectations based on our theoretical models.
In Sect.~\ref{sec:ionratios} we confirm the correspondence between the $\ion{Fe}{iii}/\ion{Fe}{ii}$ and $\ion{Ni}{iii}/\ion{Ni}{ii}$ ratios based on our measurements.

The measured wavelength difference for \ionwav{Ni}{iii}{7.35} is 0.0035~\mum, or 0.0032~\mum if corrected for the velocity of the \HII region relative to the frame of the calibrated \jwst\ data (see Sect.~\ref{sec:deltawav}).
This is much larger than the standard deviation of our \HI line wavelength offset measurements of about 0.0003~\mum. With the above information, our best empirical estimate for the \ionwav{Ni}{iii}{7.35} wavelength is $7.3524 \pm 0.0003$~\mum.
For the other [\ion{Ni}{iii}] line, with the same reasoning and velocity correction as above, the remaining offset is 0.0008~\mum and our empirical wavelength estimate is $11.0031 \pm 0.0003$~\mum.
For consistency, we will refer to the corrected lines using the wavelength from the ASD in the rest of this work.

\subsubsection{Helium recombination lines}

While many \ion{He}{i} lines are detected upon visual inspection of the \HII template spectrum, most of them are located in the blueward wing of bright \HI recombination lines (e.g., in Fig.~\ref{fig:overlap}).
In most of these cases, the brightness of the \HI line is several orders of magnitude higher than the \ion{He}{i} line, and we could not obtain reliable measurements.
An example is the subtle contribution of $^1F^o \leftarrow {^1D}$ at 6.769~\mum to the wing of the \HI 7-12 line.
Some of these transitions are not listed in the ASD, but they are present in the output of the Cloudy model and in the ``Atomic Line List'' by \citet{2018Galax...6...63V}. 
In the discussion below there are often many \ion{He}{i} lines present near the same wavelength, and we mention those with the highest Einstein coefficient.

The brightest \HI line (5-6, 7.4599~\mum) in our spectrum is contaminated by lines at 7.4561 \mum from the $^3G \leftarrow {^3H^o}$ multiplet and the $^1G \leftarrow {^1H^o}$ singlet,
and a similar group of lines at 7.4538~\mum from the $^3F^o \leftarrow {^3G}$ multiplet and $^1F^o \leftarrow {^1G}$ singlet.
Performing a fit as demonstrated in the right panel of Fig.~\ref{fig:overlap} results in a correction of about \textbf{$\sim8\%$} to the intensity of this \HI line.
There might be a similar contamination for the \HI 6-8 line at 7.503~\mum; in addition to the other \HI 8-17 (7.495~\mum) and 7-11 (7.508~\mum) lines, the Cloudy model predicts possible contributions by \ion{He}{i} multiplets at 7.499~\mum and 7.505~\mum at the 1\% to 10\% level.
With the mixing of three \HI lines and two \ion{He}{i} multiplets, there is no straightforward way to extract the small \ion{He}{i} contribution from this complex.
For this reason we do not report the intensities of most \ion{He}{i} lines, and we only use an empirical shape to correct the \HI intensity of certain lines, as discussed above.

There are a few exceptions where the \ion{He}{i} lines are sufficiently isolated to allow a measurement, as listed in Table~\ref{tab:lines}.
The line at 6.7217~\mum is part of a clearly defined double peak together with a weak \HI line.
It is dentified as a \ion{He}{i} line based on the Cloudy model, but the Atomic Line List by \citet{2018Galax...6...63V} lists many transitions around the same wavelength, with similar upper level energies, and it is unclear which would be the dominant contribution.
Therefore, we have noted the transition as ``multiple'' in Table~\ref{tab:lines}, with an  upper level energy close to all the listed upper levels.
There is another clearly defined line at 7.4334~\mum, where the $^3D \leftarrow {^3F^o}$ multiplet is present, with no interference from other nearby lines, and at 10.88~\mum the $^3S \leftarrow {^3P^o}$ multiplet is measured.

\subsubsection{Unidentified lines}

Several lines for which the origin is unclear are marked as ``UID'' in Table~\ref{tab:lines}.
We compared the observed wavelengths of these lines to the PDRs4All line list and the ASD for all elements with atomic number $< 30$ and ionization up to \ion{stage}{iv}.
Some lines have no matches, while others match individual wavelengths in these line lists, many of which are highly excited lines of H$_2$ lines or atomic recombination lines.
None of the candidates were convincing however,
and we summarize the ruled-out species and transitions below, per template region type (ionized, atomic, dissociation front), and note which unidentified lines are the brightest in each region.

In the \HII template spectrum, the brightest unidentified lines are at 5.240, 5.740, \change{5.596}, and 7.878~\mum (in order of decreasing brightness), with intensities of several \num{e-5} \lineunit.
The 5.240 and 7.878~\mum lines only appear in this region and have no matches in any of our line catalogs.
The \change{5.596}~\mum line is also exclusive to the \HII template and is a somewhat close match to the \change{recombination lines of \ion{Mg}{i} at 5.5973~\mum or \ion{C}{i} at 5.5983~\mum, but there is still a significant wavelength offset} and no recombination lines of similar expected intensity are detected.
The 5.740~\mum line is visible in all templates and matches \ion{Li}{ii} 5.7406~\mum, which is unlikely because of the same reason. 

The weaker unidentified lines in \change{the \HII region} (several \num{e-6} \lineunit) are at 5.628, 5.658,  \change{6.366, 6.498, 8.410, 10.386, and 14.333}~\mum.
\change{The line at 6.498~\mum is located close to the 6.501~\mum \HI line.} 
\change{Close wavelength matches from the line lists include a \ion{He}{i} recombination line at 6.4984~\mum
and a H$_2$ rotational line} at 6.4999~\mum ($v = 3, J = 9 \rightarrow 7)$, or a \ion{N}{i} \change{recombination} line at 6.4985~\mum.
The identification as a vibrationally excited \hmol line is unlikely because the line is not observed for the DF templates.
If the \change{6.498}~\mum line matches the \ion{N}{i} 6.4985~\mum recombination line, this could also explain another unidentified line at 10.386~\mum, which could match \ion{N}{i} 10.385~\mum. 
Since none of the other \ion{N}{i} recombination lines are detected, the remaining option is a tentative detection of the \ion{He}{i} 6.4984~\mum line.
Similarly the lines at \change{6.366 and} 8.410~\mum appear only in the ionized region, so coincidentally matching H$_2$ lines are unlikely candidates, because they do not appear in any of the molecular template spectra.
\change{The 6.366~\mum line is close to \ionwav{Ar}{iii}{6.368}, but the latter has the same upper level as \ionwav{Ar}{iii}{8.99}, while the Einstein coefficient is smaller by a factor $\sim$$10^4$, and the observed flux ratio does not match this expected ratio}.
The 8.410~\mum line matches very well in wavelength with \ion{C}{i} 8.4095~\mum, but again we do not see any \ion{C}{i} recombination lines of similar excitation that are expected to be similarly strong in the MIR spectra.
We note that \ion{C}{i} and \ion{N}{i} recombination lines have been observed in the PDRs4All NIRSpec spectra \citep{Peeters:nirspec}.
For \ion{C}{i}, the lines at 1.069 and
1.175~\mum are measurable, but these have lower upper levels and an order of magnitude higher Einstein coefficients compared to those in the MIR, and the situation is analogous for \ion{N}{i}.
\change{The line at 14.333~\mum appears in all five template spectra, and we find no matches in the catalogs used}.

In the Atomic template, the brightest lines are at 6.859, 6.498, 5.240, and 5.740~\mum (about \num{e-5} \lineunit), of which the last three are also present in the \HII template as discussed above. 
The line at 6.859~\mum is only present in the atomic region, and its wavelength matches the $v = 8, J = 20 \rightarrow 18$ transition of H$_2$.
As this line is not observed in the dissociation front templates, it is unlikely that this is the right identification.
\change{There is a \ion{N}{i} recombination line nearby at 6.8560~\mum,
but as with several other lines mentioned above, the wavelength difference remains too large to serve as a convincing identification.}

The DF1, DF2, and DF3 templates have a relatively strong feature at 18.651~\mum, for which we find no clear identification.
Its width is roughly twice that of the nearby \ion{S}{iii} line, so it is likely not a single atomic line.
The dissociation fronts exhibit weak unidentified lines at 4.928, 4.968, 5.552, 5.634, 5.740, and 6.412~\mum, of which 5.740~\mum also appears in the \HII and Atomic templates.
\change{For 5.634~\mum, the 4-3 $O(9)$ rovibrational transitions is a likely explanation (see also Sect.~ \ref{sec:h2lines})}.
Several others coincide in wavelength with a CO line of high excitation (e.g., $v = 2, J=20$), or several highly excited \hmol lines in the line list. 
The lack of other highly rotationally excited \hmol or CO lines rules out those identifications. 
\change{There is a near perfect} wavelength match with an \ion{S}{I} line at 6.4126~\mum, but this is again a non-plausible recombination line.

At around 7.0~\mum and 9.0~\mum, several strong broad features resembling emission and absorption are present.
Both of these appear near Ar lines and are particularly strong for the \HII template.
We compared spectra with a version of the data where no background subtraction was applied and confirmed that these features are not an artifact of the background subtraction.
The data of the 10 Lac calibration star (see Sect.~\ref{sec:extraction}) were also inspected at this wavelength and show no sign of similar effects.
At the time of writing it remains unclear if these particular unidentified features are another type of artifact, although it is peculiar that features of this type only appear at wavelengths near the Ar lines.

We note that certain molecules might be candidates in the DF templates.
However, we only included \hmol, HD, and CO in our search.
\change{The work by Zannese et al.\ (in preparation) investigates other molecular candidates including CH$_3^+$}, which has already been detected in \change{a disk with much denser gas} that is present in our field of view \citep{2023Natur.621...56B}.
A more in-depth and physically motivated search for unidentified line candidates, is deferred to future studies that focus on specific spatial regions and specific groups of unidentified lines.

\subsection{Intensity and uncertainty measurements}
\label{sec:intensity}

For each line, the intensity in each of the five templates is given in Table~\ref{tab:lines}.
The extraction method used for most of the line intensities is a simple local integration over the spectrum, with an estimate of the continuum subtracted.
To determine an appropriate integration window and continuum level, we first derive a value for the full width at half maximum (FWHM), using instrumental resolution curves for MIRI adapted from \citet[Fig.~9]{2021A&A...656A..57L}.
We evaluate the resolution curve at the central wavelength of the line $\lambda$, and denote the resulting FWHM as $w(\lambda)$.
After fitting a Gaussian profile to a few lines, we found that the fitted FWHM is very close to the instrumental FWHM, meaning that the lines are unresolved at the MIRI MRS velocity resolution of around $\SI{100}{\km\per\s}$.

The continuum is estimated as a local linear function, by taking the median value of the flux in the two nearby windows $[\lambda - 4w(\lambda), \lambda - 2w(\lambda)]$ and $[\lambda + 2w(\lambda), \lambda + 4w(\lambda)]$, and applying a linear interpolation between these two continuum levels.
We found that this median-based approach worked best for the large number of lines in our data, as it is more robust against local outliers and does not suffer from overfitting, as opposed to fitting a polynomial which would require manual adjustment for individual lines.
The estimated continuum levels were inspected by eye and proved reasonable for most lines.
The line intensity (in \lineunit) is then obtained by converting the flux to per-wavelength units (erg \si{\per\s\per\square\cm\per\steradian\per\micro\m}), subtracting the local linear continuum, and integrating over the wavelength window $[\lambda - 2w(\lambda), \lambda + 2w(\lambda)]$.
For the measurement of the central wavelength, see Sect.~\ref{sec:deltawav}.

\begin{figure}[tb]
    \centering
    \includegraphics{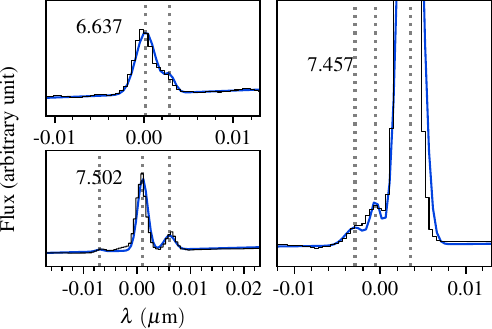}
    \caption{Examples of intensity determination for overlapping or nearby lines, in \HII template spectrum. 
    The x-axis has been shifted so the zero point matches the wavelength in \mum displayed in the top left of each panel.
    Blue curve: fitted model consisting of a linear continuum and two or three Gaussian profiles.
    Dotted lines: central wavelengths of the Gaussian profiles.
    Upper left panel: \ionwav{Ni}{ii}{6.636} and \HI\ {6.638}~\mum.
    Lower left panel: Complex of three \HI lines, with fitted wavelengths at 7.4955, 7.5030, and 7.5086~\mum.
    Right panel: \HI 5-6 and contributions by \ion{He}{I} multiplets in its left wing.}
    \label{fig:overlap}
\end{figure}

\begin{figure*}[tb]
    \centering
    \includegraphics{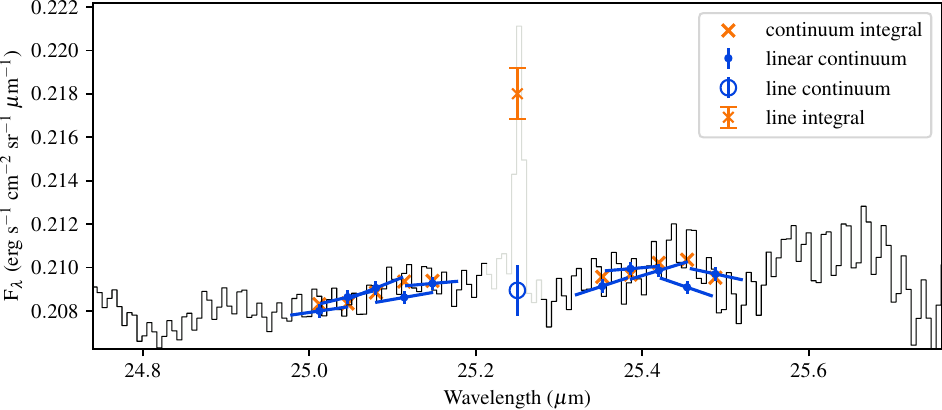}
    \caption{Demonstration of uncertainty measurement described in Sect.~\ref{sec:intensity}.
    The line in this example is \ionwav{S}{i}{25.25}, and the oscillations are an artifact of MIRI MRS Ch~4.
    Blue segments with dots: Local linear continuum estimates, with the center indicated.
    Orange crosses: Blue dot value plus the average of the continuum subtracted spectrum as integrated over the continuum window.
    The deviation between the orange cross and the blue dot samples the deviation that occurs due to local noise patterns and the linear continuum approximation.
    The median absolute deviation is used as the uncertainty on the line intensity integral.
    Blue circle and error bar: Continuum and and median absolute deviation of the above samples.
    Orange error bar: Line intensity and uncertainty, visualized approximately as an amplitude. 
}
    \label{fig:unc}
\end{figure*}

Certain groups of lines are closely spaced, and some have problematic continuum determinations with the method above.
For a total of around ten of these cases, we applied an individualized approach, fitting a functional model consisting of the sum of two or more Gaussians, on top of a first to third degree polynomial continuum.
Three of these cases are shown in Fig.~\ref{fig:overlap} to illustrate this method.
The order of the polynomial for the continuum was larger than linear for the cases where significant local curvature due to AIB emission was present.
The width of the fitting window and degree of the continuum were manually fine-tuned for each group of the overlapping lines.
The fitted amplitude of each Gaussian component is then converted to an intensity by calculating the area.
For these manually fit cases, the wavelengths were measured by fitting this model with variable central wavelengths for the Gaussian components, as opposed to the centroid method  discussed in Sect.~\ref{sec:deltawav}.

The amount of noise in the spectrum contributed by local variations in the spectral baseline depends on the noise features near each wavelength. 
The error bars on the data points of the spectrum do not suffice, as there can also be noise due to local artifacts (e.g., fringing, other oscillating patterns, or small spectral features).
Therefore we use an empirical local uncertainty estimate for each line, performed as follows.
First, we apply our linear continuum subtraction to around ten wavelength windows near the line, creating a set of samples of the local spectrum.
The continuum subtracted samples are then integrated over the same interval width that was used for the line intensity, yielding the values $\delta_i = \int_i [F_\lambda(\lambda) - \mathcal{C}_i(\lambda)]\mathrm{d}\lambda$, where $i$ is one of the wavelength intervals, $F_\lambda(\lambda)$ is the flux density per unit wavelength, and $\mathcal{C}_i(\lambda)$ the local model for the continuum.
With a perfect continuum subtraction, and a noiseless spectrum, each integral $\delta_i$ would be zero.
Since there is noise, and since a linear continuum is not a perfect representation of the underlying continuum, every $\delta_i$ deviates from zero and is a quantity in the same units as the line intensity (\lineunit).
An example is shown in Fig.~\ref{fig:unc}.
We use the median of the absolute values $|\delta_i|$ as an estimate for the uncertainty on the line intensity integral (median absolute deviation).
To visualize this uncertainty on the flux axis in Fig.~\ref{fig:unc}, the line intensity and its uncertainty were converted to amplitude values, by dividing by the Gaussian normalization factor $\sqrt{2 \pi} \sigma$, with $\sigma = (2\sqrt{2\ln 2})^{-1} w(\lambda)$, the standard deviation derived from the MIRI MRS FWHM curve.

In addition to the local noise in the spectral baseline, we also note the calibration uncertainties between the segments which we estimated as roughly 3\% (Sect.~\ref{sec:extraction}).
We expect this systematic contribution to affect all lines within one and the same segment with the same factor.
For applications where detailed uncertainty models are required, a covariance matrix can approximate this effect using the off-diagonal elements $C_{ij} = f_S^2 I_i I_j$ for lines in the same segment.
Here $f_S$ is the fractional systematic uncertainty ($\sim$0.03), and $I_i$ the intensity of line $i$.
The diagonal elements are $C_{ii} = \sigma_i^2 + f_S^2 I_i^2$, 
with $\sigma_i$ the local uncertainty estimated as described above.
For the parts of the analysis that use line ratios (Sect.~\ref{sec:analysis}), one can simply ignore the systematic calibration uncertainty for lines in the same segment, or add the fractional uncertainty to the numerator and denominator otherwise.
The latter corresponds to a $\sqrt{2 \times 0.03^2} \approx 4\%$ relative systematic uncertainty on the ratio of two line intensities, which can be added in quadrature to the local uncertainty.

\subsection{Observed wavelength offset}
\label{sec:deltawav}

\begin{figure}[tb]
    \centering
    \includegraphics{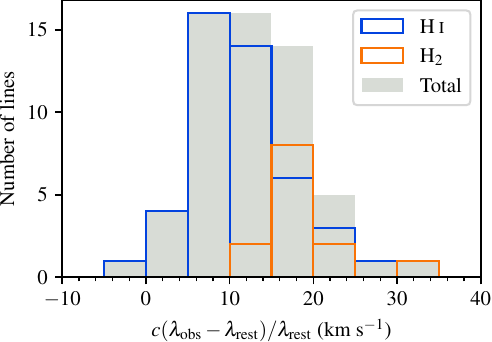}
    \caption{Histogram of fractional wavelength deviation (Sect.~\ref{sec:deltawav}).
    The \HI and \hmol lines originate from different gas layers, and the difference between the velocity averages is about \SI{7.6}{\km\per\s}.}
    \label{fig:deltawav}
\end{figure}

In addition to the intensity, we also measured the central wavelength of each line, so that the observed wavelength $\lambda_\text{obs}$ can be compared to theoretical or experimental rest-frame wavelength $\lambda_\text{rest}$.
To determine the center of each line, we calculate its centroid over a window spanning from $\lambda_\text{rest} - w(\lambda)$ to $\lambda_\text{rest} + w(\lambda)$, and the equation used is
$
\lambda_\text{obs} = [\int \lambda F_\lambda(\lambda) \text{d}\lambda] / [\int F_\lambda(\lambda) \text{d}\lambda].     
$
The measured wavelength shifts are given in a column of Table~\ref{tab:lines}, defined as $\Delta\lambda_\text{obs} = \lambda_\text{obs} - \lambda_\text{rest}$.
The most precise atomic and molecular data are those for the \HI and \hmol lines, and a histogram of the relative wavelength shift $\Delta\lambda_\text{obs} / \lambda_\text{rest}$ is shown in Fig.~\ref{fig:deltawav} for these lines.
The average relative wavelength shift for the \HI lines is \SI{11.3}{\km\per\s}, and the standard deviation of this sample is \SI{6.0}{\km\per\s}, while the \hmol lines are centered around \SI{18.9}{\km\per\s} with a standard deviation of \SI{5.4}{\km\per\s}.
Dividing the standard deviations of the distributions by the square root of the number of lines yields the uncertainty on the measurement of the averages: $11.3 \pm 0.9$ \si{\km\per\s} for \HI and $18.9 \pm 1.5$ \si{\km\per\s} for \hmol.
The difference between these two components is $7.6 \pm 1.7$ \si{\km\per\s}.

High-resolution spectroscopic observations of C and S recombination lines show that the emission lines originating from the Bar have an intrinsic width of $\sim$\SI{2.5}{\km\per\s} and a redshift corresponding to a radial velocity of $v_\text{LSR} \approx \SI{+10.6}{\km\per\s}$ \citep{2021A&A...647L...7G} with respect to the Local Standard of Rest (LSR).
In the same study, the surrounding \ion{H}{ii} region has a line width of $\sim$\SI{20}{\km\per\s} and a blueshift of $v_\text{LSR} \approx \SI{-5}{\km\per\s}$, as measured using He recombination lines.
The expected velocity difference between the \HII region and the \hmol component of the Bar is therefore about \SI{15}{\km\per\s}; the difference we observe between the two averages reported in the previous paragraph is $7.6 \pm 1.7$ \si{\km\per\s}.

The spectroscopic pipeline\footnote{\url{https://jwst-pipeline.readthedocs.io/en/latest/jwst/assign_wcs/main.html}} adjusts the wavelength grid of the spectra, so that the final products are in the barycentric (heliocentric) reference frame.
Hence the data still include the motion of the Sun, with respect to the LSR that is used to report $v_\text{LSR}$ in the literature.
Using the definition of the LSR from the ``coordinates'' module of Astropy, we find that a radial velocity of \SI{0}{\km\per\s} in the heliocentric frame corresponds to $v_\text{LSR} = \SI{-17.17}{\km\per\s}$ for the coordinates of the Orion Bar.
Converting the velocities of our \HI and \hmol components, we find $v_\text{LSR}(\text{\HI}) = \SI{-5.9}{\km\per\s}$ and $v_\text{LSR}(\text{\hmol}) = \SI{1.7}{\km\per\s}$.
The velocity we observe for the ionized region is close to the literature result of \SI{-5}{\km\per\s} cited in the previous paragraph,
while the results for the Bar differ by about \SI{9}{\km\per\s}.  
The velocity difference between the \HII and neutral layer is qualitatively similar to the previous observations, as the \HII region is also blueshifted with respect to the molecular component, but the size of the velocity difference we observe is about half the literature value.
Therefore, the H$_2$ rotational lines likely probe a different part of the molecular region than the C and S recombination lines.

The wavelength offset with respect to previous observations reported above is smaller than the reported wavelength calibration uncertainty of \SI{27}{\km\per\s} at 28~\mum, but similar to the \SI{9}{\km\per\s} accuracy that is expected near 5~\mum \citep{2023A&A...675A.111A}.
A shift of this size is smaller than one resolution element, considering that the spectral resolution of MIRI MRS varies between 3500 and 1500, from the shortest to the longest wavelengths respectively, corresponding to a radial velocity resolution of about 85-200 \si{\km\per\s} \citep{2021A&A...656A..57L}.
Our radial velocity measurements are more precise than the resolution, as the shape of the (instrumental) line profile is resolved by the detector pixels and the wavelength grid in the final products.

\section{First Analysis}
\label{sec:analysis}

\subsection{Hydrogen recombination}
\label{sec:hiratios}

\begin{figure*}
    \centering
    \includegraphics{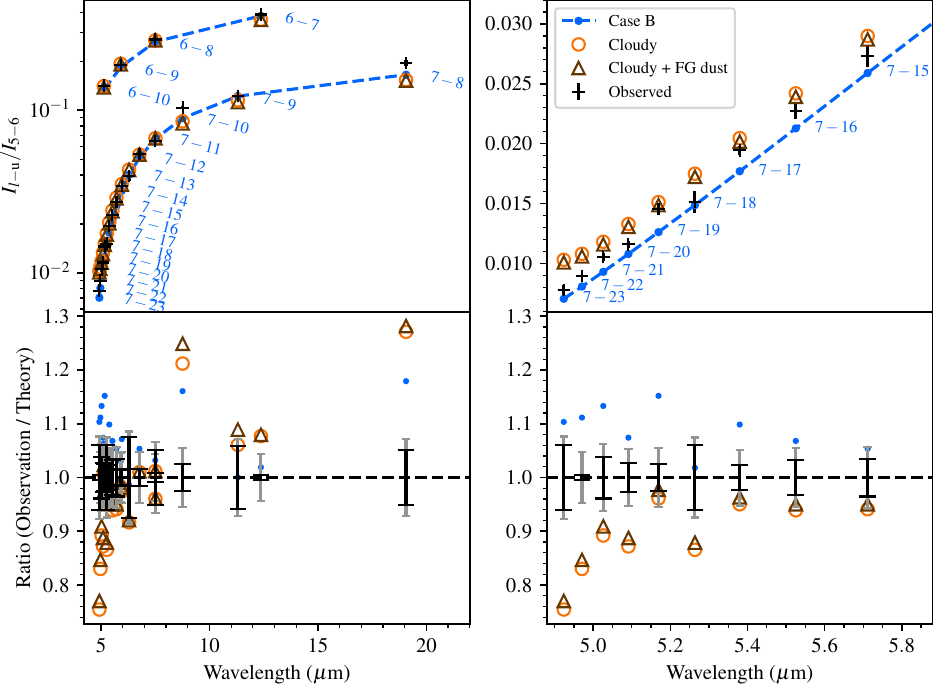}
    \caption{Analysis of \HI recombination line ratios.
    Left panel: Transitions to lower levels 6 and 7 for \HII template spectrum, comparing \HI emission line ratios between observations and theory.
    Labels next to each data point indicate the transitions.
    The Cloudy model with foreground dust correction applied (Cloudy + FG dust) is shown assuming the largest value for the Orion Bar foreground ($A(V) = 1.9$), making it clear that the effect is small.
    Right panel: Zoom-in to better show the deviations between the observations, the Case B model, and the Cloudy model.
    \textit{Bottom panels:} Observed line ratio divided by theoretical line ratio, for the three models as coded by the legend. 
    Error bars: Uncertainty due to local noise (black) and with the 3\% systematic calibration uncertainty added (gray).} 
    \label{fig:hi}
\end{figure*}

We compare the observed \HI line ratios to recombination theory \citep{1987MNRAS.224..801H, 1995MNRAS.272...41S}, using the recent Case B coefficients from \citet{2018MNRAS.478.2766P}.
The assumption of Case B theory is that all Lyman transitions are optically thick, and absorbed in the same location where they were emitted by the gas \citep{1938ApJ....88...52B}.
Practically, this is equivalent to ignoring transitions to the ground state, and case B approximates the emission well under typical nebular conditions \citep{1962ApJ...135..195O}. 
In Fig.~\ref{fig:hi}, we show the observed intensity of lines with lower state $n_l = 6$ and $n_l = 7$ over the available range of upper states $n_u$, for the \HII template.
To compare the observed intensity with the theoretical emissivity, we divide the respective values by that of the 5-6 line at $7.46$ $\mu$m.
In other words the ratio $I_\text{l-u} / I_\text{5-6}$ is shown for the observations, while the theoretical data points are the emissivity coefficient ratios $j_{l-u} / j_{5-6}$.
The lower panel of Fig.~\ref{fig:hi} shows the deviations from the theory, using the quantity ($I_\text{l-u} / I_\text{5-6}$) / ($j_\text{l-u} / j_{5-6}$), or in other words the observed line ratio divided by the theoretical line ratio.

Consistent with \citet{Peeters:nirspec}, we choose the emissivity coefficients from \citet{2018MNRAS.478.2766P} with the electron temperature $T_e = \SI{e4}{\kelvin}$ and the electron density $n_e = \SI{e3}{\per\cubic\cm}$, and zero radiation field.
These parameters work well to describe the observed emission ratios within the error bars.
We also compared with different models from the same set, for example by changing $n_e$ to \SI{e2}{\per\cubic\cm} or $T_e$ to \SI{5e3}{\kelvin}, and the differences in the observed-to-theory ratio are minor compared to the error bars. 
There is a systematic offset however, where the data points for the \HII template differ by a factor of 1.1 from Case B on average.
For the other four template apertures, the observed-to-theory ratios have a similar scatter, but they are centered at 1.0 and do not show a significant systematic offset.
The most likely explanation for the offset in the \HII template is \ion{He}{i} contamination, as these are strongest in the \HII spectrum, and nearly all \HI lines used in this analysis overlap with one or more \ion{He}{i} lines according to our Cloudy-based line list.
As discussed in Sect.~\ref{sec:intensity} and Fig.~\ref{fig:overlap}, we were able to quantify this contamination for \HI 5-6, where it is about 10\%.
For the other lines, the changes in the line profile are too subtle to separate the contamination.
Since the 5-6 intensity was corrected, but the others were not, a similar contamination level of 10\% would explain the offset by a factor of 1.1 we are seeing in Fig.~\ref{fig:hi}.

There are no deviations that resemble the typical effects of dust attenuation, where a stronger attenuation is expected at shorter wavelength, and at 10~\mum by the characteristic feature of silicate dust.
Previous work has shown that the dust in the Orion Bar has a patchy attenuation profile, strongly varying in magnitude between different areas \citep{2000A&A...364..301W, Peeters:nirspec}.
But the dust column affecting the ionized layer appears to have too little extinction in the MIR to be constrained by this set of lines.

Finally, we also compared the observations to a model developed for the planning and interpretation of the PDRs4All observations \citep[see][]{pdrs4all-pasp}, that is based on Cloudy,
a code that models the gas as a function of the depth into the medium and computes the total emission \citep{2017RMxAA..53..385F}, instead of just providing emissivity coefficients.
This is the same model from which we derived part of the theoretical line list, that we used in Sect.~\ref{sec:identification} to identify the lines.
The model parameters were taken from \citet{2009ApJ...693..285P} and \citet{2009ApJ...701..677S}, and can be briefly summarized as follows.
The main assumptions for the radiation field are an illuminating star with a Kurucz type model with $T_\text{eff} = \SI{39600}{\kelvin}$ for the illuminating star, for which the emission rate of ionizing photons is scaled to $Q_\text{LyC} = \SI{9.8e48}{\per\s}$.
The gas density assumes a constant pressure model, with an initial electronic density of $n_{0, e} = \SI{3160}{\per\cubic\cm}$.

The line intensity ratios predicted by the Cloudy model were also added to Fig.~\ref{fig:hi}, and we find that this model overestimates the line ratios at the shortest wavelengths.
Since the Cloudy model includes dust internal to the \ion{H}{ii} and the ionization front, but not the foreground dust, we test if a dust extinction correction would bring the model results more in line with the observations.
The PDRs4All NIRSpec overview by \citet{Peeters:nirspec} shows that magnitude of the foreground extinction varies between an $A(V)$ of 0.9 and 1.9, when assuming the $R(V)$-parameterized average Milky Way curve by \citet{gordon2023relation}, evaluated at $R(V) = 5.5$.
We use the same curve, and the extinction-corrected comparison is also shown in Fig.~\ref{fig:hi}, for the maximal value of $A(V) = 1.9$.
Even with this $A(V)$, the deviation from the observations remains.
The Cloudy model assumes a spherical shell geometry, and the way the observed emission is recorded in the model might not match the way we observe the ionized layer of the Orion Bar.
The dust distribution and its presence in the HII region are also assumptions that could cause deviations.
The Case B theory makes no geometrical or dust assumptions at all, and only provides emissivity coefficients which we use directly.
A good match between Case B theory and our observations, could mean that the main contribution to the \HI emission comes from an ionized gas layer subject to very little  dust or geometry effects.

\subsection{Ionic line ratios}
\label{sec:ionratios}

\begin{figure}[tb]
    \centering
    \includegraphics[]{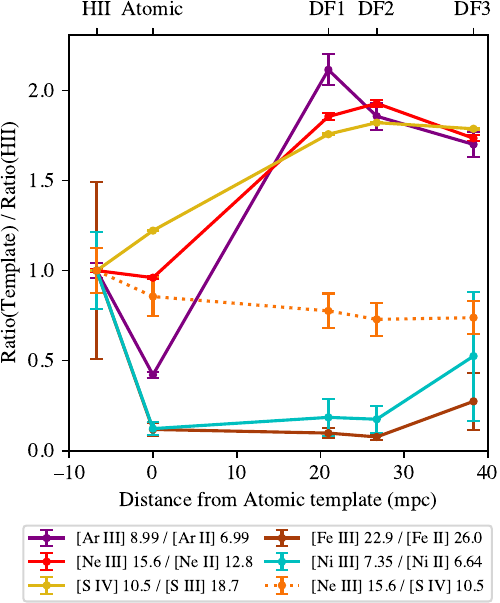}
    \caption{Line intensity ratios for several ion pairs as function of template position.
    The wavelengths are given in the legend in \mum.
    Each curve is normalized to its value at the \HII position.
    These ratios can be interpreted as indicators for radiation field hardness, sensitive to different photon energies (thresholds in Table~\ref{tab:potential}), or different layers of the gas (Fe and Ni).}
    \label{fig:ionratios}
\end{figure}

\begin{table}[b]
\centering
\setlength{\tabcolsep}{.4cm}
   \caption{Overview of ionization potentials (eV), with observed species indicated.}
   \label{tab:potential}
   \begin{tabular}{lllll}
   \hline\hline
      & \textsc{i} & \textsc{ii} & \textsc{iii} & \textsc{iv} \\
      \hline
      H & 13.6\tablefootmark{a} &&&\\
      He & 24.6\tablefootmark{a} & 54.4 &&\\
      Ne & 21.6 & 41.0\tablefootmark{a} & 63.4\tablefootmark{a} & 97.2\\
      P & 10.5 & 19.8\tablefootmark{b} & 30.2\tablefootmark{a} & 51.4\\
      S & 10.4\tablefootmark{a} & 23.3\tablefootmark{b} & 34.9\tablefootmark{a} & 47.2\tablefootmark{a}\\
      Cl & 13.0 & 23.8\tablefootmark{a} & 39.8 & 53.2\\
      Ar & 15.8 & 27.6\tablefootmark{a} & 40.7\tablefootmark{a} & 59.6\\
      Fe & 7.9 & 16.2\tablefootmark{a} & 30.7\tablefootmark{a} & 54.9\tablefootmark{b}\\
      Ni & 7.6 & 18.2\tablefootmark{a} & 35.2\tablefootmark{a} & 54.9\\
      \hline
   \end{tabular}
   \tablefoot{
    \tablefoottext{a}{Observed with MIRI (this work).}
    \tablefoottext{b}{Observed with NIRSpec \citep{Peeters:nirspec}.}
   }
\end{table}

The emission line ratios of two ions of the same element can be used to probe the hardness of the radiation field.
The spectra contain pairs of ionization stages for several elements.
Using the brightest lines for each element, we show intensity ratios in Fig.~\ref{fig:ionratios} for the ions of Ar (\ionwav{Ar}{iii}{8.99} / \ionwav{Ar}{ii}{6.99}), Ne (\ionwav{Ne}{iii}{15.6} / \ionwav{Ne}{ii}{12.8}), S (\ionwav{S}{iv}{10.5} / \ionwav{S}{iii}{18.7}), Fe (\ionwav{Fe}{iii}{22.9} / \ionwav{Fe}{ii}{26}) and Ni (\ionwav{Ni}{iii}{7.35} / \ionwav{Ni}{ii}{6.64}).
To support the discussion below, we provide an overview of the observed species and their ionization potential (from the ASD) in Table~\ref{tab:potential}.
The table shows the photon energy threshold to ionize each species to the next stage by photoionization.
For example, the energy required to produce \ion{S}{iv} is the value listed for \ion{S}{iii}.

As shown in Fig.~\ref{fig:ionratios}, there is a large difference of the line ratio profiles comparing Ar, Ne, and S versus Fe and Ni, as the ions originate from different layers along the line of sight.
The Ar, Ne and S ions reside mainly in the ionized layer, since the photon energies required to produce \ion{Ar}{ii} (15.8 eV), \ion{Ne}{ii} (21.6 eV), and \ion{S}{iii} (23.3 eV) are greater than the \ion{H}{ii} threshold of 13.6 eV.
The ratio of the $\ion{Ar}{iii}/\ion{Ar}{ii}$ pair is significantly lower in the Atomic template, and this behavior is consistent for both the available \ion{Ar}{iii} lines.
This is because the intensity of the \ionwav{Ar}{ii}{6.99} line in the Atomic template is actually stronger than that in the \HII template, as listed in Table~\ref{tab:lines}.
Because of the ionization thresholds to produce both \ion{Ar}{ii} (15.8 eV) and \ion{Ar}{iii} (27.6 eV) are significantly lower than those of the Ne and S ion pairs, the Ar ions remain present in regions with softer radiation fields, closer to the atomic layer.
The Ar ratios hence probe a wider range of physical conditions and are more complex to interpret.

The Fe and Ni ions have much lower ionization potentials, and the line ratios therefore have a very different behavior.
Since the photon energies to produce \ion{Fe}{ii} (7.9 eV) and \ion{Ni}{ii} (7.6 eV) are below 13.6 eV, the denominator in the line ratios of Fig.~\ref{fig:ionratios} will have strong contributions by the atomic H layer, as these species remain ionized past the H ionization front.
The measured intensities for \ion{Fe}{iii} and \ion{Ni}{iii} are at their highest in the \HII template, while the intensities of \ion{Fe}{ii} and \ion{Ni}{ii} reach their highest value in the Atomic template.
The energies to produce \ion{Fe}{iii} (16.3 eV) and \ion{Ni}{iii} (18.2 eV) are similar, and therefore the line ratios are expected to have a nearly identical behavior.
This is demonstrated in Fig.~\ref{fig:ionratios}, which further confirms that \ion{Ni}{iii} was identified correctly in Sect.~\ref{sec:metalionlines}.

Finally, we can use the ratios in Fig.~\ref{fig:ionratios} and the ionization potentials from Table~\ref{tab:potential} to discuss the hardness of the radiation field.
All fine-structure lines (except for \ion{S}{i}) have a maximal intensity in either the \HII or Atomic template, and then get systematically weaker for DF1, DF2, and DF3 in that order, indicating that the column densities of all ions are lower for regions further away from the star.
However, the relative abundances of the upper ionization stages (\ion{Ar}{iii}, \ion{Ne}{iii}, and \ion{S}{iv}) compared to the stages below, are observed to be larger for the DF templates.
Both the Atomic and DF template regions contain a part of the ionized layer bordering the PDR, but it appears that the radiation field affecting this ionized gas is harder at the locations of the DF templates.
The ratios for Ne and S exhibit trends that are similar to each other, as expected from their similar ionization potentials in Table~\ref{tab:potential} and existing observations of \HII regions \citep[e.g.,][]{2002ApJ...566..880G, 2002A&A...389..286M}.
We also added a line ratio probing $\ion{Ne}{iii} / \ion{S}{iv}$ to Fig.~\ref{fig:ionratios}, as these species are sensitive to radiation of around 41.0 eV and 34.9 eV respectively, and hence this ratio probes a narrower energy range.
This ratio has a slight downward trend, indicating that the radiation field hardness in the 35-41 eV range decreases mildly, while the overall hardness probed by the other line ratios over a wider energy range of roughly 20-40 eV increases strongly.

\subsection{Molecular hydrogen}

Molecular hydrogen plays an important role in the evolution of interstellar matter and star formation. Present in a variety 
of interstellar conditions (gas density and UV flux), \hmol is the most abundant molecule that efficiently cools the gas and 
may also heat it. Due to its high UV opacity, \hmol also shields itself and other molecules from photodissociation, and collisions with excited \hmol trigger endothermic chemistry \citep[see e.g.,][]{2006ApJS..164..506L}.

\subsubsection{Excitation mechanisms and heating and cooling processes}

Molecular hydrogen contributes to the gas thermal budget in the course of its excitation-deexcitation cycle that we summarize here. 
The \hmol excitation occurs mostly through radiation and collisions, with a minor contribution by the \hmol formation process with a rate 10 times lower than for UV radiation, so we will therefore not consider the latter process here. 
Radiative excitation or pumping of \hmol takes place through strong absorption bands in the FUV ($\lambda\leq 100$ nm). 
In most cases (about 9 of 10), following this UV excitation, the molecule deexcites to the ground electronic state where it cascades down through the rovibrational states. This radiative pumping thus provides both vibrational and rotational excitation of \hmol \citep{1976ApJ...203..132B, 1987ApJ...322..412B}.

On the other hand, rotational excitation of \hmol in its vibrational ground state ($v = 0$) occurs through inelastic collisions with gas species 
(mostly H, He and electrons). 
The molecule may then radiate away the energy taken up in the collision through the MIR lines, which contributes to gas cooling (see section 4.3.3). 
Collisional excitation of vibrational levels ($v \geq 1$)
is rare because it requires a high gas temperature (the typical energy between vibrational levels is 0.5 eV or $\sim$6000 K). 
At high densities ($\gtrsim 10^4$ cm$^{-3}$), collisional deexcitation of \hmol from radiatively pumped levels will lead to gas heating. 

\subsubsection{Excitation diagrams in the Orion Bar}
\label{sec:diagrams}

\begin{figure}[tb]
    \centering\includegraphics{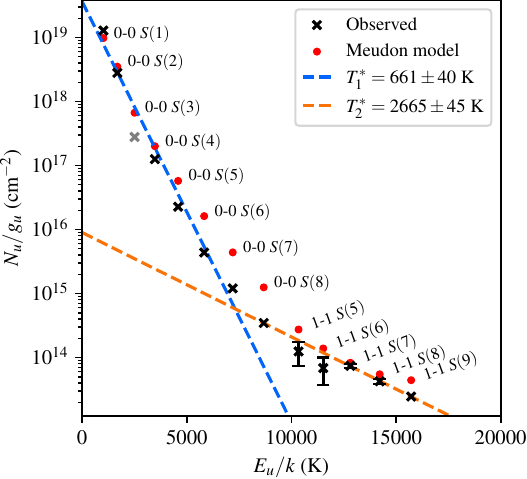}
    \caption{H$_2$ excitation diagram for DF3, with linear fits and comparison to Meudon model.
    Black crosses: column density of the upper level populations derived from the data, versus the energy of the upper level with respect to the ground state.
    Error bars are shown where they exceed 5\% (includes 3\% systematic uncertainty).
    Red circles: column densities derived from our Meudon PDR model.
    Blue line: fit including lines up to 0-0 $S(7)$, excluding the 0-0 $S(3)$ line (gray) as it is subject to extinction by the 10~\mum silicate feature.
    Orange line: fit including 0-0 $S(8)$ and 1-1 $S(5)$ to 1-1 $S(9)$. 
    Temperatures derived from the slope of each line are shown in the legend.}
    \label{fig:excitH2_DF3}
\end{figure}

\begin{table*}[bt]
\caption{H$_2$ excitation temperatures and column densities derived from the two linear components of the excitation diagram fits.} 
\label{tab:H2_excitation}
\centering
\begin{tabular}{lrrrr|rrrr}
\hline\hline
\small
 & $T_1^{*}$ & $T_2^{*}$ & $\log N_1(\text{H}_2)$ & $\log N_2(\text{H}_2)$ & $T_1^{*}$ & $T_2^{*}$ & $\log N_1(\text{H}_2)$ & $\log N_2(\text{H}_2)$\\
& (\si{\kelvin}) & (\SI{e3}{\kelvin}) & (cm$^{-2}$) & (cm$^{-2}$)& (\si{\kelvin}) & (\SI{e3}{\kelvin}) & (cm$^{-2}$) & (cm$^{-2}$)\\           
\hline
& \multicolumn{4}{c|}{No attenuation correction} & \multicolumn{4}{c}{\change{With example attenuation correction\tablefootmark{a}}} \\
HII & $659 \pm 54$ & $2.45 \pm 0.13$ & $20.10 \pm 0.20$ & $17.42 \pm 0.11$& $661 \pm 57$ & $2.44 \pm 0.16$ & $20.45 \pm 0.20$ & $17.85 \pm 0.12$ \\
Atomic & $722 \pm 50$ & $2.70 \pm 0.05$ & $19.96 \pm 0.18$ & $17.31 \pm 0.04$& $728 \pm 54$ & $2.71 \pm 0.05$ & $20.24 \pm 0.19$ & $17.65 \pm 0.04$ \\
DF1 & $706 \pm 35$ & $2.68 \pm 0.04$ & $20.44 \pm 0.15$ & $17.62 \pm 0.04$ & $713 \pm 41$ & $2.70 \pm 0.07$ & $20.95 \pm 0.16$ & $18.22 \pm 0.05$\\
DF2 & $666 \pm 36$ & $2.90 \pm 0.09$ & $20.80 \pm 0.17$ & $17.75 \pm 0.07$ & $671 \pm 40$ & $2.87 \pm 0.04$ & $21.27 \pm 0.18$ & $18.31 \pm 0.04$ \\
DF3 & $661 \pm 40$ & $2.66 \pm 0.05$ & $20.77 \pm 0.18$ & $17.82 \pm 0.05$ & $664 \pm 43$ & $2.67 \pm 0.06$ & $21.06 \pm 0.19$ & $18.16 \pm 0.06$ \\
\hline
\end{tabular}

\tablefoot{%
\tablefoottext{a}{\change{Example correction assuming a screen geometry, with an attenuation curve and 10 \mum silicate feature equal to the Milky Way extinction average of \citet{gordon2023relation} with $R(V) = 5.5$.
This results in large $A(V)$ values, e.g. $A(V) \sim 20$ for DF3.}}
}
\end{table*}

The Orion Bar is known to have a large amount of warm \hmol gas, which can be traced by pure rotational \hmol lines \citep{1991ApJ...372L..25P, 2004Habart, 2005ApJ...630..368A, 2021ApJ...919...27K}.
As a first result for these detailed \hmol data in the MIR, we present estimations of the \hmol excitation temperature based on excitation diagrams of the pure-rotational lines measured in the five template spectra.
The excitation diagram for the DF3 template is presented in Fig.~\ref{fig:excitH2_DF3}, and we analysed analogous excitation diagrams for the other four spectra.
In this type of diagram, the column density of each upper level $u$ is derived from the observed line intensities and known atomic parameters for the transitions.
The quantity on the vertical axis is $N_u / g_u = 4 \pi I_{ul} / (g_u A_{ul} [E_{u} - E_{l}])$, and plotted as a function of $E_u/k$.
In these equations, $I_{ul}$ is the measured integrated intensity of the line (\lineunit), $g_u$ is the statistical weight of the upper level, $E_u$ and $E_l$ are the energies of the upper and lower levels, $A_{ul}$ is the Einstein coefficient of the transition, and $k$ is the Boltzmann constant to express $E_u$ as a temperature.
Here we assumed the emission lines to be optically thin, given their low Einstein coefficients (Table~\ref{tab:lines}), so that an optical depth term is not used in the calculation of $N_u / g_u$ \citep[e.g.,][]{1999ApJ...517..209G}.

To describe the \hmol excitation in Fig.~\ref{fig:excitH2_DF3}, we fit two linear components to the data for which we assumed a 3\% fractional uncertainty to account for the calibration error explained in Sect.~\ref{sec:intensity}. 
The resulting parameters are presented in Table~\ref{tab:H2_excitation}, as derived from the fitted ordinate intercept and slope, based on the following linear prescription: $\ln(N_{u} / g_u) = b + a E_{u} = \ln(N^0) - E_{u} / T^*$.
With $i$ the index of the component (1 or 2), the fitted representative excitation temperature is $T^*_i = -a_i^{-1}$, and the parameter $N^0_i = e^{b_i}$ is a factor that scales with the column density. 
The total column density for each component is derived from the fit result as $N_i = \sum_u N_{u,i} = e^{b_i} \sum_u g_u \exp(E_u a_i)$, where the summation is $Z(T = -a_i^{-1})$, the rotational partition function of H$_2$ evaluated at the fitted temperature.
The equation used for Table~\ref{tab:H2_excitation} is then $N_i = e^{b_i} Z(-a_i^{-1})$, where $Z$ can be approximated as $Z(T) = 0.02477 T [1-\exp(1-6000/T)]^{-1}$ \citep{1996AJ....111.2403H}. 
While the results presented in Table~\ref{tab:H2_excitation} cover all five template spectra, the \HII and Atomic regions have one or two missing lines, as can be seen from the masked out values in Table~\ref{tab:lines}.
This results in larger uncertainties on the reported parameters for these two regions.

A more thorough \hmol analysis by Sidhu et al.\ (in preparation) will investigate the spatial variations in detail, by performing similar two-component fits on a spaxel-per-spaxel basis.
This analysis will also include the \hmol lines observed with the NIRSpec data, to provide an in-depth characterization of the rovibrational excitation.
For all five spectra, we assumed the same two-component shape for the excitation diagram, with an ortho-to-para ratio (OPR) of 3.
The OPR measured by \citet{2021ApJ...919...27K} is 2.99 based on a large number of 1-0 rovibrational lines from 1.45 to 2.45~\mum, for a similar warm molecular region of the Orion Bar PDR, somewhat further to the South-West with respect to our DF regions.
The combined NIRSpec and MIRI spectroscopy allow for the mapping of local changes in the OPR, and preliminary results from the spatially-dependent analysis by Sidhu et al.\ (in preparation) show that the OPR is very close to 3 in the areas of interest, and mostly constant over the field of view.

Although the diagram in Fig.~\ref{fig:excitH2_DF3} is curved, and suggests a continuous distribution of excitation temperature, we characterize the H$_2$ excitation with only two components. 
The $T_1^*$ component is fitted to the data of low excitation, which is rotational lines within $v=0$ from $S(1)$ to $S(7)$,
while the $T_2^*$ component is similarly fitted to rotational lines within the excited state $v=1$, from $S(5)$ to $S(9)$.
We also include the $v = 0$ $S(8)$ line in the $T_2^*$ fit, as it appears to follow the linear trend, and therefore belongs to the high excitation component, indicating that radiative pumping is also important in the $v=0$ level.
In the following, we assume that $T_1^*$ is a good proxy to the gas temperature.
While some curvature affects all transitions, causing mismatches to the linear model, the latter can still be used to quantify the average slope over the levels up to $S(7)$.
To model curved excitation diagrams in external galaxies, prescriptions based on an underlying power-law distribution for the temperature have been used \citep[e.g.,][]{2016ApJ...830...18T}, which is appropriate when a collection of PDRs and shocks is considered.
In our case, the temperature distribution is determined by the temperature gradient across the \hmol emission zone, and there is no direct argument that a power law distribution can approximate the resulting mix of temperatures.
It is clear that these data probe shape of the excitation diagram very well, and could therefore be sensitive to the continuous range of gas temperatures.
Developing a method to infer a suitable temperature distribution and reproduce the curvature of the diagram, would allow for more accurate gas mass estimations, but is outside the scope of this work.
In the next section, we further our analysis by comparing the H$_2$ line intensity data to a PDR model.

The derived $T_1^*$ or gas temperature is in the range from 660 to 720~K, \change{for all five regions.
The gas temperature of the ionized and atomic layers is expected to be much higher, while the H$_2$ rotational temperature in the \HII and Atomic apertures is similar to that of the DF apertures. 
Therefore, the H$_2$ emission may originate from a different part of the PDR, behind the \HII and Atomic regions.}
Since the lines used for this component are located in different MIRI bands, we do not need to account for correlations induced by the systematic calibration uncertainty.
The derived \hmol column density is between the orders \num{e20} and \SI{e21}{\per\square\cm}, with a significant increase for DF2 and DF3.
Comparable gas temperatures are found for the three dissociation fronts, which are at different distances from the ionization front (0.02 pc for DF1, 0.03 pc for DF2, and 0.04 pc for DF3).
This is compatible with our knowledge about the geometry of this region, where DF1, DF2, and DF3 are the edges of a terrace-like structure \citep{Habart:im, Peeters:nirspec}.
The observed temperature is somewhat higher compared to the estimations of 430 to 630~K given by \citet[Table~4]{2005ApJ...630..368A}, which used a different technique based on the $S(1) / S(2)$ and $S(2) / S(4)$ line ratios.
We note that restricting our fit to the 0-0 $S(1)$, $S(2)$ and $S(4)$ lines gives gas temperatures around 500~K, in better agreement with the \citet{2005ApJ...630..368A} results, in particular for their apertures C, D and F which are associated with dissociation fronts at a similar distance from the PDR edge, as are DF2 and DF3. 

As the energy of the upper level increases, we expect a transition from collisional to radiatively pumped \hmol excitation.
The $T_2^*$ component of our fits models the $v = 1$ lines from $S(5)$ to $S(9)$, which have much higher upper energy levels, making it unlikely that the excitation is dominated by thermal collisions for the physical conditions of the Orion Bar.
The derived excitation temperature is around 2700~K, with a column density that is significantly lower, of the order \num{e17} to \SI{e18}{\per\square\cm}.
Again, DF2 and DF3 exhibit the highest column densities of excited H$_2$.
The high rotational temperatures are compared to a model in the next section, where this behavior is discussed further.
For this fit, there are some positive correlations between the line intensities due to the calibration uncertainty, since the 1-1 $S(6)$ and $S(7)$ lines are both measured in Ch~1~MEDIUM, while 0-0 $S(8)$, 1-1 $S(8)$, and 1-1 $S(9)$ are in Ch~1~SHORT.
From the covariance matrix suggested in Sect.~\ref{sec:intensity}, only the correlations between the three lines in Ch~1~SHORT are significant, with a correlation coefficient of around 0.6 at most.
We therefore expect the correlations to have a minor effect on the fit results, given that the uncertainties for the affected lines are small, while their correlation coefficients are not very close to 1.

\change{We note that for the results discussed above, we did not apply a dust attenuation correction to the line intensities from which the excitation diagrams were derived.
Nevertheless, it can be seen in Fig.~\ref{fig:excitH2_DF3} that the $S(3)$ is affected by the silicate dust absorption feature at 10 \mum.
As will be discussed in more detail in Sect.~\ref{sec:s3attenuation}, we applied a simple attenuation correction to estimate the effect on the results.
Briefly summarized, an attenuation correction results in equivalent rotational temperatures, while for the derived column densities there could be a significant increase of around a factor $\sim$3 (right panel of Table~\ref{tab:H2_excitation}).
However, the latter factor depends heavily on the relative strength of the silicate feature in the assumed attenuation law.}

\subsubsection{Comparison to PDR model}
\label{sec:meudon}

\begin{figure*}
    \centering
    \includegraphics{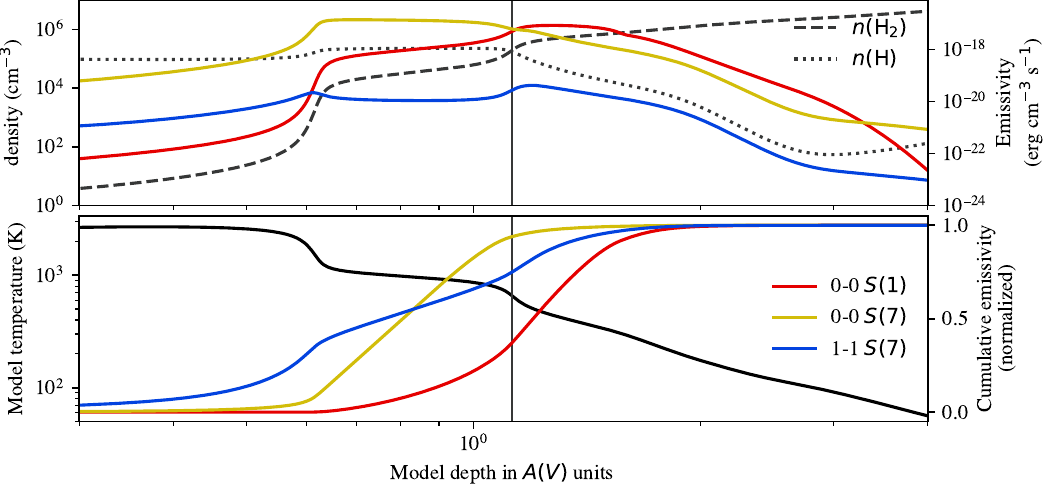}
    \caption{Top panel: Structure of the \HI/\hmol transition predicted by Meudon model as function of depth. Densities of H and \hmol, with vertical line that indicates the transition defined as $n(\text{H}) = n(\text{H}_2)$.
    The local emissivity of each gas layer is shown for four representative \hmol lines. 
    Bottom panel: Temperature and cumulative emissivity versus model depth.
    Black line and and left y-axis: gas temperature for each layer in the model.
    Colorful lines and right y-axis: cumulative of the emissivity contributions by each layer (weighted by length scale).
    By inspecting the $A(V)$ ranges over which the cumulative emissivity curves rise the most, we can see the typical range of gas temperatures from which the lines originate.
    For the 1-1 $S(7)$ line for example, the observed excitation temperature is $\sim$2700~K, while around 75\% of the emission originates from gas cooler than 1000~K, due to UV pumping.}
    \label{fig:meudon}
\end{figure*}

To interpret the observed \hmol emission, we compare our line flux data to the results of an existing Orion Bar model, computed with version 1.5.4 of the Meudon PDR Code \citep{2006ApJS..164..506L}, using parameters from the literature \citep{2018Joblin}.
This model is isobaric, with a thermal pressure $P_\text{th} = \SI{2.8e8}{\kelvin\per\cubic\cm}$, an ISRF with scaling factor $G_0 = \num{3.1e4}$, and a geometrical factor of $\sim$2 that corresponds to an inclination angle of 60°.
To model processes driven by dust such as \hmol formation and photoelectric heating, a dust size distribution of the MRN type \citep{1977ApJ...217..425M} is used with a power law exponent of -3.5 and a grain size range of 3 to 300~nm, and optical properties corresponding to a mixture of graphite (30\%) and silicate (70\%) grains.
For the attenuation of the radiation field, we used an extinction curve according to the analytical model by \citet{1990ApJS...72..163F}, assuming $R(V)=5.62$ and $N(\mathrm{H}) / E(B-V)$ = \SI{1.05e-22}{\per\square\cm}.
While a newer version of the Meudon code was used compared to \citet{2018Joblin}, most of the above parameters were adopted from the literature without further adjustments, to compare the existing knowledge to the new data.
Updating the parameters by means of fits and model grids will result in a better match to the observations, but this task is outside the scope of this work, and is part of the work by Meshaka et al.\ (in preparation).

Using the depth-dependent output of the Meudon model, we examine the gas temperature and the emissivity of several lines in each gas layer of the model, as shown in Fig.~\ref{fig:meudon}.
For each \hmol line, the most significant emissivity contributions originate from a continuous range of depths and temperatures, as illustrated by the cumulative emissivity curves.
The highly excited \hmol lines of $v = 0$ ($S(5)$ and above) peak in the first hot zones of the H/\hmol
transition, and in the model, the 0-0 $S(7)$ emission originates from the gas layer where the temperature curve has a shallow gradient as a function of depth, decreasing from around 1100 to 800~K.
A similar plateau in the emissivity curve is observed for $S(5)$, and $S(6)$ (not shown).
While the temperature decreases, the 0-0 $S(7)$ emissivity stays mostly constant (yellow curve in top panel of Fig.~\ref{fig:meudon}), suggesting an important contribution of non-thermal, radiative excitation.
On the other hand, the gas temperature in the model roughly matches the slope in the excitation diagram, so a mix of thermal collisions and radiative pumping is likely for $S(5)$ to $S(7)$.
This occurs at a depth where $n(\text{H})$ is still larger than $n(\text{H}_2)$, and there is little emissivity in these lines past the \HI/\hmol transition defined as $n(\text{H}) = n(\text{H}_2)$ (vertical line in Fig.~\ref{fig:meudon}). 

To follow the behaviour of the less excited lines (0-0 $S(4)$ and below) we show the 0-0 $S(1)$ emissivity profile: we see that it peaks around the \HI/\hmol transition and corresponds to a gas temperature around 450~K.
The curve for $S(2)$ peaks at a similar location and temperature, while $S(3)$ and $S(4)$ (not shown) have a similar profile with a peak at the \HI/\hmol transition, where the gas temperature is around 600~K.
This is in good agreement with the $T_1^*$ temperatures we found in Sect.~\ref{sec:diagrams} and supports a thermal, collisional origin for these lines. 
Using the cumulative emissivity curve (red curve in bottom panel of Fig.~\ref{fig:meudon}), we determine the depths where the 10\%, 50\%, and 90\% quantiles of the emission are reached.
We find that there are significant contributions to $S(1)$ from gas ranging from around 320 to 960~K, while the 50\% quantile corresponds to around 500~K.
This explains why the curvature of the diagram seems to smoothly continue from the high-excitation lines (driven by a mix of collisional and radiative excitation) to the low-excitation lines, even when the latter are mostly thermally driven.

Finally the emissivity profile of the highly excited 1-1 $S(7)$ line has two maxima at $A(V) \simeq 0.6$ and 1.2.
The first peak at $A(V) \simeq 0.6$ is probably due to the competing effects of the rising H$_2$ abundance (Fig.~\ref{fig:meudon} upper panel) and the decreasing radiative pumping.
The second peak points at a collisional contribution at the \HI/\hmol front, which then decreases along with the temperature, in a steeper fashion than 0-0 $S(1)$ because the collisional excitation of 1-1 $S(7)$ requires higher temperatures.
Analogous to the paragraph above, the 10\% and 90\% emission quantiles reveal a large temperature range from which 1-1 $S(7)$ originates, with significant contributions by gas roughly between 500 and 2600~K.
The above double-peaked emissivity profile and temperature range apply to all four $v = 1$ lines studied in this work. 
While there is a minor contribution from the 2600~K gas, which is close to the measured rotational temperature of around 2700~K, the bulk of the emission ($\sim$70\%) originates from gas cooler than 1000~K.
Therefore, collisional excitation can not be the main population process for the upper levels in the $v=1$ vibrational state, and pumping by UV photons is expected to be the dominant excitation mechanism.

With the temperature structure of the model described above, we now compare the observed (DF3) and theoretical excitation diagrams (black crosses and red circles in Fig.~\ref{fig:excitH2_DF3}); we note that the model intensities were not rescaled to the observations.
For the cold component of the gas, the $S(2)$ and $S(4)$ lines match well, while there is a minor mismatch with $S(1)$ that indicates a slightly different curvature.
There is a large but expected mismatch for $S(3)$, due to the 10~\mum silicate extinction (see Sect.~\ref{sec:s3attenuation}).
The $S(5)$ to $S(8)$ lines on the other hand are strongly overestimated: the level populations based on the line emission from the model are up to a factor of 3 higher than the values derived from the observations.
This deviation, in the transition region between the two linear components of the excitation diagram, is observed for all five template spectra.
Continuing to the $v = 1$ pure-rotational transitions, the model does match the data.
This means that the PDR model correctly matches the most strongly UV-excited layer, but needs parameter adjustments to model the transition into the layers where thermal \hmol excitation becomes dominant.
Despite the fact that we use literature values for the parameters instead of values optimized through fitting, this fixed model still qualitatively reproduces the transition in the slope near 0-0 $S(8)$. 

Future adjustments to the model parameters will need to slightly increase the amount of cool gas at $\sim$500~K, and more strongly reduce the amount of hot gas at $\sim$900~K, steepening the $S(1)$ to $S(8)$ part of the curve while keeping the behavior of the $v = 1$ levels unchanged. 
Changes to the model, for the Orion Bar specifically, are under development by Meshaka et al.\ (in preparation).
\change{They replace} the isotropic ISRF with scale factor $G_0$ of the model presented here, by a synthetic stellar spectrum representative of Theta 1 Orionis C.
\change{This paper will explore alternate recipes for specific processes in the PDR model, such as the excitation of \hmol that is newly formed on dust grains.}
\change{In combination with these changes, they will optimize the thermal pressure and distance with respect to the star for DF1, DF2, and DF3 individually.} 
\change{Through these efforts}, our data have the potential to reveal how well the state-of-the-art PDR codes model the excitation, heating and cooling mechanisms related to \hmol.
Alongside measurements of the AIB emission in the same spectra, these data could have important implications for understanding the thermal balance in PDRs, because the small dust grains that carry the AIB also contribute significantly to the UV extinction and gas heating via the photoelectric effect \citep{1994ApJ...427..822B}.

\subsubsection{Dust attenuation of $S(3)$}
\label{sec:s3attenuation}

We applied no dust attenuation correction for the above analysis, as we expected the extinction in the mid-IR to be small and similar for the different lines, thereby having only a minor effect on the derived results for \hmol.
\change{Despite this expectation, the 0-0 $S(3)$ line had to be excluded from the above analysis, as the attenuation due to the 10~\mum silicate dust feature appears to be rather strong}.
We measured the decrement of the $S(3)$ data with respect to an interpolated value using a local linear fit, and find values that correspond to \change{0.8 mag for HII, 1.1 mag for Atomic, 1.2 mag for DF1, 0.7 mag for DF2, and 0.9 mag for DF3 (or factors of 0.49, 0.35, 0.35, 0.51, and 0.44 respectively).}
\change{We denote these decrements as a difference between extinction values: $A(S(3)) - A(\text{other})$.}

\change{As an estimate for the typical strength of the 10 \mum silicate feature, we assume that the attenuation curve is given by} the $R(V)$-dependent average Milky Way extinction curve of \citet{gordon2023relation} evaluated for $R(V) = 5.5$, \change{in a screen geometry}.
We note that we deliberately distinguish between the words ``attenuation'' and ``extinction'' in this discussion.
As the scattering albedo at the wavelength of the $S(3)$ line is small, the difference between extinction and attenuation will \change{depend on the geometry of the dust relative to the emitting gas}.
\change{For the assumed curve,} $A(\lambda) / A(V)$ is 0.095 at the wavelength of the $S(3)$ line, while this ratio for the other MIR \hmol lines is about 0.045.
The observed \change{$S(3)$} decrement can hence be written as $A(S(3)) - A(\text{other}) \approx 0.095 A(V) - 0.045 A(V) = 0.05 A(V)$.
Using this to extrapolate the observed decrement of of $S(3)$ to the V-band extinction, \change{yields $A(V)$ values of roughly 15 (HII), 23 (Atomic), 23 (DF1), 14 (DF2), and 18 (DF3),} for the extinction that affects the emission originating from the \hmol.

\change{A recent version of the ``H$_2$ tool'' available in the PDR toolbox \citep{2023AJ....165...25P}, supports simultaneous fitting of $A(V)$ and an H$_2$ excitation diagram.
Under the same assumptions for the attenuation model, the simultaneous fit yields equivalent results for $A(V)$ as the more ad-hoc estimates presented above.
To clarify the effect of attenuation on the H$_2$ analysis of Sect.~\ref{sec:diagrams}, we applied the attenuation correction to the line intensities using the above assumptions, set up analogous excitation diagrams, and applied the same fitting method.
The results presented in the right panel of Table~\ref{tab:H2_excitation}, show that the effect on the derived temperature is minimal, while the derived column densities $\log N_i(\text{H}_2)$ increase by $\sim$0.3 to 0.5, or a factor of $\sim$2 to 3.}

\change{The results of this attenuation correction depend strongly on two assumptions.
Firstly, the correction for the H$_2$ column density presented in Table~\ref{tab:H2_excitation} is completely dependent on the assumed $A(S(3)) / A(\text{other})$ ratio.
For larger values of this ratio, the same $S(3)$ decrement can be reproduced with a smaller value of $A(\text{other})$, resulting in a smaller correction for the H$_2$ column.
Secondly, the extrapolated $A(V)$ values are highly dependent on both the silicate feature strength, and the typical V-band to MIR extinction ratio.
More precisely when the ratio of $A(S(3)) - A(\text{other})]/A(V)$ is higher, then the derived $A(V)$ will be smaller. 
Therefore, deviations from other published values are to be expected, and the $A(V)$ values presented in this section should only be interpreted as a scaling factor for the $S(3)$ decrement.
Therefore, corrected H$_2$ column densities presented in Table~\ref{tab:H2_excitation} only serve as an example of the typical impact of the attenuation correction.} 

\change{To set up a more realistic attenuation curve, several aspects of the observed system need to be considered.}
According to the diagram of the PDR structure by \citet[their Fig.~14]{Peeters:nirspec}, the foreground extinction by the ionized layer contributes $A(V) \sim 0.9 - 1.9$ mag and the atomic layer contributes $A(V) \sim 10 - 12$ mag, in front of DF1. 
In the analysis of the previous paragraph, we used a simple extrapolation to estimate the V-band extinction, based on the measured attenuation of $S(3)$ relative to the other \hmol lines and an extinction curve, which assumes a screen geometry for the dust. 
However, part of the dust is spatially mixed with the emitting \hmol gas, which affects estimates of the ``internal'' extinction in the PDR that corresponds to the integrated dust column of the atomic layer and the layer from which the \hmol emission originates.
The observed attenuation might thus differ significantly from the total extinction by the dust column along the line of sight, even in the absence of scattered light. 
To estimate the dust column, and connect the attenuation at different wavelengths more accurately, a suitable attenuation curve model needs to be set up, which is done by making assumptions about the spatial distribution of the emitting gas and the dust.
Two formalisms for the geometry were explored by \citet{Peeters:nirspec}, to determine the internal PDR extinction based on \hmol lines with the same upper level (1-0 $S(1)$ at 2.1 \mum and 1-0 $O(5)$ at 3.2 \mum).
In their ``foreground only'' case, all the dust is assumed to be in front of the \hmol emission (i.e.\ in the atomic region only), and the estimate for DF1 is $A(V) \sim 9$.
In their ``intermingled'' case, the dust is perfectly mixed with the \hmol emission and the estimated dust column increases to $A(V) \sim 37$. 
\change{This illustrates the large differences for the derived dust column, that result from geometrical assumptions.
Meshaka et al.\ (in preparation) will present a more thorough analysis of the dust attenuation, making use of both the near-IR and mid-IR H$_2$ lines, in particular the transitions that have the same upper level.}

\section{Conclusions}
\label{sec:conclusions}

The MIRI MRS data of PDRs4All allow for spatially resolved studies of the MIR lines and the diagnostics derived from those lines, for the conditions of the interstellar medium in the Orion Bar PDR.
With the updated JWST pipeline and reference files, the calibration was greatly improved since the completion of the PDRs4All observing program.
From the improved data, we extracted five template spectra that represent the \HII region, the atomic region near the ionization front, and the three dissociation fronts in the footprint of the IFU mosaic.
We provided an overview of the lines present in these regions, with measurements of the intensity, uncertainty, and central wavelength.
Thanks to the combination of spectral resolution and depth, we revealed a rich inventory of \HI recombination lines, \hmol pure-rotational lines, and metal fine-structure lines.
This opens the way to more in-depth work that makes full use of the spatial resolution of these IFU data, such as maps of the \hmol excitation by Sidhu et al.~(in preparation), or the sulfur abundance by Fuente et al.~(in preparation).
In addition, new constraints based on these lines will be applied to further improve theoretical PDR models, such as those based on the Meudon code (e.g.\ Meshaka et al., in preparation).

An initial analysis was presented, summarizing the behavior of the gas from which the lines originate.
The \HI recombination lines originating from the ionized layer bordering the PDR, were found to have ratios that closely match the emissivity coefficients from Case B recombination theory.
At MIR wavelengths, the expected dust extinction has minimal effects on these lines.
In the \HII region, there is a contamination of up to 10\% however due to \ion{He}{i} recombination lines.
For the strongest lines, the spectral resolution and S/N of these data are sufficient to separate the contaminating component.

Further diagnostics of the ionized layer are provided by the fine-structure lines of several
ionization stages of Ne, P, S, Cl, Ar, Fe, and Ni.
We provided an brief overview of the ionization conditions by showing how the ion ratios vary for \ion{Ar}{iii}/\ion{Ar}{ii}, \ion{Ne}{iii}/\ion{Ne}{ii}, \ion{S}{iv}/\ion{S}{iii}, \ion{Fe}{iii}/\ion{Fe}{ii}, and \ion{Ni}{iii}/\ion{Ni}{ii}.
The metals in the \HII layer appear more strongly ionized for the regions located in the foreground of the dissociation fronts, meaning that the radiation field is harder at these locations.
We note that the \ion{Fe}{iii}/\ion{Fe}{ii} and \ion{Ni}{iii}/\ion{Ni}{ii} ratios behave very similarly to each other, but very differently from the other line ratios, as the \ion{Fe}{ii} and \ion{Ni}{ii} lines contain significant contributions from the neutral PDR gas.

We used the rotational \hmol lines to set up excitation diagrams and find similar excitation conditions for all three dissociation fronts. 
The smooth curvature in the trend of the excitation diagrams shows the departure from the linear (single temperature) approximation, and probes a mix of two effects: firstly, the temperature distribution along the line of sight, and secondly, the gradual transition from thermal collisions to radiative pumping as the dominant excitation mechanism.
We compared the diagrams to a theoretical PDR model for which the parameters were set to values from the literature.
This model qualitatively reproduces the data, exhibiting a curved excitation diagram where the transition in the slope of the diagram occurs at the right excitation energy.
The main quantitative discrepancy is that the 0-0~$S(5)$ to $S(8)$ intensities are strongly overestimated, calling for parameter adjustments that reduce the amount of warm H$_2$ at $\sim$900~K.
On the other hand, the thermally driven 0-0 $S(1)$ to $S(4)$ lines of low excitation, and 1-1 $S(5)$ to $S(9)$ lines of vibrationally excited \hmol, driven by UV pumping, are well approximated by the model without parameter adjustments.

\begin{acknowledgements}
This work is based on observations made with the NASA/ESA/CSA James Webb Space Telescope. The data were obtained from the Mikulski Archive for Space Telescopes at the Space Telescope Science Institute, which is operated by the Association of Universities for Research in Astronomy, Inc., under NASA contract NAS 5-03127 for JWST. These observations are associated with program \#1288.
DVDP acknowledges support for program \#1288 provided by NASA through a grant from the Space Telescope Science Institute.
CB is grateful for an appointment at NASA Ames Research Center through the San Jos\'e State University Research Foundation (80NSSC22M0107).
\change{JRG thanks the Spanish MCINN for funding support under grant PID2019-106110GB-100.}
TO is supported by JSPS Bilateral Program, Grant Number 120219939.
MGW was supported in part by JWST Theory grant JWST-AR-01557.001-A.
AP would like to acknowledge financial support from Department of Science and Technology - SERB via Core Research Grant (DST-CRG) grant (SERB-CRG/2021/000907), Institutes of Eminence (IoE) incentive grant, BHU (incentive/2021- 22/32439), Banaras Hindu University, Varanasi and thanks the Inter-University Centre for Astronomy and Astrophysics, Pune for associateship. 
HZ acknowledges support from the Swedish Research Council (contract No 2020-03437)
JH is sponsored (in part) by the Chinese Academy of Sciences (CAS), through a grant to the CAS South America Center for Astronomy (CASSACA) in Santiago, Chile. 
MB acknowledges DST INSPIRE Faculty fellowship and Thanks The Inter-University Centre for Astronomy and Astrophysics for visiting associateship.
AF is grateful to the European Research Council (ERC) for funding under the Advanced Grant project SUL4LIFE, grant agreement No101096293 and to the Spanish MICINN for funding support from PID2019-106235GB-I00.
Work by YO and MR is carried out within the Collaborative Research Centre 956, sub-project C1, funded by the Deutsche Forschungsgemeinschaft (DFG) – project ID 184018867.
\end{acknowledgements} 

\clearpage
\bibliographystyle{aa}
\bibliography{00main}

\begin{thebibliography}{71}
\expandafter\ifx\csname natexlab\endcsname\relax\def\natexlab#1{#1}\fi

\bibitem[{{Allers} {et~al.}(2005){Allers}, {Jaffe}, {Lacy}, {Draine}, \&
  {Richter}}]{2005ApJ...630..368A}
{Allers}, K.~N., {Jaffe}, D.~T., {Lacy}, J.~H., {Draine}, B.~T., \& {Richter},
  M.~J. 2005, \apj, 630, 368

\bibitem[{{Argyriou} {et~al.}(2023){Argyriou}, {Glasse}, {Law}, {Labiano},
  {{\'A}lvarez-M{\'a}rquez}, {Patapis}, {Kavanagh}, {Gasman}, {Mueller},
  {Larson}, {Vandenbussche}, {Glauser}, {Royer}, {Dicken}, {Harkett},
  {Sargent}, {Engesser}, {Jones}, {Kendrew}, {Noriega-Crespo}, {Brandl},
  {Rieke}, {Wright}, {Lee}, \& {Wells}}]{2023A&A...675A.111A}
{Argyriou}, I., {Glasse}, A., {Law}, D.~R., {et~al.} 2023, \aap, 675, A111

\bibitem[{{Argyriou} {et~al.}(2020){Argyriou}, {Wells}, {Glasse}, {Lee},
  {Royer}, {Vandenbussche}, {Malumuth}, {Glauser}, {Kavanagh}, {Labiano},
  {Lahuis}, {Mueller}, \& {Patapis}}]{2020A&A...641A.150A}
{Argyriou}, I., {Wells}, M., {Glasse}, A., {et~al.} 2020, \aap, 641, A150

\bibitem[{{Baker} \& {Menzel}(1938)}]{1938ApJ....88...52B}
{Baker}, J.~G. \& {Menzel}, D.~H. 1938, \apj, 88, 52

\bibitem[{{Bakes} \& {Tielens}(1994)}]{1994ApJ...427..822B}
{Bakes}, E.~L.~O. \& {Tielens}, A.~G.~G.~M. 1994, \apj, 427, 822

\bibitem[{{Bernard-Salas} {et~al.}(2009){Bernard-Salas}, {Spoon},
  {Charmandaris}, {Lebouteiller}, {Farrah}, {Devost}, {Brandl}, {Wu}, {Armus},
  {Hao}, {Sloan}, {Weedman}, \& {Houck}}]{2009ApJS..184..230B}
{Bernard-Salas}, J., {Spoon}, H.~W.~W., {Charmandaris}, V., {et~al.} 2009,
  \apjs, 184, 230

\bibitem[{{Bern{\'e}} {et~al.}(2022){Bern{\'e}}, {Habart}, {Peeters},
  {Abergel}, {Bergin}, {Bernard-Salas}, {Bron}, {Cami}, {Dartois}, {Fuente},
  {Goicoechea}, {Gordon}, {Okada}, {Onaka}, {Robberto}, {R{\"o}llig},
  {Tielens}, {Vicente}, {Wolfire}, {Alarc{\'o}n}, {Boersma}, {Canin}, {Chown},
  {Dicken}, {Languignon}, {Le Gal}, {Pound}, {Trahin}, {Simmer}, {Sidhu}, {Van
  De Putte}, {Cuadrado}, {Guilloteau}, {Maragkoudakis}, {Schefter}, {Schirmer},
  {Cazaux}, {Aleman}, {Allamandola}, {Auchettl}, {Baratta}, {Bejaoui}, {Bera},
  {Bilalbegovi{\'c}}, {Black}, {Boulanger}, {Bouwman}, {Brandl}, {Brechignac},
  {Br{\"u}nken}, {Burkhardt}, {Candian}, {Cernicharo}, {Chabot}, {Chakraborty},
  {Champion}, {Colgan}, {Cooke}, {Coutens}, {Cox}, {Demyk}, {Donovan Meyer},
  {Engrand}, {Foschino}, {Garc{\'\i}a-Lario}, {Gavilan}, {Gerin}, {Godard},
  {Gottlieb}, {Guillard}, {Gusdorf}, {Hartigan}, {He}, {Herbst}, {Hornekaer},
  {J{\"a}ger}, {Janot-Pacheco}, {Joblin}, {Kaufman}, {Kemper}, {Kendrew},
  {Kirsanova}, {Klaassen}, {Knight}, {Kwok}, {Labiano}, {Lai}, {Lee},
  {Lefloch}, {Le Petit}, {Li}, {Linz}, {Mackie}, {Madden}, {Mascetti},
  {McGuire}, {Merino}, {Micelotta}, {Misselt}, {Morse}, {Mulas}, {Neelamkodan},
  {Ohsawa}, {Omont}, {Paladini}, {Palumbo}, {Pathak}, {Pendleton},
  {Petrignani}, {Pino}, {Puga}, {Rangwala}, {Rapacioli}, {Ricca},
  {Roman-Duval}, {Roser}, {Roueff}, {Rouill{\'e}}, {Salama}, {Sales},
  {Sandstrom}, {Sarre}, {Sciamma-O'Brien}, {Sellgren}, {Shannon}, {Shenoy},
  {Teyssier}, {Thomas}, {Togi}, {Verstraete}, {Witt}, {Wootten}, {Ysard},
  {Zettergren}, {Zhang}, {Zhang}, \& {Zhen}}]{pdrs4all-pasp}
{Bern{\'e}}, O., {Habart}, {\'E}., {Peeters}, E., {et~al.} 2022, \pasp, 134,
  054301

\bibitem[{{Bern{\'e}} {et~al.}(2023){Bern{\'e}}, {Martin-Drumel}, {Schroetter},
  {Goicoechea}, {Jacovella}, {Gans}, {Dartois}, {Coudert}, {Bergin}, {Alarcon},
  {Cami}, {Roueff}, {Black}, {Asvany}, {Habart}, {Peeters}, {Canin}, {Trahin},
  {Joblin}, {Schlemmer}, {Thorwirth}, {Cernicharo}, {Gerin}, {Tielens},
  {Zannese}, {Abergel}, {Bernard-Salas}, {Boersma}, {Bron}, {Chown},
  {Cuadrado}, {Dicken}, {Elyajouri}, {Fuente}, {Gordon}, {Issa}, {Kannavou},
  {Khan}, {Lacinbala}, {Languignon}, {Le Gal}, {Maragkoudakis}, {Meshaka},
  {Okada}, {Onaka}, {Pasquini}, {Pound}, {Robberto}, {R{\"o}llig}, {Schefter},
  {Schirmer}, {Sidhu}, {Tabone}, {Van De Putte}, {Vicente}, \&
  {Wolfire}}]{2023Natur.621...56B}
{Bern{\'e}}, O., {Martin-Drumel}, M.-A., {Schroetter}, I., {et~al.} 2023, \nat,
  621, 56

\bibitem[{{Bern{\'e} et al.}(2024)}]{Berne:disk2024}
{Bern{\'e} et al.} 2024, Science

\bibitem[{{Black} \& {Dalgarno}(1976)}]{1976ApJ...203..132B}
{Black}, J.~H. \& {Dalgarno}, A. 1976, \apj, 203, 132

\bibitem[{{Black} \& {van Dishoeck}(1987)}]{1987ApJ...322..412B}
{Black}, J.~H. \& {van Dishoeck}, E.~F. 1987, \apj, 322, 412

\bibitem[{{Bohlin} {et~al.}(2014){Bohlin}, {Gordon}, \&
  {Tremblay}}]{2014PASP..126..711B}
{Bohlin}, R.~C., {Gordon}, K.~D., \& {Tremblay}, P.~E. 2014, \pasp, 126, 711

\bibitem[{{Chown} {et~al.}(2023){Chown}, {Sidhu}, {Peeters}, {Tielens}, {Cami},
  {Bern{\'e}}, {Habart}, {Alarc{\'o}n}, {Canin}, {Schroetter}, {Trahin}, {Van
  De Putte}, {Abergel}, {Bergin}, {Bernard-Salas}, {Boersma}, {Bron},
  {Cuadrado}, {Dartois}, {Dicken}, {El-Yajouri}, {Fuente}, {Goicoechea},
  {Gordon}, {Issa}, {Joblin}, {Kannavou}, {Khan}, {Lacinbala}, {Languignon},
  {Le Gal}, {Maragkoudakis}, {Meshaka}, {Okada}, {Onaka}, {Pasquini}, {Pound},
  {Robberto}, {R{\"o}llig}, {Schefter}, {Schirmer}, {Vicente}, {Wolfire},
  {Zannese}, {Aleman}, {Allamandola}, {Auchettl}, {Baratta}, {Bejaoui}, {Bera},
  {Black}, {Boulanger}, {Bouwman}, {Brandl}, {Brechignac}, {Br{\"u}nken},
  {Buragohain}, {Burkhardt}, {Candian}, {Cazaux}, {Cernicharo}, {Chabot},
  {Chakraborty}, {Champion}, {Colgan}, {Cooke}, {Coutens}, {Cox}, {Demyk},
  {Donovan Meyer}, {Foschino}, {Garc{\'\i}a-Lario}, {Gavilan}, {Gerin},
  {Gottlieb}, {Guillard}, {Gusdorf}, {Hartigan}, {He}, {Herbst}, {Hornekaer},
  {J{\"a}ger}, {Janot-Pacheco}, {Kaufman}, {Kemper}, {Kendrew}, {Kirsanova},
  {Klaassen}, {Kwok}, {Labiano}, {Lai}, {Lee}, {Lefloch}, {Le Petit}, {Li},
  {Linz}, {Mackie}, {Madden}, {Mascetti}, {McGuire}, {Merino}, {Micelotta},
  {Misselt}, {Morse}, {Mulas}, {Neelamkodan}, {Ohsawa}, {Omont}, {Paladini},
  {Palumbo}, {Pathak}, {Pendleton}, {Petrignani}, {Pino}, {Puga}, {Rangwala},
  {Rapacioli}, {Ricca}, {Roman-Duval}, {Roser}, {Roueff}, {Rouill{\'e}},
  {Salama}, {Sales}, {Sandstrom}, {Sarre}, {Sciamma-O'Brien}, {Sellgren},
  {Shenoy}, {Teyssier}, {Thomas}, {Togi}, {Verstraete}, {Witt}, {Wootten},
  {Zettergren}, {Zhang}, {Zhang}, \& {Zhen}}]{Chown:AIB}
{Chown}, R., {Sidhu}, A., {Peeters}, E., {et~al.} 2023, arXiv e-prints,
  arXiv:2308.16733

\bibitem[{{Cormier} {et~al.}(2012){Cormier}, {Lebouteiller}, {Madden}, {Abel},
  {Hony}, {Galliano}, {Baes}, {Barlow}, {Cooray}, {De Looze}, {Galametz},
  {Karczewski}, {Parkin}, {R{\'e}my}, {Sauvage}, {Spinoglio}, {Wilson}, \&
  {Wu}}]{2012A&A...548A..20C}
{Cormier}, D., {Lebouteiller}, V., {Madden}, S.~C., {et~al.} 2012, \aap, 548,
  A20

\bibitem[{{Elyajouri} {et~al.}(2024){Elyajouri}, {Ysard}, {Abergel}, {Habart},
  {Verstraete}, {Jones}, {Juvela}, {Schirmer}, {Meshaka}, {Dartois},
  {Lebourlot}, {Rouille}, {Onaka}, {Peeters}, {Berne}, {Alarcon},
  {Bernard-Salas}, {Buragohain}, {Cami}, {Canin}, {Chown}, {Demyk}, {Gordon},
  {Kannavou}, {Kirsanova}, {Madden}, {Paladini}, {Pendleton}, {Salama},
  {Schroetter}, {Sidhu}, {Rollig}, {Trahin}, \& {Van De
  Putte}}]{2024arXiv240101221E}
{Elyajouri}, M., {Ysard}, N., {Abergel}, A., {et~al.} 2024, arXiv e-prints,
  arXiv:2401.01221

\bibitem[{{Farrah} {et~al.}(2007){Farrah}, {Bernard-Salas}, {Spoon}, {Soifer},
  {Armus}, {Brandl}, {Charmandaris}, {Desai}, {Higdon}, {Devost}, \&
  {Houck}}]{2007ApJ...667..149F}
{Farrah}, D., {Bernard-Salas}, J., {Spoon}, H.~W.~W., {et~al.} 2007, \apj, 667,
  149

\bibitem[{{Ferland} {et~al.}(2017){Ferland}, {Chatzikos}, {Guzm{\'a}n},
  {Lykins}, {van Hoof}, {Williams}, {Abel}, {Badnell}, {Keenan}, {Porter}, \&
  {Stancil}}]{2017RMxAA..53..385F}
{Ferland}, G.~J., {Chatzikos}, M., {Guzm{\'a}n}, F., {et~al.} 2017, \rmxaa, 53,
  385

\bibitem[{{Fitzpatrick} \& {Massa}(1990)}]{1990ApJS...72..163F}
{Fitzpatrick}, E.~L. \& {Massa}, D. 1990, \apjs, 72, 163

\bibitem[{{Gaia Collaboration} {et~al.}(2023){Gaia Collaboration}, {Vallenari},
  {Brown}, {Prusti}, {de Bruijne}, {Arenou}, {Babusiaux}, {Biermann},
  {Creevey}, {Ducourant}, {Evans}, {Eyer}, {Guerra}, {Hutton}, {Jordi},
  {Klioner}, {Lammers}, {Lindegren}, {Luri}, {Mignard}, {Panem}, {Pourbaix},
  {Randich}, {Sartoretti}, {Soubiran}, {Tanga}, {Walton}, {Bailer-Jones},
  {Bastian}, {Drimmel}, {Jansen}, {Katz}, {Lattanzi}, {van Leeuwen}, {Bakker},
  {Cacciari}, {Casta{\~n}eda}, {De Angeli}, {Fabricius}, {Fouesneau},
  {Fr{\'e}mat}, {Galluccio}, {Guerrier}, {Heiter}, {Masana}, {Messineo},
  {Mowlavi}, {Nicolas}, {Nienartowicz}, {Pailler}, {Panuzzo}, {Riclet}, {Roux},
  {Seabroke}, {Sordo}, {Th{\'e}venin}, {Gracia-Abril}, {Portell}, {Teyssier},
  {Altmann}, {Andrae}, {Audard}, {Bellas-Velidis}, {Benson}, {Berthier},
  {Blomme}, {Burgess}, {Busonero}, {Busso}, {C{\'a}novas}, {Carry}, {Cellino},
  {Cheek}, {Clementini}, {Damerdji}, {Davidson}, {de Teodoro}, {Nu{\~n}ez
  Campos}, {Delchambre}, {Dell'Oro}, {Esquej}, {Fern{\'a}ndez-Hern{\'a}ndez},
  {Fraile}, {Garabato}, {Garc{\'\i}a-Lario}, {Gosset}, {Haigron}, {Halbwachs},
  {Hambly}, {Harrison}, {Hern{\'a}ndez}, {Hestroffer}, {Hodgkin}, {Holl},
  {Jan{\ss}en}, {Jevardat de Fombelle}, {Jordan}, {Krone-Martins}, {Lanzafame},
  {L{\"o}ffler}, {Marchal}, {Marrese}, {Moitinho}, {Muinonen}, {Osborne},
  {Pancino}, {Pauwels}, {Recio-Blanco}, {Reyl{\'e}}, {Riello}, {Rimoldini},
  {Roegiers}, {Rybizki}, {Sarro}, {Siopis}, {Smith}, {Sozzetti}, {Utrilla},
  {van Leeuwen}, {Abbas}, {{\'A}brah{\'a}m}, {Abreu Aramburu}, {Aerts},
  {Aguado}, {Ajaj}, {Aldea-Montero}, {Altavilla}, {{\'A}lvarez}, {Alves},
  {Anders}, {Anderson}, {Anglada Varela}, {Antoja}, {Baines}, {Baker},
  {Balaguer-N{\'u}{\~n}ez}, {Balbinot}, {Balog}, {Barache}, {Barbato},
  {Barros}, {Barstow}, {Bartolom{\'e}}, {Bassilana}, {Bauchet}, {Becciani},
  {Bellazzini}, {Berihuete}, {Bernet}, {Bertone}, {Bianchi}, {Binnenfeld},
  {Blanco-Cuaresma}, {Blazere}, {Boch}, {Bombrun}, {Bossini}, {Bouquillon},
  {Bragaglia}, {Bramante}, {Breedt}, {Bressan}, {Brouillet}, {Brugaletta},
  {Bucciarelli}, {Burlacu}, {Butkevich}, {Buzzi}, {Caffau}, {Cancelliere},
  {Cantat-Gaudin}, {Carballo}, {Carlucci}, {Carnerero}, {Carrasco},
  {Casamiquela}, {Castellani}, {Castro-Ginard}, {Chaoul}, {Charlot}, {Chemin},
  {Chiaramida}, {Chiavassa}, {Chornay}, {Comoretto}, {Contursi}, {Cooper},
  {Cornez}, {Cowell}, {Crifo}, {Cropper}, {Crosta}, {Crowley}, {Dafonte},
  {Dapergolas}, {David}, {David}, {de Laverny}, {De Luise}, {De March}, {De
  Ridder}, {de Souza}, {de Torres}, {del Peloso}, {del Pozo}, {Delbo},
  {Delgado}, {Delisle}, {Demouchy}, {Dharmawardena}, {Di Matteo}, {Diakite},
  {Diener}, {Distefano}, {Dolding}, {Edvardsson}, {Enke}, {Fabre}, {Fabrizio},
  {Faigler}, {Fedorets}, {Fernique}, {Fienga}, {Figueras}, {Fournier},
  {Fouron}, {Fragkoudi}, {Gai}, {Garcia-Gutierrez}, {Garcia-Reinaldos},
  {Garc{\'\i}a-Torres}, {Garofalo}, {Gavel}, {Gavras}, {Gerlach}, {Geyer},
  {Giacobbe}, {Gilmore}, {Girona}, {Giuffrida}, {Gomel}, {Gomez},
  {Gonz{\'a}lez-N{\'u}{\~n}ez}, {Gonz{\'a}lez-Santamar{\'\i}a},
  {Gonz{\'a}lez-Vidal}, {Granvik}, {Guillout}, {Guiraud},
  {Guti{\'e}rrez-S{\'a}nchez}, {Guy}, {Hatzidimitriou}, {Hauser}, {Haywood},
  {Helmer}, {Helmi}, {Sarmiento}, {Hidalgo}, {Hilger}, {H{\l}adczuk}, {Hobbs},
  {Holland}, {Huckle}, {Jardine}, {Jasniewicz}, {Jean-Antoine Piccolo},
  {Jim{\'e}nez-Arranz}, {Jorissen}, {Juaristi Campillo}, {Julbe}, {Karbevska},
  {Kervella}, {Khanna}, {Kontizas}, {Kordopatis}, {Korn}, {K{\'o}sp{\'a}l},
  {Kostrzewa-Rutkowska}, {Kruszy{\'n}ska}, {Kun}, {Laizeau}, {Lambert},
  {Lanza}, {Lasne}, {Le Campion}, {Lebreton}, {Lebzelter}, {Leccia}, {Leclerc},
  {Lecoeur-Taibi}, {Liao}, {Licata}, {Lindstr{\o}m}, {Lister}, {Livanou},
  {Lobel}, {Lorca}, {Loup}, {Madrero Pardo}, {Magdaleno Romeo}, {Managau},
  {Mann}, {Manteiga}, {Marchant}, {Marconi}, {Marcos}, {Marcos Santos},
  {Mar{\'\i}n Pina}, {Marinoni}, {Marocco}, {Marshall}, {Martin Polo},
  {Mart{\'\i}n-Fleitas}, {Marton}, {Mary}, {Masip}, {Massari},
  {Mastrobuono-Battisti}, {Mazeh}, {McMillan}, {Messina}, {Michalik}, {Millar},
  {Mints}, {Molina}, {Molinaro}, {Moln{\'a}r}, {Monari}, {Mongui{\'o}},
  {Montegriffo}, {Montero}, {Mor}, {Mora}, {Morbidelli}, {Morel}, {Morris},
  {Muraveva}, {Murphy}, {Musella}, {Nagy}, {Noval}, {Oca{\~n}a}, {Ogden},
  {Ordenovic}, {Osinde}, {Pagani}, {Pagano}, {Palaversa}, {Palicio},
  {Pallas-Quintela}, {Panahi}, {Payne-Wardenaar}, {Pe{\~n}alosa Esteller},
  {Penttil{\"a}}, {Pichon}, {Piersimoni}, {Pineau}, {Plachy}, {Plum}, {Poggio},
  {Pr{\v{s}}a}, {Pulone}, {Racero}, {Ragaini}, {Rainer}, {Raiteri}, {Rambaux},
  {Ramos}, {Ramos-Lerate}, {Re Fiorentin}, {Regibo}, {Richards}, {Rios Diaz},
  {Ripepi}, {Riva}, {Rix}, {Rixon}, {Robichon}, {Robin}, {Robin}, {Roelens},
  {Rogues}, {Rohrbasser}, {Romero-G{\'o}mez}, {Rowell}, {Royer}, {Ruz Mieres},
  {Rybicki}, {Sadowski}, {S{\'a}ez N{\'u}{\~n}ez}, {Sagrist{\`a} Sell{\'e}s},
  {Sahlmann}, {Salguero}, {Samaras}, {Sanchez Gimenez}, {Sanna},
  {Santove{\~n}a}, {Sarasso}, {Schultheis}, {Sciacca}, {Segol}, {Segovia},
  {S{\'e}gransan}, {Semeux}, {Shahaf}, {Siddiqui}, {Siebert}, {Siltala},
  {Silvelo}, {Slezak}, {Slezak}, {Smart}, {Snaith}, {Solano}, {Solitro},
  {Souami}, {Souchay}, {Spagna}, {Spina}, {Spoto}, {Steele},
  {Steidelm{\"u}ller}, {Stephenson}, {S{\"u}veges}, {Surdej}, {Szabados},
  {Szegedi-Elek}, {Taris}, {Taylor}, {Teixeira}, {Tolomei}, {Tonello}, {Torra},
  {Torra}, {Torralba Elipe}, {Trabucchi}, {Tsounis}, {Turon}, {Ulla}, {Unger},
  {Vaillant}, {van Dillen}, {van Reeven}, {Vanel}, {Vecchiato}, {Viala},
  {Vicente}, {Voutsinas}, {Weiler}, {Wevers}, {Wyrzykowski}, {Yoldas}, {Yvard},
  {Zhao}, {Zorec}, {Zucker}, \& {Zwitter}}]{2023A&A...674A...1G}
{Gaia Collaboration}, {Vallenari}, A., {Brown}, A.~G.~A., {et~al.} 2023, \aap,
  674, A1

\bibitem[{{Giveon} {et~al.}(2002){Giveon}, {Sternberg}, {Lutz}, {Feuchtgruber},
  \& {Pauldrach}}]{2002ApJ...566..880G}
{Giveon}, U., {Sternberg}, A., {Lutz}, D., {Feuchtgruber}, H., \& {Pauldrach},
  A.~W.~A. 2002, \apj, 566, 880

\bibitem[{{Goicoechea} \& {Cuadrado}(2021)}]{2021A&A...647L...7G}
{Goicoechea}, J.~R. \& {Cuadrado}, S. 2021, \aap, 647, L7

\bibitem[{{Goicoechea} {et~al.}(2016){Goicoechea}, {Pety}, {Cuadrado},
  {Cernicharo}, {Chapillon}, {Fuente}, {Gerin}, {Joblin}, {Marcelino}, \&
  {Pilleri}}]{2016Natur.537..207G}
{Goicoechea}, J.~R., {Pety}, J., {Cuadrado}, S., {et~al.} 2016, \nat, 537, 207

\bibitem[{{Goldsmith} \& {Langer}(1999)}]{1999ApJ...517..209G}
{Goldsmith}, P.~F. \& {Langer}, W.~D. 1999, \apj, 517, 209

\bibitem[{{Gordon} {et~al.}(2022){Gordon}, {Bohlin}, {Sloan}, {Rieke}, {Volk},
  {Boyer}, {Muzerolle}, {Schlawin}, {Deustua}, {Hines}, {Kraemer}, {Mullally},
  \& {Su}}]{gordon2022fluxcal}
{Gordon}, K.~D., {Bohlin}, R., {Sloan}, G.~C., {et~al.} 2022, \aj, 163, 267

\bibitem[{Gordon {et~al.}(2023)Gordon, Clayton, Decleir, Fitzpatrick, Massa,
  Misselt, \& Tollerud}]{gordon2023relation}
Gordon, K.~D., Clayton, G.~C., Decleir, M., {et~al.} 2023, One Relation for All
  Wavelengths: The Far-Ultraviolet to Mid-Infrared Milky Way Spectroscopic R(V)
  Dependent Dust Extinction Relationship

\bibitem[{{Habart} {et~al.}(2004){Habart}, {Boulanger}, {Verstraete},
  {Walmsley}, \& {Pineau des For{\^e}ts}}]{2004Habart}
{Habart}, E., {Boulanger}, F., {Verstraete}, L., {Walmsley}, C.~M., \& {Pineau
  des For{\^e}ts}, G. 2004, \aap, 414, 531

\bibitem[{{Habart} {et~al.}(2023{\natexlab{a}}){Habart}, {Le Gal}, {Alvarez},
  {Peeters}, {Bern{\'e}}, {Wolfire}, {Goicoechea}, {Schirmer}, {Bron}, \&
  {R{\"o}llig}}]{2023A&A...673A.149H}
{Habart}, E., {Le Gal}, R., {Alvarez}, C., {et~al.} 2023{\natexlab{a}}, \aap,
  673, A149

\bibitem[{{Habart} {et~al.}(2023{\natexlab{b}}){Habart}, {Peeters},
  {Bern{\'e}}, {Trahin}, {Canin}, {Chown}, {Sidhu}, {Van De Putte},
  {Alarc{\'o}n}, {Schroetter}, {Dartois}, {Vicente}, {Abergel}, {Bergin},
  {Bernard-Salas}, {Boersma}, {Bron}, {Cami}, {Cuadrado}, {Dicken},
  {Elyajouri}, {Fuente}, {Goicoechea}, {Gordon}, {Issa}, {Joblin}, {Kannavou},
  {Khan}, {Lacinbala}, {Languignon}, {Le Gal}, {Maragkoudakis}, {Meshaka},
  {Okada}, {Onaka}, {Pasquini}, {Pound}, {Robberto}, {R{\"o}llig}, {Schefter},
  {Schirmer}, {Tabone}, {Tielens}, {Wolfire}, {Zannese}, {Ysard},
  {Miville-Deschenes}, {Aleman}, {Allamandola}, {Auchettl}, {Baratta},
  {Bejaoui}, {Bera}, {Black}, {Boulanger}, {Bouwman}, {Brandl}, {Brechignac},
  {Br{\"u}nken}, {Buragohain}, {Burkhardt}, {Candian}, {Cazaux}, {Cernicharo},
  {Chabot}, {Chakraborty}, {Champion}, {Colgan}, {Cooke}, {Coutens}, {Cox},
  {Demyk}, {Donovan Meyer}, {Foschino}, {Garc{\'\i}a-Lario}, {Gavilan},
  {Gerin}, {Gottlieb}, {Guillard}, {Gusdorf}, {Hartigan}, {He}, {Herbst},
  {Hornekaer}, {J{\"a}ger}, {Janot-Pacheco}, {Kaufman}, {Kemper}, {Kendrew},
  {Kirsanova}, {Klaassen}, {Kwok}, {Labiano}, {Lai}, {Lee}, {Lefloch}, {Le
  Petit}, {Li}, {Linz}, {Mackie}, {Madden}, {Mascetti}, {McGuire}, {Merino},
  {Micelotta}, {Misselt}, {Morse}, {Mulas}, {Neelamkodan}, {Ohsawa}, {Omont},
  {Paladini}, {Palumbo}, {Pathak}, {Pendleton}, {Petrignani}, {Pino}, {Puga},
  {Rangwala}, {Rapacioli}, {Ricca}, {Roman-Duval}, {Roser}, {Roueff},
  {Rouill{\'e}}, {Salama}, {Sales}, {Sandstrom}, {Sarre}, {Sciamma-O'Brien},
  {Sellgren}, {Shenoy}, {Teyssier}, {Thomas}, {Togi}, {Verstraete}, {Witt},
  {Wootten}, {Zettergren}, {Zhang}, {Zhang}, \& {Zhen}}]{Habart:im}
{Habart}, E., {Peeters}, E., {Bern{\'e}}, O., {et~al.} 2023{\natexlab{b}},
  arXiv e-prints, arXiv:2308.16732

\bibitem[{{Habart} {et~al.}(2005){Habart}, {Walmsley}, {Verstraete}, {Cazaux},
  {Maiolino}, {Cox}, {Boulanger}, \& {Pineau des
  For{\^e}ts}}]{2005SSRv..119...71H}
{Habart}, E., {Walmsley}, M., {Verstraete}, L., {et~al.} 2005, \ssr, 119, 71

\bibitem[{{Habing}(1968)}]{1968BAN....19..421H}
{Habing}, H.~J. 1968, \bain, 19, 421

\bibitem[{{Herbst} {et~al.}(1996){Herbst}, {Beckwith}, {Glindemann},
  {Tacconi-Garman}, {Kroker}, \& {Krabbe}}]{1996AJ....111.2403H}
{Herbst}, T.~M., {Beckwith}, S.~V.~W., {Glindemann}, A., {et~al.} 1996, \aj,
  111, 2403

\bibitem[{{Hogerheijde} {et~al.}(1995){Hogerheijde}, {Jansen}, \& {van
  Dishoeck}}]{1995A&A...294..792H}
{Hogerheijde}, M.~R., {Jansen}, D.~J., \& {van Dishoeck}, E.~F. 1995, \aap,
  294, 792

\bibitem[{{Hollenbach} \& {Tielens}(1997)}]{1997ARA&A..35..179H}
{Hollenbach}, D.~J. \& {Tielens}, A.~G.~G.~M. 1997, \araa, 35, 179

\bibitem[{{Hummer} \& {Storey}(1987)}]{1987MNRAS.224..801H}
{Hummer}, D.~G. \& {Storey}, P.~J. 1987, \mnras, 224, 801

\bibitem[{{Jansen} {et~al.}(1995){Jansen}, {Spaans}, {Hogerheijde}, \& {van
  Dishoeck}}]{1995A&A...303..541J}
{Jansen}, D.~J., {Spaans}, M., {Hogerheijde}, M.~R., \& {van Dishoeck}, E.~F.
  1995, \aap, 303, 541

\bibitem[{{Joblin} {et~al.}(2018){Joblin}, {Bron}, {Pinto}, {Pilleri}, {Le
  Petit}, {Gerin}, {Le Bourlot}, {Fuente}, {Berne}, {Goicoechea}, {Habart},
  {K{\"o}hler}, {Teyssier}, {Nagy}, {Montillaud}, {Vastel}, {Cernicharo},
  {R{\"o}llig}, {Ossenkopf-Okada}, \& {Bergin}}]{2018Joblin}
{Joblin}, C., {Bron}, E., {Pinto}, C., {et~al.} 2018, \aap, 615, A129

\bibitem[{{Kaplan} {et~al.}(2021){Kaplan}, {Dinerstein}, {Kim}, \&
  {Jaffe}}]{2021ApJ...919...27K}
{Kaplan}, K.~F., {Dinerstein}, H.~L., {Kim}, H., \& {Jaffe}, D.~T. 2021, \apj,
  919, 27

\bibitem[{{Kaplan} {et~al.}(2017){Kaplan}, {Dinerstein}, {Oh}, {Mace}, {Kim},
  {Sokal}, {Pavel}, {Lee}, {Pak}, {Park}, {Sok Oh}, \&
  {Jaffe}}]{2017ApJ...838..152K}
{Kaplan}, K.~F., {Dinerstein}, H.~L., {Oh}, H., {et~al.} 2017, \apj, 838, 152

\bibitem[{{Kassis} {et~al.}(2006){Kassis}, {Adams}, {Campbell}, {Deutsch},
  {Hora}, {Jackson}, \& {Tollestrup}}]{2006ApJ...637..823K}
{Kassis}, M., {Adams}, J.~D., {Campbell}, M.~F., {et~al.} 2006, \apj, 637, 823

\bibitem[{{Kaufman} {et~al.}(2006){Kaufman}, {Wolfire}, \&
  {Hollenbach}}]{2006ApJ...644..283K}
{Kaufman}, M.~J., {Wolfire}, M.~G., \& {Hollenbach}, D.~J. 2006, \apj, 644, 283

\bibitem[{{Koo} {et~al.}(2016){Koo}, {Raymond}, \& {Kim}}]{2016JKAS...49..109K}
{Koo}, B.-C., {Raymond}, J.~C., \& {Kim}, H.-J. 2016, Journal of Korean
  Astronomical Society, 49, 109

\bibitem[{Kramida {et~al.}(2022)Kramida, {Yu.~Ralchenko}, Reader, \& {and NIST
  ASD Team}}]{NIST_ASD}
Kramida, A., {Yu.~Ralchenko}, Reader, J., \& {and NIST ASD Team}. 2022, {NIST
  Atomic Spectra Database (ver. 5.10), [Online]. Available:
  {\tt{https://physics.nist.gov/asd}} [2023, March 9]. National Institute of
  Standards and Technology, Gaithersburg, MD.}

\bibitem[{{Labiano} {et~al.}(2021){Labiano}, {Argyriou},
  {{\'A}lvarez-M{\'a}rquez}, {Glasse}, {Glauser}, {Patapis}, {Law}, {Brandl},
  {Justtanont}, {Lahuis}, {Mart{\'\i}nez-Galarza}, {Mueller}, {Noriega-Crespo},
  {Royer}, {Shaughnessy}, \& {Vandenbussche}}]{2021A&A...656A..57L}
{Labiano}, A., {Argyriou}, I., {{\'A}lvarez-M{\'a}rquez}, J., {et~al.} 2021,
  \aap, 656, A57

\bibitem[{{Law} {et~al.}(2023){Law}, {E. Morrison}, {Argyriou}, {Patapis},
  {{\'A}lvarez-M{\'a}rquez}, {Labiano}, \&
  {Vandenbussche}}]{2023AJ....166...45L}
{Law}, D.~R., {E. Morrison}, J., {Argyriou}, I., {et~al.} 2023, \aj, 166, 45

\bibitem[{{Le Petit} {et~al.}(2006){Le Petit}, {Nehm{\'e}}, {Le Bourlot}, \&
  {Roueff}}]{2006ApJS..164..506L}
{Le Petit}, F., {Nehm{\'e}}, C., {Le Bourlot}, J., \& {Roueff}, E. 2006, \apjs,
  164, 506

\bibitem[{{Lutz} {et~al.}(1996){Lutz}, {Feuchtgruber}, {Genzel}, {Kunze},
  {Rigopoulou}, {Spoon}, {Wright}, {Egami}, {Katterloher}, {Sturm},
  {Wieprecht}, {Sternberg}, {Moorwood}, \& {de Graauw}}]{1996A&A...315L.269L}
{Lutz}, D., {Feuchtgruber}, H., {Genzel}, R., {et~al.} 1996, \aap, 315, L269

\bibitem[{{Mart{\'\i}n-Hern{\'a}ndez}
  {et~al.}(2002){Mart{\'\i}n-Hern{\'a}ndez}, {Vermeij}, {Tielens}, {van der
  Hulst}, \& {Peeters}}]{2002A&A...389..286M}
{Mart{\'\i}n-Hern{\'a}ndez}, N.~L., {Vermeij}, R., {Tielens}, A.~G.~G.~M., {van
  der Hulst}, J.~M., \& {Peeters}, E. 2002, \aap, 389, 286

\bibitem[{{Mathis} {et~al.}(1977){Mathis}, {Rumpl}, \&
  {Nordsieck}}]{1977ApJ...217..425M}
{Mathis}, J.~S., {Rumpl}, W., \& {Nordsieck}, K.~H. 1977, \apj, 217, 425

\bibitem[{{Menten} {et~al.}(2007){Menten}, {Reid}, {Forbrich}, \&
  {Brunthaler}}]{2007A&A...474..515M}
{Menten}, K.~M., {Reid}, M.~J., {Forbrich}, J., \& {Brunthaler}, A. 2007, \aap,
  474, 515

\bibitem[{{Osterbrock}(1962)}]{1962ApJ...135..195O}
{Osterbrock}, D.~E. 1962, \apj, 135, 195

\bibitem[{{Parmar} {et~al.}(1991){Parmar}, {Lacy}, \&
  {Achtermann}}]{1991ApJ...372L..25P}
{Parmar}, P.~S., {Lacy}, J.~H., \& {Achtermann}, J.~M. 1991, \apjl, 372, L25

\bibitem[{{Patapis} {et~al.}(2023){Patapis}, {Argyriou}, {Law}, {Glauser},
  {Glasse}, {Labiano}, {{\'A}lvarez-M{\'a}rquez}, {Kavanagh}, {Gasman},
  {Mueller}, {Larson}, {Vandenbussche}, {Klaassen}, {Guillard}, \&
  {Wright}}]{2023arXiv230701025P}
{Patapis}, P., {Argyriou}, I., {Law}, D.~R., {et~al.} 2023, arXiv e-prints,
  arXiv:2307.01025

\bibitem[{{Peeters} {et~al.}(2023){Peeters}, {Habart}, {Berne}, {Sidhu},
  {Chown}, {Van De Putte}, {Trahin}, {Schroetter}, {Canin}, {Alarcon},
  {Schefter}, {Khan}, {Pasquini}, {Tielens}, {Wolfire}, {Dartois},
  {Goicoechea}, {Maragkoudakis}, {Onaka}, {Pound}, {Vicente}, {Abergel},
  {Bergin}, {Bernard-Salas}, {Boersma}, {Bron}, {Cami}, {Cuadrado}, {Dicken},
  {Elyajour}, {Fuente}, {Gordon}, {Issa}, {Joblin}, {Kannavou}, {Lacinbala},
  {Languignon}, {Le Gal}, {Meshaka}, {Okada}, {Robberto}, {Roellig},
  {Schirmer}, {Tabone}, {Zannese}, {Aleman}, {Allamandola}, {Auchettl},
  {Baratta}, {Bejaoui}, {Bera}, {Black}, {Boulanger}, {Bouwman}, {Brandl},
  {Brechignac}, {Brunken}, {Buragohain}, {Burkhardt}, {Candian}, {Cazaux},
  {Cernicharo}, {Chabot}, {Chakraborty}, {Champion}, {Colgan}, {Cooke},
  {Coutens}, {Cox}, {Demyk}, {Donovan Meyer}, {Foschino}, {Garcia-Lario},
  {Gerin}, {Gottlieb}, {Guillard}, {Gusdorf}, {Hartigan}, {He}, {Herbst},
  {Hornekaer}, {Jager}, {Janot-Pacheco}, {Kaufman}, {Kendrew}, {Kirsanova},
  {Klaassen}, {Kwok}, {Labiano}, {Lai}, {Lee}, {Lefloch}, {Le Petit}, {Li},
  {Linz}, {Mackie}, {Madden}, {Mascetti}, {McGuire}, {Merino}, {Micelotta},
  {Misselt}, {Morse}, {Mulas}, {Neelamkodan}, {Ohsawa}, {Paladini}, {Palumbo},
  {Pathak}, {Pendleton}, {Petrignani}, {Pino}, {Puga}, {Rangwala}, {Rapacioli},
  {Ricca}, {Roman-Duval}, {Roser}, {Roueff}, {Rouille}, {Salama}, {Sales},
  {Sandstrom}, {Sarre}, {Sciamma-O'Brien}, {Sellgren}, {Shenoy}, {Teyssier},
  {Thomas}, {Togi}, {Verstraete}, {Witt}, {Wootten}, {Ysard}, {Zettergren},
  {Zhang}, {Zhang}, \& {Zhen}}]{Peeters:nirspec}
{Peeters}, E., {Habart}, E., {Berne}, O., {et~al.} 2023, arXiv e-prints,
  arXiv:2310.08720

\bibitem[{{Pellegrini} {et~al.}(2009){Pellegrini}, {Baldwin}, {Ferland},
  {Shaw}, \& {Heathcote}}]{2009ApJ...693..285P}
{Pellegrini}, E.~W., {Baldwin}, J.~A., {Ferland}, G.~J., {Shaw}, G., \&
  {Heathcote}, S. 2009, \apj, 693, 285

\bibitem[{{Pound} \& {Wolfire}(2023)}]{2023AJ....165...25P}
{Pound}, M.~W. \& {Wolfire}, M.~G. 2023, \aj, 165, 25

\bibitem[{{Prozesky} \& {Smits}(2018)}]{2018MNRAS.478.2766P}
{Prozesky}, A. \& {Smits}, D.~P. 2018, \mnras, 478, 2766

\bibitem[{{Rosenthal} {et~al.}(2000){Rosenthal}, {Bertoldi}, \&
  {Drapatz}}]{2000A&A...356..705R}
{Rosenthal}, D., {Bertoldi}, F., \& {Drapatz}, S. 2000, \aap, 356, 705

\bibitem[{{Rubin} {et~al.}(2007){Rubin}, {Simpson}, {Colgan}, {Dufour}, {Ray},
  {Erickson}, {Haas}, {Pauldrach}, \& {Citron}}]{2007MNRAS.377.1407R}
{Rubin}, R.~H., {Simpson}, J.~P., {Colgan}, S. W.~J., {et~al.} 2007, \mnras,
  377, 1407

\bibitem[{{Rubin} {et~al.}(2011){Rubin}, {Simpson}, {O'Dell}, {McNabb},
  {Colgan}, {Zhuge}, {Ferland}, \& {Hidalgo}}]{2011MNRAS.410.1320R}
{Rubin}, R.~H., {Simpson}, J.~P., {O'Dell}, C.~R., {et~al.} 2011, \mnras, 410,
  1320

\bibitem[{{Shaw} {et~al.}(2009){Shaw}, {Ferland}, {Henney}, {Stancil}, {Abel},
  {Pellegrini}, {Baldwin}, \& {van Hoof}}]{2009ApJ...701..677S}
{Shaw}, G., {Ferland}, G.~J., {Henney}, W.~J., {et~al.} 2009, \apj, 701, 677

\bibitem[{{Sheffer} {et~al.}(2011){Sheffer}, {Wolfire}, {Hollenbach},
  {Kaufman}, \& {Cordier}}]{2011ApJ...741...45S}
{Sheffer}, Y., {Wolfire}, M.~G., {Hollenbach}, D.~J., {Kaufman}, M.~J., \&
  {Cordier}, M. 2011, \apj, 741, 45

\bibitem[{{Storey} \& {Hummer}(1995)}]{1995MNRAS.272...41S}
{Storey}, P.~J. \& {Hummer}, D.~G. 1995, \mnras, 272, 41

\bibitem[{{Tielens} {et~al.}(1993){Tielens}, {Meixner}, {van der Werf},
  {Bregman}, {Tauber}, {Stutzki}, \& {Rank}}]{1993Sci...262...86T}
{Tielens}, A.~G.~G.~M., {Meixner}, M.~M., {van der Werf}, P.~P., {et~al.} 1993,
  Science, 262, 86

\bibitem[{{Togi} \& {Smith}(2016)}]{2016ApJ...830...18T}
{Togi}, A. \& {Smith}, J.~D.~T. 2016, \apj, 830, 18

\bibitem[{{van der Werf} {et~al.}(1996){van der Werf}, {Stutzki}, {Sternberg},
  \& {Krabbe}}]{1996A&A...313..633V}
{van der Werf}, P.~P., {Stutzki}, J., {Sternberg}, A., \& {Krabbe}, A. 1996,
  \aap, 313, 633

\bibitem[{{van Hoof}(2018)}]{2018Galax...6...63V}
{van Hoof}, P. A.~M. 2018, Galaxies, 6, 63

\bibitem[{{Verma} {et~al.}(2003){Verma}, {Lutz}, {Sturm}, {Sternberg},
  {Genzel}, \& {Vacca}}]{2003A&A...403..829V}
{Verma}, A., {Lutz}, D., {Sturm}, E., {et~al.} 2003, \aap, 403, 829

\bibitem[{{Walmsley} {et~al.}(2000){Walmsley}, {Natta}, {Oliva}, \&
  {Testi}}]{2000A&A...364..301W}
{Walmsley}, C.~M., {Natta}, A., {Oliva}, E., \& {Testi}, L. 2000, \aap, 364,
  301

\bibitem[{{Wolfire} {et~al.}(1990){Wolfire}, {Tielens}, \&
  {Hollenbach}}]{1990ApJ...358..116W}
{Wolfire}, M.~G., {Tielens}, A.~G.~G.~M., \& {Hollenbach}, D. 1990, \apj, 358,
  116

\bibitem[{{Wolfire} {et~al.}(2022){Wolfire}, {Vallini}, \&
  {Chevance}}]{2022ARA&A..60..247W}
{Wolfire}, M.~G., {Vallini}, L., \& {Chevance}, M. 2022, \araa, 60, 247

\bibitem[{{Young Owl} {et~al.}(2000){Young Owl}, {Meixner}, {Wolfire},
  {Tielens}, \& {Tauber}}]{2000ApJ...540..886Y}
{Young Owl}, R.~C., {Meixner}, M.~M., {Wolfire}, M., {Tielens}, A.~G.~G.~M., \&
  {Tauber}, J. 2000, \apj, 540, 886

\end{thebibliography}

\setlength{\tabcolsep}{4pt}
\onecolumn
\begin{landscape}
\begin{longtable}{ll rl rrr rl rl rl rl rl}
\caption{Identification, measured wavelength offset, and intensity measurement\tablefootmark{a} $I$ with uncertainty\tablefootmark{b} $\sigma$ of the most prominent lines in each region.}
\label{tab:lines}\\
\hline\hline
Species & Transition &  $\lambda_{\text{rest}}$ &  $\Delta\lambda_{\text{obs}}$ &      $A$ &  $E_l$ &  $E_u$ & $I_\text{HII}$ & $\sigma_\text{HII}$ & $I_\text{Atomic}$ & $\sigma_\text{Atomic}$ & $I_\text{DF1}$ & $\sigma_\text{DF1}$ & $I_\text{DF2}$ & $\sigma_\text{DF2}$ & $I_\text{DF3}$ & $\sigma_\text{DF3}$ \\
        &            & (\mum) & ($10^{-4}$\mum)     & (\si{\per\s}) & (\si{\per\cm}) & (\si{\per\cm}) &  \multicolumn{10}{|c}{(\num{e-5} \lineunit)}\\
\hline
\endfirsthead
\caption{Continued.}\\
\hline\hline
Species & Transition &  $\lambda_{\text{rest}}$ &  $\Delta\lambda_{\text{obs}}$ &      $A$ &  $E_l$ &  $E_u$ & $I_\text{HII}$ & $\sigma_\text{HII}$ & $I_\text{Atomic}$ & $\sigma_\text{Atomic}$ & $I_\text{DF1}$ & $\sigma_\text{DF1}$ & $I_\text{DF2}$ & $\sigma_\text{DF2}$ & $I_\text{DF3}$ & $\sigma_\text{DF3}$ \\
        &            & (\mum) & ($10^{-4}$\mum)     & (\si{\per\s}) & (\si{\per\cm}) & (\si{\per\cm}) &  \multicolumn{10}{|c}{(\num{e-5} \lineunit)}\\
\hline
\endhead
\endfoot
H I & 7-23 & 4.9237 & 2 & 3.62e+02 & 107440 & 109471 & 1.85 & 0.10 & 1.16 & 0.17 & 0.58 & 0.06 & 0.55 & 0.32 & 0.36 & 0.25 \\
UID & -- & 4.9280 & 0 & -- & -- & -- & -- & -- & -- & -- & 0.23 & 0.17 & 0.46 & 0.34 & 0.42 & 0.22 \\
H$_2$ & 1-1 $S(9)$ & 4.9541 & 3 & 4.37e-07 & 8908 & 10927 & 0.69 & 0.11 & 0.78 & 0.12 & 1.54 & 0.09 & 2.72 & 0.10 & 2.38 & 0.09 \\
UID & -- & 4.9680 & 0 & -- & -- & -- & -- & -- & -- & -- & 0.20 & 0.22 & 0.28 & 0.14 & 0.28 & 0.14 \\
H I & 7-22 & 4.9709 & 2 & 4.57e+02 & 107440 & 109452 & 2.13 & 0.01 & 1.70 & 0.06 & 0.71 & 0.13 & 0.38 & 0.05 & 0.39 & 0.14 \\
H I & 7-21 & 5.0261 & 1 & 5.82e+02 & 107440 & 109430 & 2.50 & 0.09 & 1.93 & 0.19 & 0.73 & 0.02 & 0.56 & 0.15 & 0.78 & 0.10 \\
H$_2$ & 0-0 $S(8)$ & 5.0531 & 3 & 3.24e-07 & 4052 & 6031 & 2.48 & 0.06 & 2.35 & 0.09 & 4.66 & 0.10 & 7.12 & 0.05 & 7.38 & 0.17 \\
H I & 7-20 & 5.0913 & 2 & 7.52e+02 & 107440 & 109404 & 2.75 & 0.07 & 2.02 & 0.18 & 0.89 & 0.11 & 0.82 & 0.15 & 0.63 & 0.22 \\
H I & 6-10 & 5.1287 & 2 & 3.69e+04 & 106632 & 108582 & 33.66 & 0.13 & 25.58 & 0.35 & 10.24 & 0.13 & 8.92 & 0.13 & 7.46 & 0.16 \\
H I & 7-19 & 5.1693 & 1 & 9.85e+02 & 107440 & 109375 & 3.45 & 0.08 & 2.21 & 0.19 & 1.02 & 0.27 & 0.82 & 0.19 & 0.97 & 0.30 \\
UID & -- & 5.2400 & 0 & -- & -- & -- & 3.76 & 0.11 & 2.00 & 0.26 & -- & -- & -- & -- & -- & -- \\
H I & 7-18 & 5.2637 & 1 & 1.31e+03 & 107440 & 109340 & 3.60 & 0.21 & 2.66 & 0.23 & 1.02 & 0.04 & 1.10 & 0.13 & 0.87 & 0.05 \\
H$_2$ & 1-1 $S(8)$ & 5.3300 & 2 & 2.92e-07 & 8008 & 9884 & -- & -- & -- & -- & 0.35 & 0.23 & 0.91 & 0.03 & 0.78 & 0.07 \\
Fe II & $a^6D\! -\! a^4F$ 9/2 - 9/2 & 5.3402 & 3 & -- & 0 & 1873 & 20.43 & 0.02 & 50.52 & 0.40 & 22.48 & 0.53 & 16.45 & 0.18 & 6.76 & 0.11 \\
H I & 7-17 & 5.3798 & 3 & 1.78e+03 & 107440 & 109299 & 4.62 & 0.10 & 3.18 & 0.22 & 1.29 & 0.20 & 1.16 & 0.07 & 1.00 & 0.10 \\
H$_2$ & 0-0 $S(7)$ & 5.5112 & 4 & 2.00e-07 & 3187 & 5002 & 11.16 & 0.45 & 11.43 & 0.23 & 27.17 & 0.13 & 42.13 & 0.11 & 39.45 & 0.26 \\
H I & 7-16 & 5.5252 & 2 & 2.47e+03 & 107440 & 109250 & 5.40 & 0.16 & 4.23 & 0.12 & 1.83 & 0.13 & 1.84 & 0.02 & 1.35 & 0.04 \\
UID & -- & 5.5515 & 0 & -- & -- & -- & -- & -- & -- & -- & 0.76 & 0.18 & 1.78 & 0.07 & -- & 0.07 \\
UID & -- & 5.5960 & 0 & -- & -- & -- & 1.33 & 0.19 & 0.14 & 0.21 & -- & -- & -- & -- & -- & -- \\
UID & -- & 5.6280 & 0 & -- & -- & -- & 0.67 & 0.14 & -- & -- & -- & -- & -- & -- & -- & -- \\
UID & -- & 5.6340 & 0 & -- & -- & -- & -- & -- & -- & -- & 0.19 & 0.31 & 0.46 & 0.27 & 0.20 & 0.11 \\
UID & -- & 5.6580 & 0 & -- & -- & -- & 0.83 & 0.06 & 1.72 & 0.21 & 0.57 & 0.21 & 0.35 & 0.24 & 0.31 & 0.08 \\
H I & 7-15 & 5.7115 & 1 & 3.52e+03 & 107440 & 109191 & 6.48 & 0.22 & 4.46 & 0.35 & 1.74 & 0.27 & 1.80 & 0.15 & 1.51 & 0.09 \\
UID & -- & 5.7400 & 0 & -- & -- & -- & 3.17 & 0.27 & 1.76 & 0.33 & 0.46 & 0.28 & 0.58 & 0.14 & 0.49 & 0.09 \\
H$_2$ & 1-1 $S(7)$ & 5.8109 & 4 & 1.82e-07 & 7188 & 8908 & 0.42 & 0.11 & 0.07 & 0.32 & 1.07 & 0.30 & 2.64 & 0.10 & 2.10 & 0.16 \\
H I & 6-9 & 5.9082 & 1 & 7.06e+04 & 106632 & 108325 & 45.15 & 0.62 & 34.98 & 0.78 & 13.38 & 0.46 & 11.73 & 0.32 & 9.92 & 0.14 \\
H I & 7-14 & 5.9568 & 1 & 5.16e+03 & 107440 & 109119 & 8.14 & 0.12 & 6.43 & 0.23 & 2.54 & 0.16 & 2.05 & 0.20 & 1.75 & 0.07 \\
H$_2$ & 0-0 $S(6)$ & 6.1086 & 4 & 1.14e-07 & 2415 & 4052 & 5.67 & 0.32 & 5.63 & 0.57 & 15.81 & 0.88 & 25.42 & 0.30 & 22.03 & 0.23 \\
H I & 7-13 & 6.2919 & 2 & 7.85e+03 & 107440 & 109030 & 9.39 & 0.71 & 5.50 & 1.50 & 1.78 & 0.38 & 2.01 & 0.31 & 1.85 & 0.23 \\
H I & 8-29 & 6.3159 & 4 & 9.84e+01 & 107965 & 109548 & 0.56 & 0.60 & -- & -- & -- & -- & -- & -- & -- & -- \\
H I & 8-28 & 6.3540 & 2 & 1.18e+02 & 107965 & 109539 & 0.81 & 0.57 & -- & -- & -- & -- & -- & -- & -- & -- \\
UID & -- & 6.3660 & 0 & -- & -- & -- & 0.59 & 0.16 & -- & -- & -- & -- & -- & -- & -- & -- \\
H I & 8-27 & 6.3969 & 0 & 1.42e+02 & 107965 & 109528 & 0.79 & 0.10 & -- & -- & -- & -- & -- & -- & -- & -- \\
UID & -- & 6.4120 & 0 & -- & -- & -- & -- & -- & -- & -- & 1.77 & 0.64 & 2.13 & 0.38 & 0.82 & 0.20 \\
H$_2$ & 1-1 $S(6)$ & 6.4383 & 3 & 1.05e-07 & 6454 & 8008 & 0.24 & 0.14 & 0.19 & 0.25 & 0.09 & 0.63 & 0.32 & 0.23 & 0.30 & 0.14 \\
H I & 8-26 & 6.4455 & 2 & 1.73e+02 & 107965 & 109516 & 0.84 & 0.03 & -- & -- & -- & -- & -- & -- & -- & -- \\
UID & -- & 6.4976 & 0 & -- & -- & -- & 0.67 & 0.28 & 2.60 & 0.41 & -- & -- & -- & -- & -- & -- \\
H I & 8-25 & 6.5010 & 1 & 2.12e+02 & 107965 & 109503 & 0.93 & 0.12 & 1.11 & 0.40 & 0.38 & 0.08 & 0.14 & 0.10 & 0.26 & 0.07 \\
H I & 8-24 & 6.5647 & 2 & 2.63e+02 & 107965 & 109488 & 1.23 & 0.15 & 0.82 & 0.16 & 0.05 & 0.11 & 0.12 & 0.10 & 0.13 & 0.10 \\
Ni II & $^2D\! -\! ^2D$ 5/2 - 3/2 & 6.6360 & 3 & 5.54e-02 & 0 & 1507 & 3.70 & 0.22 & 17.96 & 0.83 & 5.57 & 0.67 & 4.65 & 0.48 & 1.44 & 0.26 \\
H I & 8-23 & 6.6384 & -10 & 3.28e+02 & 107965 & 109471 & 0.78 & 0.34 & -- & -- & -- & -- & -- & -- & -- & -- \\
He I & multiple & 6.7217 & 1 & -- & 195114 & 196601 & 1.30 & 0.15 & 3.23 & 0.26 & 1.69 & 0.17 & 1.15 & 0.09 & 0.54 & 0.02 \\
H I & 8-22 & 6.7245 & 4 & 4.15e+02 & 107965 & 109452 & 1.38 & 0.10 & 1.14 & 0.07 & 0.48 & 0.14 & 0.37 & 0.09 & 0.40 & 0.03 \\
H I & 7-12 & 6.7720 & 2 & 1.25e+04 & 107440 & 108917 & 12.82 & 0.20 & 9.30 & 0.37 & 3.23 & 0.35 & 2.66 & 0.30 & 2.41 & 0.21 \\
H I & 8-21 & 6.8259 & 2 & 5.31e+02 & 107965 & 109430 & 1.34 & 0.16 & 0.26 & 0.55 & -- & -- & -- & -- & -- & -- \\
UID & -- & 6.8590 & 0 & -- & -- & -- & -- & -- & 3.38 & 0.40 & -- & -- & -- & -- & -- & -- \\
H$_2$ & 0-0 $S(5)$ & 6.9095 & 3 & 5.88e-08 & 1740 & 3187 & 29.93 & 0.43 & 29.75 & 0.83 & 98.77 & 1.11 & 161.79 & 1.11 & 137.70 & 1.08 \\
H I & 8-20 & 6.9468 & 0 & 6.89e+02 & 107965 & 109404 & 2.18 & 1.57 & 1.38 & 1.89 & 0.31 & 0.76 & 0.28 & 0.58 & 0.57 & 0.64 \\
Ar II & $^2P^o\! -\! ^2P^o$ 3/2 - 1/2 & 6.9853 & 4 & 4.23e-02 & 0 & 1432 & 1105.30 & 10.23 & 1423.34 & 8.85 & 174.53 & 1.25 & 167.85 & 0.99 & 147.86 & 0.94 \\
H I & 8-19 & 7.0927 & 3 & 9.07e+02 & 107965 & 109375 & 2.41 & 0.22 & -- & -- & -- & -- & -- & -- & -- & -- \\
H I & 8-18 & 7.2717 & 4 & 1.22e+03 & 107965 & 109340 & 2.62 & 0.13 & 1.09 & 0.53 & 0.09 & 0.22 & 0.18 & 0.29 & 0.20 & 0.20 \\
H$_2$ & 1-1 $S(5)$ & 7.2801 & 4 & 5.44e-08 & 5814 & 7187 & 0.16 & 0.19 & 0.12 & 0.79 & 0.40 & 0.31 & 0.91 & 0.43 & 0.66 & 0.28 \\
Ni III & $^3F\! -\! ^3F$ 4-3 & 7.3492 & 35 & 6.50e-02 & 0 & 1361 & 12.66 & 0.19 & 7.54 & 0.52 & 3.53 & 0.38 & 2.78 & 0.21 & 2.59 & 0.21 \\
He I & $^3D\! -\! ^3F^o$ multiplet & 7.4334 & 3 & -- & 193917 & 195262 & 0.40 & 0.02 & -- & -- & -- & -- & -- & -- & -- & -- \\
H I & 5-6 & 7.4599 & 2 & 1.02e+06 & 105292 & 106632 & 237.55 & 0.71 & 185.67 & 1.39 & 69.24 & 0.63 & 61.44 & 0.23 & 50.09 & 0.13 \\
H I & 8-17 & 7.4951 & 4 & 1.66e+03 & 107965 & 109299 & 1.48 & 0.52 & -- & -- & -- & -- & -- & -- & -- & -- \\
H I & 6-8 & 7.5025 & 4 & 1.56e+05 & 106632 & 107965 & 64.75 & 0.49 & 48.65 & 1.36 & 18.14 & 0.74 & 16.22 & 0.42 & 13.29 & 0.22 \\
H I & 7-11 & 7.5081 & 3 & 2.12e+04 & 107440 & 108772 & 15.37 & 0.81 & 11.08 & 3.11 & 3.95 & 2.03 & 3.73 & 1.47 & 3.09 & 0.60 \\
H I & 8-16 & 7.7804 & 2 & 2.33e+03 & 107965 & 109250 & 4.21 & 0.28 & 5.03 & 1.20 & -- & -- & -- & -- & -- & -- \\
UID & -- & 7.8780 & 0 & -- & -- & -- & 2.54 & 0.41 & -- & -- & -- & -- & -- & -- & -- & -- \\
H$_2$ & 0-0 $S(4)$ & 8.0250 & 5 & 2.64e-08 & 1169 & 2415 & 16.95 & 0.16 & 16.26 & 0.64 & 57.39 & 0.47 & 96.98 & 0.68 & 85.43 & 0.34 \\
H I & 8-15 & 8.1549 & 4 & 3.36e+03 & 107965 & 109191 & 4.85 & 0.13 & 3.45 & 0.61 & 1.36 & 0.36 & 1.31 & 0.37 & 0.88 & 0.18 \\
H I & 9-27 & 8.3084 & 0 & 1.30e+02 & 108325 & 109528 & 0.92 & 0.15 & -- & -- & -- & -- & -- & -- & -- & -- \\
H I & 9-26 & 8.3907 & 4 & 1.59e+02 & 108325 & 109516 & 0.54 & 0.11 & -- & -- & -- & -- & -- & -- & -- & -- \\
UID & -- & 8.4095 & 0 & -- & -- & -- & 0.76 & 0.31 & -- & -- & -- & -- & -- & -- & -- & -- \\
H I & 9-25 & 8.4849 & 3 & 1.95e+02 & 108325 & 109503 & 0.57 & 0.40 & -- & -- & -- & -- & -- & -- & -- & -- \\
H I & 9-24 & 8.5938 & -4 & 2.42e+02 & 108325 & 109488 & 1.04 & 0.48 & -- & -- & -- & -- & -- & -- & -- & -- \\
H I & 8-14 & 8.6645 & 3 & 5.01e+03 & 107965 & 109119 & 4.38 & 0.55 & 2.75 & 0.95 & -- & 0.77 & 0.19 & 0.53 & 0.26 & 0.44 \\
H I & 9-23 & 8.7206 & 3 & 3.04e+02 & 108325 & 109471 & 0.74 & 0.33 & -- & -- & -- & -- & -- & -- & -- & -- \\
H I & 7-10 & 8.7601 & 3 & 3.90e+04 & 107440 & 108582 & 24.65 & 0.53 & 18.23 & 0.54 & 8.08 & 0.66 & 6.76 & 0.65 & 5.25 & 0.35 \\
H I & 9-22 & 8.8697 & 5 & 3.85e+02 & 108325 & 109452 & 0.72 & 0.24 & -- & -- & -- & -- & -- & -- & -- & -- \\
Ar III & $^3P\! -\! ^3P$ 2 - 1 & 8.9914 & -1 & 3.10e-02 & 0 & 1112 & 1466.00 & 1.67 & 795.10 & 1.49 & 489.49 & 0.92 & 413.35 & 0.68 & 333.44 & 0.42 \\
H I & 9-21 & 9.0470 & 4 & 4.95e+02 & 108325 & 109430 & 0.76 & 0.65 & 0.40 & 0.63 & -- & -- & -- & -- & -- & -- \\
H I & 9-20 & 9.2605 & 1 & 6.45e+02 & 108325 & 109404 & 0.81 & 0.20 & 0.58 & 0.07 & 0.47 & 0.11 & 0.43 & 0.06 & 0.30 & 0.08 \\
H I & 8-13 & 9.3920 & 4 & 7.80e+03 & 107965 & 109030 & 7.22 & 0.29 & 5.24 & 0.08 & 2.31 & 0.14 & 2.09 & 0.15 & 1.64 & 0.13 \\
H I & 9-19 & 9.5217 & 4 & 8.56e+02 & 108325 & 109375 & 0.83 & 0.56 & 0.68 & 0.41 & -- & -- & -- & -- & -- & -- \\
H$_2$ & 0-0 $S(3)$ & 9.6649 & 8 & 9.84e-09 & 705 & 1740 & 38.46 & 0.21 & 22.03 & 0.47 & 63.15 & 0.32 & 199.96 & 0.16 & 148.41 & 0.17 \\
H I & 9-18 & 9.8470 & 9 & 1.16e+03 & 108325 & 109340 & 1.42 & 0.31 & 0.99 & 0.29 & -- & -- & -- & -- & -- & -- \\
H I & 9-17 & 10.2613 & 5 & 1.60e+03 & 108325 & 109299 & 2.60 & 0.25 & 1.91 & 0.33 & 0.84 & 0.18 & 0.67 & 0.12 & 0.59 & 0.13 \\
UID & -- & 10.3845 & 0 & -- & -- & -- & 1.40 & 0.16 & 0.46 & 0.19 & 0.07 & 0.20 & 0.29 & 0.08 & 0.08 & 0.11 \\
S IV & $^2P^o\! -\! ^2P^o$ 1/2 - 3/2 & 10.5105 & 2 & 7.30e-03 & 0 & 951 & 492.10 & 0.56 & 413.84 & 0.78 & 271.81 & 0.56 & 251.25 & 0.39 & 190.24 & 0.30 \\
Ni II & $^4F\! -\! ^4F$ 9/2 - 7/2 & 10.6822 & 8 & 2.71e-02 & 8394 & 9330 & 2.20 & 0.40 & 3.97 & 0.32 & 0.58 & 0.27 & 0.54 & 0.11 & 0.47 & 0.17 \\
H I & 9-16 & 10.8036 & 4 & 2.27e+03 & 108325 & 109250 & 2.79 & 0.20 & 1.98 & 0.28 & 0.57 & 0.20 & 0.50 & 0.11 & 0.54 & 0.05 \\
He I & $^3S\! -\! ^3P^o$ multiplet & 10.8820 & 5 & -- & -- & -- & 0.62 & 0.22 & 0.51 & 0.59 & -- & -- & -- & -- & -- & -- \\
Ni III & $^3F\! -\! ^3F$ 3 - 2 & 11.0023 & 14 & 2.70e-02 & 1361 & 2270 & 2.17 & 0.80 & 1.96 & 2.93 & 0.89 & 1.96 & 0.55 & 0.94 & 0.52 & 0.48 \\
H I & 7-9 & 11.3087 & 6 & 8.24e+04 & 107440 & 108325 & 29.05 & 1.69 & 22.42 & 2.75 & 9.41 & 2.17 & 7.86 & 1.15 & 6.27 & 1.06 \\
H I & 9-15 & 11.5395 & 6 & 3.32e+03 & 108325 & 109191 & 2.76 & 0.29 & 1.15 & 0.81 & 0.14 & 0.40 & 0.15 & 0.27 & 0.17 & 0.20 \\
H I & 10-20 & 12.1568 & 2 & 6.17e+02 & 108582 & 109404 & 0.99 & 0.19 & -- & -- & -- & -- & -- & -- & -- & -- \\
H$_2$ & 0-0 $S(2)$ & 12.2786 & 7 & 2.76e-09 & 354 & 1169 & 24.67 & 0.24 & 20.48 & 0.26 & 56.56 & 0.15 & 107.33 & 0.12 & 89.38 & 0.05 \\
H I & 6-7 & 12.3719 & 9 & 4.56e+05 & 106632 & 107440 & 92.29 & 0.42 & 68.00 & 0.38 & 27.00 & 0.40 & 23.62 & 0.31 & 19.23 & 0.23 \\
H I & 8-11 & 12.3872 & 9 & 2.30e+04 & 107965 & 108772 & 10.66 & 0.47 & 7.55 & 0.72 & 3.22 & 0.59 & 3.10 & 0.40 & 2.25 & 0.29 \\
H I & 9-14 & 12.5871 & 8 & 5.08e+03 & 108325 & 109119 & 4.29 & 0.06 & 2.92 & 0.27 & 0.75 & 0.08 & 0.99 & 0.15 & 0.82 & 0.02 \\
H I & 10-19 & 12.6110 & 6 & 8.25e+02 & 108582 & 109375 & 1.35 & 0.26 & -- & -- & -- & -- & -- & -- & -- & -- \\
Ne II & $^2P^o\! -\! ^2P^o$ 3/2 - 1/2 & 12.8135 & 7 & 8.32e-03 & 0 & 780 & 6156.40 & 0.79 & 4612.99 & 1.96 & 1422.69 & 1.40 & 1188.46 & 1.42 & 1013.34 & 0.65 \\
H I & 10-18 & 13.1880 & 5 & 1.13e+03 & 108582 & 109340 & 1.47 & 0.23 & -- & -- & -- & -- & -- & -- & -- & -- \\
H I & 10-17 & 13.9418 & -1 & 1.58e+03 & 108582 & 109299 & 1.48 & 0.20 & -- & -- & -- & -- & -- & -- & -- & -- \\
H I & 9-13 & 14.1831 & 4 & 8.19e+03 & 108325 & 109030 & 5.02 & 0.31 & 3.14 & 0.82 & 0.95 & 0.50 & 0.87 & 0.29 & 0.70 & 0.18 \\
UID & -- & 14.3330 & 0 & -- & -- & -- & 0.93 & 0.49 & 0.90 & 0.59 & 0.50 & 0.32 & 0.56 & 0.36 & 0.30 & 0.12 \\
Cl II & $^3P\! -\! ^3P$ 2 - 1 & 14.3678 & 3 & 7.62e-03 & 0 & 696 & 20.66 & 0.61 & 82.39 & 0.68 & 36.86 & 0.35 & 30.19 & 0.30 & 10.67 & 0.22 \\
H I & 10-16 & 14.9623 & 5 & 2.28e+03 & 108582 & 109250 & 2.11 & 0.22 & 1.51 & 0.40 & -- & -- & -- & -- & -- & -- \\
Ne III & $^3P\! -\! ^3P$ 2 - 1 & 15.5550 & 10 & 5.85e-03 & 0 & 643 & 2039.97 & 0.74 & 1468.13 & 1.15 & 874.50 & 0.72 & 759.39 & 0.57 & 582.45 & 0.43 \\
H I & 8-10 & 16.2091 & 0 & 4.68e+04 & 107965 & 108582 & 16.53 & 0.55 & 10.92 & 0.89 & 4.44 & 0.73 & 4.50 & 0.53 & 3.33 & 0.41 \\
H I & 10-15 & 16.4117 & -2 & 3.42e+03 & 108582 & 109191 & 4.28 & 1.01 & 6.84 & 5.23 & -- & -- & -- & -- & -- & -- \\
H I & 9-12 & 16.8806 & 12 & 1.43e+04 & 108325 & 108917 & 6.16 & 1.09 & 4.15 & 1.25 & 1.73 & 0.59 & 1.53 & 0.64 & 1.16 & 0.49 \\
H$_2$ & 0-0 $S(1)$ & 17.0348 & 19 & 4.76e-10 & 118 & 706 & 24.97 & 1.00 & 19.89 & 1.60 & 46.92 & 1.29 & 114.43 & 1.05 & 119.29 & 0.70 \\
P III & $^2P^o\! -\! ^2P^o$ 1/2 - 3/2 & 17.8846 & 4 & 1.47e-03 & 0 & 559 & 39.80 & 0.22 & 26.08 & 1.44 & 14.54 & 1.15 & 13.53 & 0.90 & 10.63 & 0.87 \\
Fe II & $a^4F\! -\! a^4F$ 9/2 - 7/2 & 17.9360 & -0 & 5.84e-03 & 1873 & 2430 & 15.17 & 0.93 & 30.99 & 0.22 & 6.09 & 0.25 & 5.05 & 0.35 & 3.02 & 0.16 \\
UID & -- & 18.6510 & 0 & -- & -- & -- & -- & -- & -- & -- & 7.69 & 0.98 & 9.40 & 0.73 & 7.54 & 0.66 \\
S III & $^3P\! -\! ^3P$ 1 - 2 & 18.7130 & -1 & 1.43e-03 & 299 & 833 & 5996.73 & 1.28 & 4127.26 & 1.75 & 1884.99 & 1.39 & 1681.10 & 0.73 & 1297.20 & 0.96 \\
H I & 7-8 & 19.0619 & 4 & 2.27e+05 & 107440 & 107965 & 46.39 & 2.04 & 35.43 & 3.73 & 14.22 & 1.88 & 12.01 & 1.30 & 9.97 & 1.11 \\
Ar III & $^3P\! -\! ^3P$ 1 - 0 & 21.8291 & -2 & 5.19e-03 & 1112 & 1570 & 94.13 & 1.38 & 48.27 & 2.02 & 27.58 & 0.68 & 24.18 & 0.60 & 19.10 & 0.70 \\
H I & 9-11 & 22.3405 & -26 & 2.81e+04 & 108325 & 108772 & 13.28 & 0.68 & -- & -- & -- & -- & -- & -- & -- & -- \\
Fe III & $^5D\! -\! ^5D$ 4 - 3 & 22.9250 & 15 & 2.80e-03 & 0 & 436 & 141.05 & 3.71 & 88.91 & 3.20 & 39.08 & 2.11 & 29.57 & 1.23 & 25.94 & 1.21 \\
S I & $^3P\! -\! ^3P$ 2 - 1 & 25.2490 & 44 & 1.40e-03 & 0 & 396 & -- & -- & -- & -- & 6.64 & 4.04 & 7.57 & 3.49 & 16.38 & 1.94 \\
Fe II & $a^6D\! -\! a^6D$ 9/2 - 7/2 & 25.9884 & 70 & 2.13e-03 & 0 & 385 & 55.80 & 10.65 & 298.31 & 33.29 & 158.19 & 14.68 & 152.99 & 11.69 & 37.46 & 8.31 \\
\\
\hline
\end{longtable}
\tablefoot{%
    \tablefoottext{a}{Intensities omitted where line is not present or too weak for  reliable measurement.}
    \tablefoottext{b}{Reported uncertainties $\sigma$ are based on local noise measurement only (Fig.~\ref{fig:unc}), suitable to determine S/N.
    We suggest a 3\% fractional uncertainty to represent systematic calibration uncertainties. See Sect.~\ref{sec:extraction} for details.}
}
\end{landscape}

\end{document}